\documentclass[a4paper,fleqn]{cas-dc}

\usepackage{soul} 
\usepackage{color, xcolor} 
\usepackage[numbers]{natbib}
\usepackage{tabularx}
\usepackage{bbding}
\usepackage{makecell}
\usepackage{amsmath} 
\usepackage{amssymb} 
\usepackage{amsfonts} 

\usepackage{hyperref}
\hypersetup{
    colorlinks=true,
    linkcolor=blue,
    urlcolor=blue,
    anchorcolor=blue,
    citecolor=blue
}

\usepackage[switch]{lineno}

\begin{document}
\shorttitle{Digital Fingerprinting on Multimedia: A Survey} 
\shortauthors{W. Chen \textit{et al.}}

\title [mode = title]{Digital Fingerprinting on Multimedia: A Survey}   
\author[1]{Wendi Chen}
\ead{wendichan114@gmail.com}
\address[1]{College of Cyber Security, Jinan University, Guangzhou 510632, China}

\author[1]{Wensheng Gan}
\cortext[cor1]{Corresponding author}
\ead{wsgan001@gmail.com}
\cormark[1]

\author[4]{Philip S. Yu}
\ead{psyu@uic.edu}
\address[4]{Department of Computer Science, University of Illinois Chicago, Chicago, IL 60607, USA}

\begin{abstract}
   The explosive growth of multimedia content in the digital economy era has brought challenges in content recognition, copyright protection, and data management. As an emerging content management technology, perceptual hash-based digital fingerprints, serving as compact summaries of multimedia content, have been widely adopted for efficient multimedia content identification and retrieval across different modalities (e.g., text, image, video, audio), attracting significant attention from both academia and industry. Despite the increasing applications of digital fingerprints, there is a lack of systematic and comprehensive literature review on multimedia digital fingerprints. This survey aims to fill this gap and provide an important resource for researchers studying the details and related advancements of multimedia digital fingerprints. The survey first introduces the definition, characteristics, and related concepts (including hash functions, granularity, similarity measures, etc.) of digital fingerprints. It then focuses on analyzing and summarizing the algorithms for extracting unimodal fingerprints of different types of digital content, including text fingerprints, image fingerprints, video fingerprints, and audio fingerprints. Particularly, it provides an in-depth review and summary of deep learning-based fingerprints. Additionally, the survey elaborates on the various practical applications of digital fingerprints and outlines the challenges and potential future research directions. The goal is to promote the continued development of multimedia digital fingerprint research.
\end{abstract}


\makeatletter\def\Hy@Warning#1{}\makeatother
\maketitle

\section{Introduction}  
The generation of multimedia content has experienced explosive growth in the past decade \cite{gan2023web,wu2023multimodal}. The ubiquity of portable devices such as smartphones and cameras has enabled everyone to easily create and share high-quality photos, videos, and audio files \cite{li2007new,sun2023internet}. Meanwhile, social media platforms like Meta, Instagram, and TikTok, with their user-friendly interfaces and powerful dissemination mechanisms, have further driven the generation and consumption of multimedia content. Simultaneously, artificial intelligence (AI) technology \cite{anantrasirichai2022artificial} has significantly enhanced the efficiency of content creation. Statistics show that YouTube has approximately 5 billion video views per day and 3.7 million new videos uploaded. The thriving of media platforms like YouTube and Spotify has made content easily accessible, but has also brought about many serious issues, such as content plagiarism \cite{borkar2021music}. It has also made it very challenging to retrieve specific content from the massive multimedia resources on the web \cite{chen2022deep}. Most traditional and common retrieval methods rely on some form of metadata (e.g., titles, keywords, or descriptions) added to the content to enable retrieval based on annotated words \cite{yildizer2012efficient}. However, the traditional manual keyword labeling not only is time-consuming but also struggles to capture the diversity and ambiguity of the content \cite{wang2013new}. The explosion of multimedia content has also led to the duplication of files, consuming precious storage space and network bandwidth \cite{zhang2018secure,xia2016comprehensive}. Facing new challenges such as content copying, content search, copyright verification, and content ranking, traditional data management methods, algorithms, frameworks, and tools are no longer sufficient to handle such massive data volumes and provide effective solutions for data growth management \cite{bello2016social}. As digital content, such as artificial intelligence-generated content (AIGC) \cite{wu2023ai}, has become increasingly pervasive in human life, the effective identification, authentication, and regulation of these digital contents have become an urgent problem to be solved. Traditional copyright protection \cite{subramanya2006digital} and content regulation measures are already struggling to meet the increasingly complex demands. Digital fingerprint \cite{lu2009video} has emerged as an important technology to solve this problem, with its computational and spatial efficiency. It provides a better solution for organizing, managing, differentiating, identifying, and retrieving large-scale multimedia \cite{arunakumari2023fingerprint}. Digital fingerprints serve as a novel approach to resource management and protection. They provide a distinct characterization of content identity through its signal representation, known as "content-based identification" \cite{becker2003digital}, similar to how human fingerprints uniquely distinguish individual identities. Compared to traditional digital watermarking techniques, digital fingerprints do not alter the content of multimedia files. Instead, they represent the content of multimedia files by extracting stable features and generating a compact summary, which can then be used for efficient content management, copyright protection, and retrieval applications by comparing the generated compact summaries.

\begin{figure}[ht]
    \centering
    \includegraphics[clip,scale=0.26]{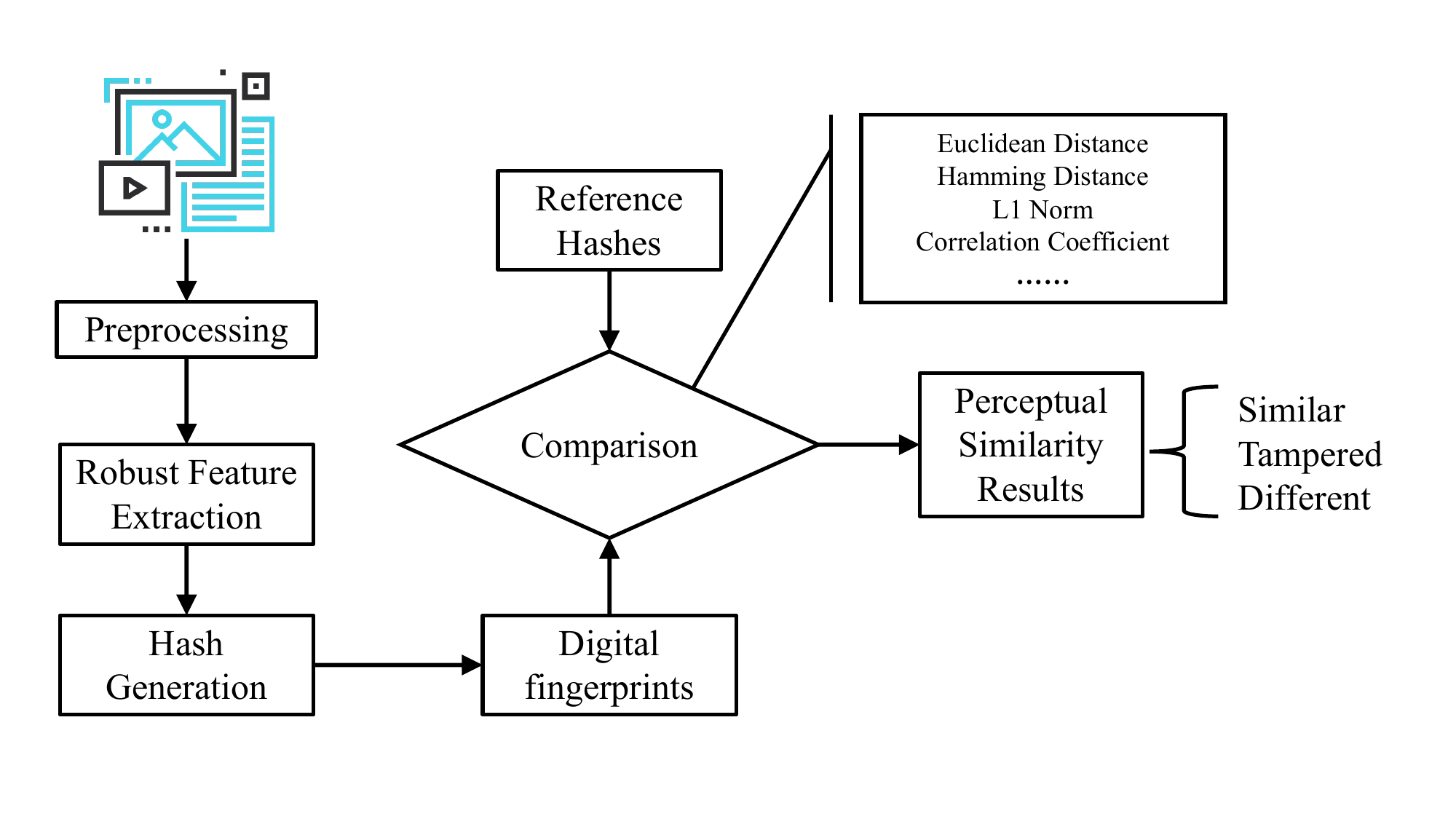}
    \caption{A general framework of digital fingerprinting systems.}
    \label{fig: frame}
\end{figure}

As the scale of multimedia data continues to expand, fingerprint-based recognition applications, such as image search, song identification, and content duplication detection, have been widely adopted \cite{arunakumari2023fingerprint}. The main advantage of fingerprints is that they are highly compressed representations of the original data, containing only relevant features \cite{irtaza2014embedding}, effectively reducing the search space. Compared to unstructured and cumbersome browsing, fingerprints can greatly reduce image retrieval time \cite{akgul2011content}. Fingerprint-based deduplication can simplify data organization and optimize the management, retrieval, and access of content assets. By creating unique fingerprints for different content segments, fingerprint identification technology can achieve more accurate deduplication and partial matching recognition \cite{chaudhari2024hash}. This technology has been widely applied to content plagiarism monitoring across multimedia types such as text, images, videos, and audios \cite{schleimer2003winnowing,rashid2023sampling}. Its copyright infringement detection functionality is crucial for protecting the intellectual property of artists' works \cite{ansori2023hades}. A general framework of the digital fingerprinting system is shown in Figure \ref{fig: frame}.

Digital fingerprint algorithms are an important research direction in the field of digital signal processing and pattern recognition \cite{samanta2021analysis}. By analyzing and extracting feature information from digital content, they generate unique digital fingerprint codes for applications such as content identification, authentication, and tracing. As digital content has become widely used and permeated in human life, digital fingerprint technology has played an increasingly important role in areas such as copyright protection, media regulation, and information security. Research on digital fingerprint algorithms began in the early 1990s, and after more than 30 years of development, a relatively mature technical system has been formed. Currently, digital fingerprints have been widely applied in various digital content types, including audio, video, images, and text, and the fingerprint extraction algorithms for different types of digital content also have their own characteristics. Meanwhile, with the rapid development of AI technology \cite{winston1984artificial}, deep learning-based digital fingerprint extraction and recognition methods have also emerged, further enhancing the performance and applicability of digital fingerprint technology.

The so-called multimedia-based digital fingerprint \cite{li2019robust} refers to the use of digital content containing multiple media modalities (such as audio, video, images, text, etc.) to generate digital fingerprints. Compared to single-media types, multimedia digital fingerprints can leverage the complementarity of different modality information to improve the robustness and discriminability of fingerprints, and support more complex application scenarios. The diversity and complexity of multimedia content also pose new challenges to the extraction and matching of digital fingerprints, as detailed below. First, existing fingerprint extraction and matching algorithms have high computational costs and long processing times when dealing with large-scale multimedia content \cite{esmaeili2010robust2}. These algorithms need optimization in processing speed and computational complexity to meet the higher requirements for real-time and computational resources in scenarios such as streaming media and real-time monitoring. In addition, the accuracy and robustness of feature extraction is also a major issue, as noise, compression, and other processing operations may affect the effectiveness of feature extraction, and there is a particular need to enhance robustness against composite attacks and adversarial attacks. As the volume of multimedia content grows exponentially, developing compact and efficient fingerprint representation methods and deduplication technologies to reduce storage requirements and bandwidth consumption have become increasingly important. Finally, privacy protection \cite{gan2018privacy,li2022frequent} is also a major challenge. The content information contained in perceptual hashes may lead to privacy leakage \cite{jain2023deep}, and there is an urgent need to design privacy protection mechanisms that balance computational complexity and efficiency.

\begin{figure*}[ht]
    \centering
    \includegraphics[trim=0 80 20 40,clip,scale=0.5]{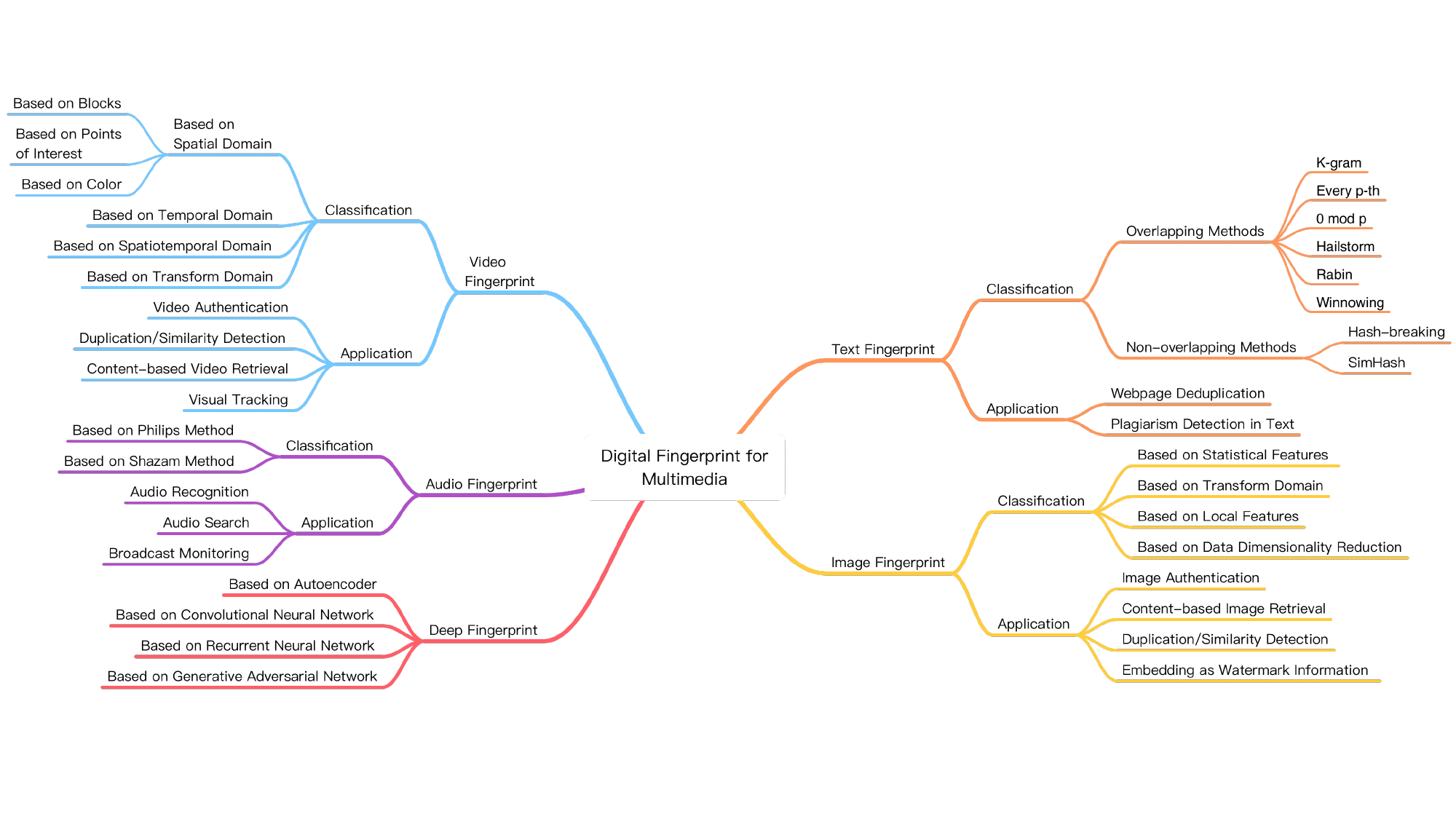}
    \caption{Digital fingerprint for multimedia.}
    \label{fig: DigitalFingerprint}
\end{figure*}

In the digital watermarking field, the term "fingerprint" has also been used to describe a technology that deters unauthorized redistribution of multimedia content by embedding a distinct identifying signal in each authorized copy \cite{varna2009fingerprinting}. The fingerprint identification method is fundamentally different from the watermark concept \cite{herre2003content}. The digital fingerprint technology discussed in this paper is a content-based compressed signature for summarizing multimedia content \cite{arunakumari2023fingerprint}. Over the past decades, there have been many surveys and reviews in the field of digital fingerprints, but there is currently no comprehensive review from a multimedia perspective. A comparison with related reviews is presented in Table \ref{table:contribution}. This paper provides a comprehensive survey of digital fingerprint algorithms based on different multimedia types. The outline is shown in Figure \ref{fig: DigitalFingerprint}. The main contributions of this paper are as follows:

\begin{itemize}
    \item Comprehensive survey of digital fingerprints based on different multimedia types. This survey systematically reviews the research status and development trends of multimedia-based digital fingerprints. It first introduces the concept, characteristics, and applications of digital fingerprints in various fields (Section \ref{sec:concepts}). It then focuses on analyzing the unimodal fingerprint extraction algorithms for different types of digital content, including text fingerprints, image fingerprints, video fingerprints, and audio fingerprints (Sections \ref{sec:textdf}-\ref{sec:audiodf}). 

    \item Exploration of the most advanced deep digital fingerprint algorithms. We discuss in detail the applications and algorithms of deep learning techniques in digital fingerprints, including methods based on convolutional neural networks (CNNs), recurrent neural networks (RNNs), generative adversarial networks (GANs), and others (Section \ref{sec:deephashing}).

    \item Practical applications of digital fingerprints (Section \ref{section:applications}). The paper summarizes a wide range of real-world application cases of digital fingerprints, covering areas from content recognition to copyright protection and copy detection, and introduces the practical applications of digital fingerprints in different scenarios, especially large model-based applications.

    \item Challenges and future development directions (Section \ref{sec:oppotrunities}). The paper delves into the key challenges faced by multimedia digital fingerprint technology in practical applications, such as robustness, security, computational complexity, and privacy protection. Additionally, the paper outlines the future directions of digital fingerprints and proposes potential research topics, providing a comprehensive reference for researchers and practitioners in the related fields.
\end{itemize}

\begin{table*}[ht]
    \centering
    \footnotesize
    \caption{Contributions and gaps of existing papers.}
    \label{table:contribution}
    \begin{tabularx}{\textwidth}
    {|m{0.6cm}<{\centering}
    |m{0.6cm}<{\centering}
    |m{1.5cm}<{\centering}
    |m{0.6cm}<{\centering}
    |m{3.3cm}<{\raggedright}
    |m{3.4cm}<{\raggedright}
    |m{3.4cm}<{\raggedright}|}
        \toprule
        \hline
        \multicolumn{1}{|c|}{\textbf{Ref.}} & \multicolumn{1}{c|}{\textbf{Year}} & \multicolumn{1}{c|}{\textbf{\makecell{Multimedia\\Types}}} & \multicolumn{1}{c|}{\textbf{Comparison}} & \multicolumn{1}{c|}{\textbf{Applications}} & \multicolumn{1}{c|}{\textbf{Threat concerns}} & \multicolumn{1}{c|}{\textbf{Trends}}\\
        \hline
        \hline

        \cite{farid2021overview} & 2021 & \makecell[c]{Image,\\Video,\\Audio} & \multicolumn{1}{c|}{\XSolidBrush} & Limiting redistribution of terrorist content, copyright-infringing materials, {CSAM}, etc. & Handling multiple content variants, computational efficiency, balancing privacy and security &Enhancing technical performance, creating industry-wide shared databases\\
        \hline
        
        \cite{jekabsons2020evaluation} & 2022 & Text & \multicolumn{1}{c|}{\XSolidBrush} & Plagiarism detection, source detection, information flow analysis & Comparison and evaluation of multiple algorithms, handling the complexity of real text reuse & Frequency approximation methods, system integration, solving algorithm limitations \\
        \hline
        
        \cite{du2020perceptual} & 2020 & Image & \multicolumn{1}{c|}{\Checkmark} & \multicolumn{1}{c|}{\XSolidBrush} & Resisting only one or several types of attacks, general security evaluation methods & Advanced information, combining different image features, high-dimensional decision methods, and analyzing security by modifying related content based on hashing \\
        \hline

        \cite{roy2023various} & 2023 & Image & \multicolumn{1}{c|}{\Checkmark} & \multicolumn{1}{c|}{\XSolidBrush} & Multiple types of attacks, computational time of learning-based methods, and the limitation of application domains & \multicolumn{1}{c|}{\XSolidBrush} \\
        \hline
        
        \cite{allouche2022video} & 2022 & Video & \multicolumn{1}{c|}{\XSolidBrush} & Visual fake news, autonomous driving & video content size and typology, complexity of near-duplicated copies, compressed stream extraction, and energy consumption reduction & Stronger constraints on video fingerprinting properties, emerging application domains\\
        \hline

        \cite{kekre2013review} & 2023 & Audio & \multicolumn{1}{c|}{\Checkmark} & Broadcast monitoring, audience measurement, name that tune, metadata collection, find duplicates, added value services & \multicolumn{1}{c|}{\XSolidBrush} & \multicolumn{1}{c|}{\XSolidBrush} \\
        \hline

        Our work & 2024 & \makecell[c]{Text,\\Image,\\Video,\\Audio} &\multicolumn{1}{c|}{\Checkmark} & Reducing computational cost, reducing storage requirements, composite attacks, adversarial attacks, privacy leakage, accuracy and robustness of algorithms & Computational resources, model efficiency, data-related, practical applications, fairness, privacy, accountability, patient autonomy & Deep digital fingerprint, fusion of multiple feature information, robustness and attack-resistance, practicality, algorithm performance evaluation schemes, emerging application scenarios, model copy detection, etc. \\
        \hline
    \end{tabularx}
\end{table*}

\begin{table}[ht]
    \footnotesize
    \centering
    \caption{Important terms of acronyms and corresponding full form.}
    \label{table:acronyms}
    \begin{tabular}{|m{1.5cm}<{\raggedright}|m{5.3cm}<{\raggedright}|}
        \hline
        \multicolumn{1}{|c|}{\textbf{Acronym}} & \multicolumn{1}{c|}{\textbf{Full Form}} \\  
        \hline
        AI & Artificial Intelligence\\
        \hline
        CAE & Convolutional Autoencoder\\
        \hline
        CNN & Convolutional Neural Network\\
        \hline
        DCT & Discrete Cosine Transform\\
        \hline
        DFT & Discrete Fourier Transform\\
        \hline
        DNN & Deep Neural Network\\
        \hline
        DWT & Discrete Wavelet Transform\\
        \hline
        GAN & Generative Adversarial Network\\
        \hline
        GPT & Generative Pre-trained Transformer \\
        \hline
        LLM & Large Language Model \\
        \hline
        LSTM & Long Short-term Memory Network \\
        \hline
        MFCC & Mel-Frequency Cepstral Coefficients\\
        \hline
        RNN & Recurrent Neural Network \\
        \hline
        SIFT & Scale-Invariant Feature Transform\\
        \hline
        SVD & Singular Value Decomposition\\
        \hline         
    \end{tabular}
\end{table}

\section{Related Concepts} \label{sec:concepts}
\subsection{Definition of Digital Fingerprint}

Currently, there are two main definitions of digital fingerprints: (1) Digital watermarking \cite{mohanarathinam2020digital} is used for copyright protection. By embedding unique information related to the buyer's identity in copies of digital works, when illegal copies are found, the digital work's vendor can identify the original purchaser of the illegal copy based on the embedded information.  (2) Perceptual hash-based digital fingerprints \cite{wang2015visual} is applied to multimedia content identification. This technique maps the features of multimedia data to a digital fingerprint summary. The latter is the meaning of the term "fingerprint" in this paper. In general, altering the content by incorporating identifiers or signatures, which results in changes to the original content, is regarded as a form of watermarking; extracting identifiers or signatures from the content without changing the content is considered a fingerprint \cite{lu2009video}. Note that this paper focuses on the perceptual hash-based digital fingerprints in multimedia data.

Fingerprints are also called perceptual hashing, content-based digital signatures, and message digests. Fingerprint recognition is also commonly referred to as copy detection or content-based copy detection (CBCD) \cite{farid2021overview}. Digital fingerprints are a unique identifier used to identify and authenticate digital content. It is obtained by analyzing the inherent features of digital content, extracting and encoding them into a unique digital sequence, which serves as the identity identifier for the content. Unlike biometric fingerprints, digital fingerprints are unique "fingerprints" for digitized content. Digital fingerprints have the following main characteristics:

\begin{itemize}
    \item Uniqueness: Each digital content has its unique digital fingerprint code, which will not repeat with other content.

    \item Stability: As long as the digital content itself does not change, its digital fingerprint code will not change. Even with a certain degree of processing, the fingerprint code can remain relatively stable.

    \item Extractability: The digital fingerprint code can be automatically extracted from the digital content using specific algorithms.

    \item Compactness: The digital fingerprint code is usually a short binary sequence, occupying small space and with high computational efficiency.
\end{itemize}

Based on these characteristics, digital fingerprints are widely used in areas such as copyright protection, content regulation, and tracing, providing effective protection for the security and integrity of digital content. Digital fingerprints have become one of the key technologies for digital content security management, with widespread applications across various industries. As the application of digital content continues to expand, the role of digital fingerprints will become increasingly important. Digital fingerprint technology has been widely applied in multiple fields, mainly including:

\begin{itemize}
    \item Copyright protection \cite{sun2022deep}: By adding a unique digital fingerprint code to digital content, unauthorized copying and distribution can be effectively prevented. Once illegal use is detected, the fingerprint information can be used for tracing and evidence.

    \item Content regulation \cite{steinebach2023analysis}: Embedding digital fingerprints in audio, video, image, and other media content can be used for the detection and control of illegal content, such as the identification and blocking of harmful information like pornography and violence.

    \item Identity authentication \cite{du2020perceptual}: Digital fingerprints can verify the source and integrity of digital content, ensuring the authenticity and reliability of the content, such as electronic contracts and digital certificates.

    \item Personalized services \cite{gabryel2020browser}: Embedding the user's digital fingerprint in personal devices or applications can provide personalized content recommendations, privacy protection, and other services.

    \item Data management \cite{zhang2018secure,zhang2015memory}: Digital fingerprint technology can be applied in big data management, enabling indexing, deduplication, and version control of massive data.
\end{itemize}

\subsection{Hash Functions}
\subsubsection{Cryptographic Hash and Perceptual Hash}

A hash function compresses a message of arbitrary length into a fixed-length message digest. Towards multimedia information processing, some studies have borrowed the hash concept to represent the perceptual summary of multimedia data and proposed the concept of perceptual hashing, also known as robust hashing. Traditional cryptographic hash functions exhibit high sensitivity to input data, where even a single bit alteration can trigger the avalanche effect, resulting in entirely different hash values. Therefore, perceptual hash functions are generally used for multimedia content integrity authentication and analysis \cite{du2020perceptual}. Perceptual hashing maps the perceptual content of multimedia data to a concise binary string, which can be seen as a digital summary of the multimedia content, and is essentially a mapping that adheres to specific constraints \cite{wang2015visual}. Let the perceptual hash function be denoted as \textit{PH}: $M$ $\rightarrow$ $H_p$, where $M$ is the set of digital representation sequences of multimedia data, and $H_p$ is the set of perceptual digital summaries (fingerprints). $n$ contrast to cryptographic hash functions, perceptual hash functions are designed to be less sensitive to minor variations in the input data \cite{monga2006clustering}. They use perceptual features to generate their hash values, which are easily affected by the perceptual feature differences of the comparison objects. For perceptual hashing, it must be robust to content-preserving operations such as affine transformation, frame rate change, and format change, but very sensitive to content tampering. Traditional cryptographic hash functions are commonly used in applications that require strong anti-collision capabilities, such as password authentication, and file or data identification. Perceptual hash functions are widely used in multimedia content authentication, tampering detection, and retrieval \cite{wang2015visual}.

\subsubsection{Characteristics of Perceptual Hash}

The performance evaluation of perceptual hashing is generally performed from six aspects: robustness, discrimination, tampering detection, security, computational efficiency, and hash length \cite{li2022unified}. Effective perceptual hashing should have the following characteristics \cite{wang2015visual}:

\textbf{(1) Robustness}, also known as stability or resilience, means that the perceptual hashes of multimedia files with similar or close content should be the same or similar after hashing. Ideally, the generated hash value should be similar to the original multimedia file's hash value after content-preserving operations.

\textbf{(2) Discrimination} means that multimedia files with different content should have significantly different perceptual hashes after hashing. Generally, the perceptual robustness and discrimination of image hashing are inversely related, where the former requires stability to minor disturbances, and the latter requires sensitivity to minor malicious modifications \cite{monga2006perceptual}. Therefore, a balance between the two must be considered in practice.

\textbf{(3) One-wayness} means that it should be mathematically impossible to derive the content of the multimedia file or its features from the generated perceptual hash. The process of deriving the perceptual hash from the hash function should be one-way, without an inverse function.

\textbf{(4) Compactness} indicates that the perceptual hash must be significantly smaller in size compared to the original multimedia file, with its length minimized as much as possible.

\subsubsection{Granularity}

In digital fingerprint technology, granularity refers to the level of detail and precision involved in analyzing and representing multimedia content as digital fingerprints \cite{lulu2016overview}. The importance of granularity lies in the fact that it directly affects the effectiveness and accuracy of the hash algorithm \cite{monga2006clustering}. It determines the sensitivity of the hash algorithm to changes and the level of detail considered in the fingerprint generation process. In text fingerprints, the size of substrings defines the granularity of the fingerprint. The spatial granularity of video fingerprints can range from the entire video frame to subdivided blocks, and even to interest points within the frame \cite{lu2009video}. Temporal granularity can be defined by keyframes, groups of frames, downsampled individual frames, or even by every single frame. In audio fingerprints, the fingerprint calculated for each frame is called a subfingerprint, and a subfingerprint usually does not contain enough information for audio recognition. The audio unit (composed of a certain number of frames) that is sufficient to identify the entire unknown audio is called the fingerprint granularity, or the fingerprint block.

Granularity can be divided into fine-grained and coarse-grained types, which are suitable for different application scenarios \cite{stein2007fingerprint}. Fine-grained fingerprints contain highly detailed information about the content and can capture subtle changes and differences. This is very important in applications that require precise content identification and differentiation, such as duplicate detection and fine-grained content verification. In these applications, fine granularity helps identify near-duplicate images or videos by capturing detailed differences and ensures the integrity of multimedia content. However, the high sensitivity of fine granularity may lead to false positives, so a balance needs to be considered when designing the algorithm \cite{lulu2016overview}. Coarse-grained fingerprints provide a more general representation, focusing on larger features and ignoring subtle changes. Coarse granularity is very useful in applications where precise matching is not critical, such as general content classification or large-scale indexing. In such cases, coarse granularity helps achieve faster multimedia content retrieval by focusing on larger features but may miss some important details.

\subsubsection{Similarity Measurement Methods}

Digital fingerprints are an effective representation of multimedia file content, so the similarity between digital fingerprints can be used to determine whether the two corresponding multimedia files have similar content, and thus whether one is a copy of the other. Since perceptually similar multimedia content generates fingerprints with higher similarity, the distance between them is smaller, while the distance between perceptually different multimedia content is larger. If the distance is greater than a certain empirically determined threshold, the multimedia content is likely different or has undergone content modification.

Some commonly used similarity measures include Hamming distance or normalized Hamming distance, Minkowski distance, Euclidean distance and its weighted versions, bit error rate (BER), peak cross-correlation (PCC), Hausdorff distance, Mallows distance, Kullback-Leibler divergence, etc \cite{samanta2021analysis}. The choice of distance type depends on the type of generated hash vector. For example, if the hash vector is composed only of "0" and "1", Hamming distance (HD) or normalized Hamming distance (NHD) can be used; for integers or real values, Euclidean distance and its weighted versions can be used \cite{roy2023various}.

By combining these different similarity measurement methods, digital fingerprint technology can provide efficient and accurate identification and comparison capabilities in different application scenarios to meet various needs and challenges \cite{datta2008image}. Specifically, Euclidean distance and its weighted versions are suitable for global feature comparison but sensitive to local changes; Hausdorff distance is suitable for shape and contour comparison, with certain robustness but computational complexity; Mallows distance is used for comparing sets of images, suitable for high-precision tasks; Kullback-Leibler divergence is used for comparing probability distributions, effectively measuring distribution differences but with high computational complexity. Bit error rate and peak cross-correlation are also commonly used similarity measurement methods, where the former is suitable for comparing binary hash vectors and the latter performs well in image matching.

Assuming \(h_1\) and \(h_2\) are the hash vectors generated from two multimedia files, and \(h_1(i)\) and \(h_2(i)\) are the i-th elements in the hash vectors \(h_1\) and \(h_2\) respectively, the definitions of two similarity measurement are as follows: 

\textbf{(1) Hamming distance} \cite{ong2016improved} is used to calculate the similarity between binary hash sequences. For non-binary hash forms, they need to be converted to binary hash sequences before using Hamming distance to measure their similarity. Hamming distance can be calculated with the number of different bits in the binary hash sequence: 
\begin{equation}
d(h_1, h_2) = \sum_{i=1}^{N} \left| h_1(i) - h_2(i) \right|
\label{eq:formula1}
\end{equation}
The smaller the Hamming distance, the smaller difference between the robust hashes, and the more similar the multimedia file contents represented by the robust hashes. Conversely, if the Hamming distance is large, the contents of two multimedia files are different and there is no copy \cite{samanta2021analysis}. 

\textbf{(2) Minkowski distance} \cite{roche2015minkowski}, also called Minkowski metric or Lp norm, is defined as: 
\begin{equation}
L_P(h_1, h_2) = \left( \sum_{i=1}^{N} \left| h_1(i) - h_2(i) \right|^P \right)^{\frac{1}{P}}
\label{eq:formula2}
\end{equation}

Minkowski distance can be categorized into Manhattan distance, Euclidean distance, and Chebyshev distance, depending on the value of \( P\). When \( P = 1 \), the above formula is called Manhattan distance or city block distance, i.e., the L1 norm; when \( P = 2 \), it becomes Euclidean distance, i.e., the L2 norm; when \( P \to \infty \), it becomes Chebyshev distance. Similar to the Hamming distance, the smaller the Minkowski distance between two hash vectors, the higher the similarity between them.

\section{Text-Based Digital Fingerprints} \label{sec:textdf}

In data mining and information retrieval, detecting local text reuse is a critical task. One effective method for accomplishing this is the use of text-based digital fingerprint identification technology. Text-based digital fingerprint technology first divides the document into a sequence of terms (i.e., k-grams or shingles), and then uses a specific text block selection strategy to extract features from the text words or words. The extracted string is the "fingerprint" of the text. These feature strings are then converted into digital fingerprints through hash functions. A comparison of classic digital fingerprint algorithms for text is provided in Table \ref{table:text-based}.

\begin{table*}[ht]
    \centering
    \footnotesize
    \caption{Classic text-based digital fingerprint algorithms.}
    \label{table:text-based}
    \begin{tabularx}{\textwidth}
    {|m{1.9cm}<{\centering}
    |m{4.3cm}<{\centering}
    |m{2.2cm}<{\centering}
    |m{3.4cm}<{\centering}
    |m{3.4cm}<{\centering}|}
        \toprule
        \hline
        \multicolumn{1}{|c|}{\textbf{Technique}} & \multicolumn{1}{c|}{\textbf{Description}} & \multicolumn{1}{c|}{\textbf{\makecell{Selection\\Mechanism}}} & \multicolumn{1}{c|}{\textbf{Advantages}} & \multicolumn{1}{c|}{\textbf{Disadvantages}}\\
        \hline
        \hline

        K-gram & Divides text into contiguous substrings of length k, using all possible substrings as fingerprints. & All possible substrings are used & High accuracy due to comprehensive coverage & Generates a large number of fingerprints, leading to high computational cost \\
        \hline
        Winnowing & Uses a sliding window on k-grams to select a minimal subset of fingerprints with the smallest hash values within each window & Selects minimum hash value within a window & Reduces the number of fingerprints, efficient in finding matches & May miss some matches if they fall below the noise threshold\\
        \hline
        Hailstorm & An enhancement of winnowing, providing broader coverage by selecting the lowest or highest hash value within each window & Selecting extreme hash values for broader coverage & Provides total coverage with locality property, robust in detecting overlaps & More complex, potentially higher computational overhead than basic winnowing \\
        \hline
        {DCT} & Applies a mathematical transformation to the fingerprint values to emphasize lower-frequency components, which are less sensitive to small changes. & Transformation applied to the sequence of hash values & Robust against small textual changes, good for detecting near duplicates & Less efficient with extensive text modifications, higher computational cost \\
        \hline
        Hash-breaking & Divides text into segments where hash values meet a specific condition (e.g., divisible by a parameter) & Breakpoints at hash values meeting specific conditions & Efficient in segmenting text for reuse detection & Sensitive to small edits, which can alter hash values and fingerprint matches\\
        \hline
    \end{tabularx}
\end{table*}

\subsection{Preprocessing}

To generate a series of fingerprints from a document's text, several preprocessing steps may be applied, including the removal of special characters like punctuation, tokenizing the text into individual words, converting all words to lowercase, and eliminating stop words. Subsequently, stemming or lemmatization is applied to the words to reduce them to their base forms, either as stems or lemmas, effectively minimizing inflectional and derivational variations \cite{jekabsons2020evaluation}. The result is a clean, standardized sequence of terms. Several other aspects need to be considered in designing an accurate and effective fingerprint identification technique \cite{lulu2016overview}:

\textbf{(1) Text granularity selection}. Text granularity generally refers to the length of the text selected from the document to generate the digital fingerprint. The selection of text granularity can be roughly divided into overlapping and non-overlapping. Overlapping selection means that the adjacent text granules have the same text characters, and the Winnowing algorithm is a method that uses sliding windows for overlapping selection. In the overlapping selection method, changes to the text will affect the fingerprint values of the adjacent windows. In contrast, in the non-overlapping selection method, there is no overlap in the content of the selected text blocks. In the non-overlapping selection method, changes to the text only affect the fingerprint value of the current text block and do not affect the fingerprint values of other text blocks.

\textbf{(2) Digital fingerprint mapping}. Digital fingerprint mapping uses hash functions to convert strings into numerical sequences. Hash functions commonly used in digital fingerprint technology include MD5 \cite{rivest1992md5}, Karp-Rabin algorithm \cite{karp1987efficient}, and Simhash algorithm \cite{charikar2002similarity}.

\textbf{(3) Digital fingerprint selection}. Digital fingerprint selection refers to the strategy of selecting all or part of the hash values from the hash value sequence as the text digital fingerprint. The selection strategies include full fingerprint selection, hash breakpoint method, min-hash method, frequency-based selection, structure-based selection, and position-based selection.

\subsection{Text-Based Digital Fingerprint Algorithms}

Text-based fingerprint algorithms play an important role in multimedia content identification. By hashing consecutive word or character sequences, they can efficiently detect text reuse and are widely used in plagiarism detection, source tracing, and information flow analysis. Various fingerprint selection algorithms, including K-gram \cite{bondarenko2005documents}, every p-th \cite{jekabsons2020evaluation}, 0 mod p \cite{manber1994finding}, Winnowing \cite{schleimer2003winnowing}, and its improved versions \cite{sun2013near}, and Hailstorm \cite{abdel2009detecting}, provide diverse choices. Winnowing and 0 mod p algorithms perform particularly well under different fingerprint selection sizes, maintaining high detection quality even at lower fingerprint selection ratios. The MinHash algorithm \cite{broder1998min} is suitable for large-scale and high-dimensional text reuse detection \cite{hassanian2019pruning}. However, these algorithms also have some shortcomings, such as the high storage and computational cost of the full fingerprint strategy, the sensitivity of the Every p-th algorithm to position changes, the possible occurrence of unselected long intervals in the 0 mod p algorithm, the dependence on frequency tables in the FBW and MFBW algorithms, and the lower accuracy of the MinHash algorithm, mainly sacrificing accuracy to improve efficiency. Therefore, properly selecting the appropriate fingerprint algorithm can help find the best balance between performance and cost, and improve the accuracy and efficiency of multimedia content identification. Hash-based text-based digital fingerprint technologies can be roughly divided into two categories: overlapping methods and non-overlapping methods.

\subsubsection{Overlapping Methods}

Overlapping selection generally uses the content segmentation method, first setting a block size (i.e., window size), which is the size of the text granularity, and then reading the text content from the beginning of the file, selecting the text through the sliding of the window. The size of each window movement (i.e., the offset) is smaller than the window size, so that the adjacent windows have overlapping content. This method based on sliding windows is called the overlapping method.

The standardized word sequence is segmented into overlapping groups of k consecutive words, known as k-grams or shingles. Each k-gram is then transformed into an integer using a specific hashing algorithm, such as MD5 \cite{rivest1992md5} or Rabin \cite{rabin1981fingerprinting}. For a document containing \( l \) words, a total of \( m = l - k + 1 \) k-grams can be generated. The simplest fingerprint strategy is to use all k-grams (i.e., their calculated hash values) as the fingerprint of the document. Although the "full fingerprinting" strategy can find the most correct text reuse sources, its storage space and fingerprint comparison costs are too high to be used for large document collections. To optimize the balance between result quality, storage demands, and computational costs, a range of fingerprint selection algorithms has been developed.

\textbf{(1) K-gram method }\cite{bondarenko2005documents}: K-gram refers to a fragment consisting of K consecutive words or characters. After text preprocessing, according to the selected K value, the text is divided into multiple K-gram segments using a sliding window of size k. For each K-gram segment, a hash function is used to convert it into a fixed-length binary value or integer code. A weight is assigned to each feature to reflect its importance. The weighted feature vectors of all K-gram segments are then summed or XOR-ed to obtain the digital fingerprint of the text.

\textbf{(2) Every p-th method }\cite{jekabsons2020evaluation}: It selects the p-th k-gram in the document, for a total of m/p k-grams. This algorithm is highly sensitive to changes in the order of k-grams, as well as to insertions, and deletions, and it lacks position independence.

\textbf{(3) 0 mod p method }\cite{manber1994finding}: The 0 mod p method is based on the K-gram method, but only selects the k-gram blocks whose hash values are divisible by p  \cite{rabin1981fingerprinting}. The 0 mod p method typically selects m/p k-grams on average. The selection of a k-gram is solely based on its individual characteristics, independent of other k-grams within the document. This guarantees that each specific k-gram is either consistently selected or never selected across all documents containing it. However, a potential drawback of this algorithm is the unbounded maximum gap between two selected k-grams, which could lead to an arbitrary number of consecutive unselected k-grams. Consequently, this may create extended intervals of unselected k-grams, hindering the detection of any matches within those intervals \cite{wang2015visual}.

\textbf{(4) Hailstorm method} \cite{abdel2009detecting}: The preprocessed text is segmented into an overlapping sequence of k consecutive words, and a hash function is applied to compute the hash value of each k-gram. If the hash value of either the leftmost or rightmost word in a k-gram happens to be the smallest among the k words in the entire k-gram, then the hash value of this k-gram will be selected as part of the fingerprint. \cite{abdel2009detecting}. Similar to the 0 mod p algorithm, the Hailstorm algorithm also has context independence, i.e., whether a k-gram is selected depends only on the words of the k-gram itself, not on the other k-grams in the document. Any given k-gram is either selected in all documents containing it or is never selected. The Hailstorm also guarantees that each word in the document appears in at least one selected k-gram. It is competitive when selecting a larger proportion of fingerprints, but its performance deteriorates significantly when the number of fingerprints is reduced to below 40\%-45\%. Overall, this algorithm performs better in detecting longer unchanged text passages but is less effective in handling short text reuse and rewrites.

\textbf{(5) Rabin algorithm} \cite{rabin1981fingerprinting}: In 1981, Professor Rabin of Harvard University proposed the Rabin fingerprint calculation method \cite{rabin1981fingerprinting}, which has been widely used in file similarity detection. The calculation of Rabin fingerprints occurs in the finite field \( GF(2^n) \), assuming \( A(a_1, a_2, a_3, \ldots, a_m) \) is a binary string of \( m \) characters, then a corresponding \( (m-1) \)-degree polynomial (where \( t \) is an indeterminate) can be constructed based on \( A \) as follows: 
\begin{equation}
A(t) = (a_1 t^{m-1} + a_2 t^{m-2} + \cdots + a_{m-1} t + a_m)
\label{eq:A(t)}
\end{equation}
Assuming \( P(t) \) is an irreducible polynomial of degree \( k \) over the finite field \( GF(2^n) \):
\begin{equation}
P(t) = (b_1 t^k + b_2 t^{k-1} + \cdots + b_{k-1} t + b_k)
\label{eq:P(t)}
\end{equation}
Then the remainder $f(t)$ of $A(t)$ divided by $P(t)$ has a degree of ($k$-1). For a given string $A$, the fingerprint $f(A)$ of $A$ is defined as $f(A)$ = $A(t)$ $\mod$ $P(t)$. Similar to hash functions, the Rabin fingerprint algorithm is mainly used for fingerprint generation, with fast calculation and easy implementation. The fingerprint values $f(A)$ and $f(B)$ of different strings $A$ and $B$ are also different. The Rabin algorithm also satisfies the distributive law, i.e., $f(A+B)$ = $f(A)$ + $f(B)$. Another algorithm often used to generate digital fingerprints is the Karp-Rabin algorithm \cite{karp1987efficient}, proposed in 1987, which uses hash values for efficient string matching.

\textbf{(6) Winnowing algorithm \cite{schleimer2003winnowing} and its extensions \cite{sun2013near}}: The Winnowing algorithm is used for text similarity comparison and deduplication. It was originally proposed by Schleimer et al. in 2003 \cite{schleimer2003winnowing}. Based on the Rabin algorithm, the Winnowing algorithm adds a denoising function. Whereas in the sliding window selection process, Winnowing only retains the fragments with the minimum hash value and discards the other fragments, to remove some interfering characters. This can greatly reduce the size of the text representation while retaining the fragments related to the key information of the text. Based on Winnowing, frequency-biased winnowing (FBW) \cite{sun2013near} is an improved fingerprint selection method for text reuse detection. Its main feature is to select the k-grams with the lowest frequency in the document collection as fingerprints, thus reducing the number of matches with erroneous sources. Specifically, FBW selects the k-gram with the lowest frequency in the sliding window, and if there are multiple k-grams with the same frequency, it selects them in alphabetical order. This strategy of selecting rare k-grams improves the accuracy of detection and the sensitivity to minor text changes, while also reducing search time. The modified frequency-biased winnowing (MFBW) algorithm \cite{jekabsons2020evaluation} is an optimization, that addresses the shortcomings of FBW in handling zero-frequency k-grams in the query document. MFBW treats zero-frequency k-grams as having infinite frequency, making them the least preferred choice, thus avoiding selecting k-grams that are not in the collection. In each sliding window, MFBW selects the n-gram with the lowest non-zero frequency as the fingerprint, and if all k-grams have a frequency of zero, it selects them in alphabetical order. This enhances the algorithm's robustness and matching accuracy while retaining the advantages of FBW, such as high precision and reduced search time.

\subsubsection{Non-Overlapping Methods}

In non-overlapping approaches, the document is divided into distinct text segments rather than overlapping blocks. This segmentation process, known as text breaking, identifies specific word positions where these divisions occur, referred to as breakpoints.

\textbf{(1) Hash-breaking algorithm \cite{brin1995copy}}: This method does not rely on k-grams or shingles; instead, it divides the document into non-overlapping blocks. Each word in the document is hashed, and the text is split at any point where the hash value is divisible by a specified parameter $p$. The generated text blocks are then hashed and used as the fingerprint of the document. The Hash-breaking algorithm typically selects text blocks with an average length of $p$ tokens. However, in practical scenarios, the actual length of these blocks can vary significantly, becoming either much shorter or longer depending on the hash value distribution. Notably, when a sequence is extremely brief and composed of highly common words (such as "a" or "the"), this can lead to either excessive segmentation or insufficient segmentation, which may subsequently impact the algorithm's overall effectiveness. Since the text blocks selected by hash-breaking can also be very long, the method is very sensitive to changes. The variable length of the text blocks generated by the Hash-breaking algorithm may also make it difficult to align and match in the comparison of similar texts. Seo and Croft \cite{seo2008local} improved Hash-breaking by introducing a minimum length limit and using discrete cosine transform (DCT) to improve the robustness and accuracy.

\textbf{(2) SimHash algorithm} \cite{charikar2002similarity}: The core idea of locality-sensitive hashing (LSH) \cite{kulis2011kernelized} is to map neighboring points in the original space to the new space while maintaining their proximity. This idea has been widely applied in web deduplication, document similarity, image retrieval, fingerprint matching, and music retrieval. For text retrieval, the core ideas of Minhash (2000) \cite{broder1998min,wu2020review} and SimHash (2002) \cite{charikar2002similarity} are to incorporate dimensionality reduction techniques based on LSH. SimHash maps a high-dimensional feature vector to a lower-dimensional space to preserve the features as much as possible. Using this algorithm, similar texts generate similar digital fingerprints, and then the texts similarity can be obtained by calculating the Hamming distance.

\begin{itemize}
    \item Feature vector representation: For each feature item, a suitable hash function is used to convert it into a fixed-length binary value.

    \item Weight calculation: To better distinguish the importance of different feature items, a weight can be assigned to each feature item. Common methods include using TF-IDF (term frequency-inverse document frequency) weights or other relevant algorithms to calculate the weights.

    \item Feature weighting: Multiply the binary vector of each feature item by the corresponding weight to obtain the weighted feature vector.

    \item Feature aggregation: For the weighted feature vectors, perform a weighted summation operation on each bit.

    \item Dimensionality reduction: For the aggregated feature vector, if the weighted sum of a certain bit is greater than or equal to 0, that bit is set to 1; otherwise it is set to 0. The resulting SimHash fingerprint is a fixed-length binary value.
\end{itemize}

\section{Image-Based Digital Fingerprints}

Image fingerprints (also known as image hashes, image digests, or image authentication codes) are based on the characteristics of the human visual system (HVS) \cite{gao2010image}, analyzing and extracting robust features for encoding, and mapping images into short binary sequences. In contrast to watermarks, content-based image fingerprints do not depend on embedding markers within the image. Instead, they utilize the inherent features of the multimedia content to establish ownership. Until now, perceptual image hashing has widespread applications across various image-related fields, such as image authentication \cite{zhao2012robust}, content-based image retrieval, image indexing, copy or near-duplicate detection, image forgery detection, and image quality assessment \cite{xia2023perceptual}.

\subsection{Construction Methods and Classification of Digital Image Fingerprints}

The goal of perceptual image hashing is to simulate the human visual system, evaluating images based on the underlying scene content, rather than pure numerical comparison of pixel values. Digital image fingerprints extract a concise, unique, and perceptually meaningful identifier feature from the large-scale pixel space representation. Digital image fingerprints are robust to image modifications, including compression, color shift, cropping, rotation, the addition of logos or overlaid text, or any other modifications that do not fundamentally change the underlying content but alter the underlying pixel values. To ensure both robustness and security in image hashing, the majority of current schemes adhere to a three-step framework for hash generation: Step 1) image preprocessing: including resizing, color space conversion, dimensionality reduction, filtering, etc., to enhance the robustness of the features; Step 2) perceptual feature extraction; Step 3) compression or encoding of the quantized feature vectors. Among them, the feature extraction stage is the most challenging and important part of the generic image hashing framework.

In image feature extraction, statistical feature-based algorithms provide good robustness by extracting global features, but they lack sensitivity to local tampering and have poor uniqueness. Transform domain-based methods can effectively deal with geometric attacks, but their higher computational complexity may limit practical application. Local feature point methods perform well under geometric transformations but are sensitive to noise, have high computational costs, and cannot prove the uniqueness of image content. While dimensionality reduction-based techniques have advantages in reducing computational complexity, information loss may affect the discrimination and robustness of the fingerprint. Currently, most digital image fingerprint algorithms mainly focus on feature extraction and classification performance, with fewer algorithms focusing on human visual characteristics. Hash algorithms have high recognition rates for large-area content tampering operations, but their performance in detecting local small-area content tampering needs to be improved. The following introduces how to extract identifying features from the pixel space to generate digital image fingerprints from the perspectives of image statistical properties, image transform domain, image local feature points, and image data dimensionality reduction.

\subsection{Hash Construction Based on Image Statistical Properties}

Early methods for extracting image hashes mostly used some simple and effective global statistical properties to characterize the image hash. These statistical properties mainly include: various histograms of the image (including brightness histogram, gradient histogram, color histogram, cumulative histogram, and cross-histogram), as well as mean, variance, color aggregation vector (CCV), and color moments, etc. \cite{tang2015robust2,huang2018robustness,hosny2018robust}. While methods that utilize the global statistical properties of images can generate hashes with a degree of robustness against conventional operations like JPEG compression and filtering, they often lack the sensitivity needed to detect subtle illegal alterations. Moreover, the statistical properties of images are not very unique to a large extent, so the fingerprints constructed based on image statistical properties have poor discrimination, and cannot be effectively applied to image authentication. In addition, the hashes extracted by this method are too dependent on the image's content, resulting in the extracted hashes being very long, which wastes a lot of storage space and is not conducive to the effective utilization of computer storage resources. Tang et al. \cite{tang2018image} used DCT compression of the histogram of the inner ring CVA to generate hashes. It can resist large-angle rotation operations. To enhance the robustness, Tang et al. \cite{tang2019robust} introduced a hashing algorithm that determines the saliency map using the Fourier transform phase spectrum (PFT) combined with the ring partition (RP) technique. This method extracts the statistical features, specifically the mean and variance, from the image rings, which significantly enhances the robustness against rotation. In a separate study, Zhao et al. \cite{zhao2020perceptual} developed a hash algorithm leveraging three-dimensional color structure features along with brightness gradient features. This approach not only offers improved classification performance, a shorter hash length, and reduced average hash generation time but also demonstrates strong detection capabilities in both image copy detection and image tampering scenarios.

\subsection{Hash Construction Based on Image Transform Domain}

Methods utilizing invariant feature transforms first extract robust features from the transform domain and subsequently use these coefficients to generate the final hash value. The transform domain can simultaneously express the detailed features and global contours of the image. Different transform domains can characterize different properties of the image, so the appropriate transform domain can be selected according to the needs for feature extraction. Image fingerprint algorithms that classify based on the domain where the feature vector extraction is located can be divided into two main categories: transform domain-based and time domain-based. Transform domain-based algorithms can be further subdivided into image fingerprint algorithms in the DCT domain, DFT domain, DWT domain, and Radon domain \cite{roy2013spatial}. The transform domain has the advantage of capturing the global characteristics of the image and resisting image noise, so most image fingerprint algorithms are based on the transform domain.

The fingerprints extracted through the transform domain usually have better robustness and natural resistance to conventional image operations. Furthermore, this fingerprint extraction algorithm can also combine multiple transform domains to jointly extract image features, laying a solid foundation for proposing more practical image fingerprint algorithms in the future. Tang et al. \cite{tang2014robust2} utilized key DCT coefficients from each image block to compute the hash value, resulting in an algorithm with strong resilience against brightness adjustment, JPEG compression, and contrast modification. Drawing inspiration from this approach, Liu and Huang \cite{liu2019efficient} employed invariant moments alongside 2D DCT for hash value calculation, enhancing the algorithm's resistance to geometric transformation distortions. Additionally, Tang et al. \cite{tang2014robust1} developed a method to derive discriminative hash values from color images by integrating DWT with the color vector angle (CVA). In a subsequent study, Tang et al. \cite{tang2018perceptual} applied an edge detector to extract edges and processed these results using 2D DWT. By calculating the weighted DWT coefficients across different subbands, they obtained a perceptual hash value, which proved effective in reducing reference image quality assessment artifacts. The characteristics of image digital fingerprint algorithms based on different transform domains are presented in Table \ref{table:imagetransformdomains}.

\begin{table}[ht]
    \footnotesize
    \centering
    \caption{Image fingerprints based on image transform domain.}
    \label{table:imagetransformdomains}
    \begin{tabular}{|m{1cm}<{\centering}|m{4.3cm}<{\raggedright}|m{1.7cm}<{\raggedright}|}
        \hline
        \multicolumn{1}{|c|}{\textbf{Category}} & \multicolumn{1}{c|}{\textbf{Brief description}} & \multicolumn{1}{c|}{\textbf{Papers}} \\  
        \hline
        DWT & Demonstrates strong robustness against non-malicious distortions, such as low-pass and high-pass filtering & \cite{tang2018perceptual,ahmed2010secure,lu2005geometric,lu2004robust,monga2006perceptual,karsh2017robust}\\
        \hline
        DCT& Excels in the classification and detection of duplicate copies & \cite{tang2016robust,de2005robust,tang2014robust2,tang2011lexicographical}\\
        \hline
        Radon & Maintains stability under rotation, scaling, and translation attacks, but shows weaker resistance to certain geometric transformations & \cite{paul2020image,lei2011robust,seo2004robust,wu2009novel}\\
        \hline
        DFT & Maintains a low collision probability while being resilient to alterations that preserve the content & \cite{qin2013robust,swaminathan2004image,swaminathan2006robust}\\
        \hline
            
        \end{tabular}
\end{table}

\subsection{Hash Construction Based on Image Feature Points and Edges (Local Feature Points)}

The image hash extraction methods introduced above are mostly aimed at various conventional digital image processing operations other than image rotation, but it is difficult for them to resist image rotation operations. Therefore, many studies have focused on the study of image hash extraction methods that are robust to image rotation operations. This type of extraction method mainly relies on the robust local feature points of the image (such as SURF, Harris \cite{wang2012image}, SIFT \cite{lv2012perceptual,yan2016multi}, end-stopped, etc.) to construct the image hash, mainly including the Radon transform method and the Harris corner method. The method based on local feature points relies on the invariance of feature extraction techniques to improve classification. However, key points may be repeated in near-duplicate images, reducing the identification capability. Lv et al. \cite{lv2012perceptual} developed a hash algorithm that integrates shape context with SIFT and Harris points, enabling effective detection and localization of tampered images. Building on this, Pun et al. \cite{pun2018robust} introduced a progressive feature point selection algorithm that leverages SIFT features while considering both the structural and color characteristics of the image. Paul et al. \cite{paul2019image} proposed an image hashing method based on shape context with accelerated robust features (SURF). This approach is significantly faster than SIFT-based algorithms and offers enhanced resistance to various image modifications, effectively addressing rotation sensitivity issues. Singh et al. \cite{singh2021new} utilized singular value decomposition (SVD) combined with KAZE features to generate hashes, resulting in an algorithm that merges global and local features, demonstrating strong robustness against gamma correction and geometric attacks.

\subsection{Hash Construction Based on Image Data Dimensionality Reduction}

Dimensionality reduction techniques play a crucial role in transforming low-level features from high-dimensional spaces into more manageable low-dimensional representations. These techniques, including singular value decomposition (SVD) \cite{singh2021new,mihccak2001new}, non-negative matrix factorization (NMF) \cite{monga2007robust}, fast Johnson-Lindenstrauss transform (FJLT) \cite{lv2009extended}, local linear embedding (LLE) \cite{tang2015robust1}, and compressed sensing (CS) \cite{sun2014secure}, have proven effective in capturing essential features that remain invariant to many image processing attacks. Their application to image hashing \cite{ghouti2014robust,tagliasacchi2009hash} has been particularly successful. While dimensionality reduction methods are typically effective in resisting geometric attacks, the challenge lies in balancing the extraction of robust features with the need to maintain a compact hash size. For instance, Tang et al. \cite{tang2016robust} constructed a high-dimensional matrix using DCT coefficients and utilized LLE to compute the hash value from this matrix. This approach offers good discrimination and resistance to common digital operations, though it is limited in its ability to counter small-angle rotations. In a subsequent study, Tang et al. \cite{tang2017robust} developed a method using a feature matrix constructed from the discrete Fourier transform (DFT) and log-polar transform, with the compact hash learned through multidimensional scaling (MDS). This method demonstrates robustness to various content-preserving operations, including arbitrary-angle rotation, while achieving high discrimination. However, Qin et al. \cite{qin2018perceptual} proposed combining SVD with color vector angle (CVA) for image hashing, but this approach is hampered by its high computational cost. Table \ref{table:Image-Based} provides a comparison of different image-based fingerprint extraction techniques.

\begin{table*}[ht]
    \centering
    \footnotesize
    \caption{Image-Based fingerprint extraction techniques.}
    \label{table:Image-Based}
    \begin{tabularx}{\textwidth}{|m{2cm}<{\centering}|m{3.2cm}<{\raggedright}|m{2.4cm}<{\raggedright}|m{3.8cm}<{\raggedright}|m{3.8cm}<{\raggedright}|}
        \hline
        \multicolumn{1}{|c|}{\textbf{Category}} & 
        \multicolumn{1}{c|}{\textbf{Summary}} & 
        \multicolumn{1}{c|}{\textbf{Methods}} & 
        \multicolumn{1}{c|}{\textbf{Advantages}} & 
        \multicolumn{1}{c|}{\textbf{Disadvantages}} \\  
        \hline

        Based on image statistical features & Extract hash features by calculating the global statistical properties of images & Histograms, mean, variance, color aggregation vector (CCV), color moments &  Robust against noise, blurring, and compression distortion; easy to implement. & The extracted hash length is too large, wasting storage space; global statistical features cannot represent significant image features, leading to insufficient security. \\  \hline
        
        Based on image transform domains & Robust features are first extracted from transform domains, followed by utilizing the resulting coefficients to produce the final hash values & DFT, DWT, DCT, and Radon & Robust against certain distortions and attacks; capable of capturing global image features with inherent resistance to image noise. & Cannot effectively reflect image content and has poor robustness against geometric attacks. \\  \hline
        
        Based on image feature points and edges & Robust local feature points of images are extracted to construct hashing & Radon, Harris corner & Capable of reflecting image content with good performance on JPEG compression and geometric operations. & The generated hash length varies, and the method is ineffective for images without significant corner points. \\  \hline
        
        Based on image dimensionality reduction & Map the low-level features from high dimensional space into a reduced-dimensional representation & Non-negative matrix factorization, SVD & The generated hash are compact and capable of reflecting local image content. SVD's invariance ensures image hashing's robustness with good resistance to geometric attacks. & Some degree of loss occurs during the extraction of robust features. \\   \hline
        
        Based on deep learning & The global features of image semantic level are extracted by DNN & CNN, CAE, GAN &  Automatically learn fundamental features from training samples, minimizing the need for complex, manually crafted features. & High hardware requirements; model overfitting; and high computational cost. It is less suitable for real-world scenarios that require faster detection and classification. \\
        \hline
    \end{tabularx}
\end{table*}

\section{Video Digital Fingerprint}
\subsection{Fundamentals of Video Digital Fingerprints}

The concept of video fingerprinting was first proposed by Indyk and Shivakumar at Stanford University in 1999 \cite{indyk1999finding}. The initial application's purpose was to detect pirated videos on the internet. With the rapid development of internet technologies, especially the rise of P2P technologies and user-generated content (UGC) websites \cite{gan2023web}, like YouTube, the research on video fingerprinting technology has been further advanced. A video fingerprint is a feature vector that uniquely distinguishes one video segment from another \cite{lee2008robust}. Specifically, a video fingerprint is a compact binary code (hash) generated from the perceptual features (e.g., color, texture, and motion) of the video content. Video fingerprints are robust to changes in video resolution, frame rate, and format, while also maintaining good discriminability between different video contents. The simplest way to extend perceptual hashing from images to videos is to apply an arbitrarily selected perceptual image hash to each reference video frame. However, this method has two drawbacks: first, the large number of video frames will result in thousands or even tens of thousands of separate hash codes for each video; second, this method isolates the changing video sequence into static images, ignoring the temporal relationships within the video signal, and thus cannot fully and accurately describe the perceptual content of the video.

Video fingerprint algorithms, as a key technology in multimedia content identification, have the advantage of effectively handling the identification and matching tasks of large-scale video data \cite{hampapur2001comparison}. Even if the video content undergoes various transformations such as compression, editing, and noise addition, they can maintain a high recognition accuracy. Moreover, these algorithms have good real-time performance, allowing for rapid retrieval of similar content in large-scale video databases. However, video fingerprint algorithms also have certain limitations. First, as video resolution and complexity increase, the algorithm's computational complexity and resource consumption also increase, especially when dealing with ultra-high-definition and complex scenarios. Second, while the robustness of these algorithms can handle common transformations, they may still be vulnerable to advanced adversarial attacks (such as adversarial generative network attacks). Furthermore, their performance may be inferior to that of specially designed multimodal recognition algorithms when processing multimodal videos (such as video + audio).

\subsection{Construction Methods and Classification of Video Digital Fingerprints}

The core of video fingerprint technology is to extract robust and discriminative video features from the video content. Based on the form of feature extraction, the construction methods of fingerprints can be roughly divided into three categories: spatial, temporal, and transform domain. Due to the more diverse and complex information provided by videos compared to images, several studies in video hashing algorithm design have mainly focused on learning the spatial features of videos. Spatial domain fingerprints describe the spatial features of video frames, calculated independently of other frames. Spatial features encompass brightness patterns, gradient or differential brightness patterns, and edges. Temporal domain fingerprints, on the other hand, capture the temporal characteristics of videos by measuring differences between consecutive frames. These temporal features include frame difference measures, motion vector patterns, and shot duration. Transform domain fingerprints are derived from the coefficients of image or video transforms, such as DCT and wavelet transforms. These fingerprints offer alternative descriptions and representations of certain spatial and temporal features within the transform domain \cite{lu2009video}.

\subsubsection{Spatial Fingerprint Construction Methods}

Spatial features play a crucial role in video copy detection and recognition, as they can locally or globally locate salient points in the spatial domain and withstand typical video processing steps, including lossy compression, resizing, and frame rate alterations, as well as resist geometric attacks such as scaling and rotation \cite{neelima2017collusion}. In the spatial domain, video fingerprints are implemented by extracting hash codes or feature vectors from each keyframe or every frame of the video \cite{wary2019review}. Since videos typically contain a large number of frames, processing all frames as input would result in enormous computational overhead. To solve this problem, key frame extraction is usually performed in the pre-processing step to reduce the computational load. Identifying keyframes that can effectively represent the video content is a critical task in the spatial domain, as these frames serve as important anchors in the video sequence \cite{neelima2017collusion}. Spatial domain methods primarily rely on extracting perceptual visual features, such as brightness, color, edge contours, texture, and corners, from video frames to construct video fingerprints. These image features highlight the differences between specific frames and other frames, allowing the video to be distinguished from other videos. However, a limitation of spatial domain methods is that they do not consider the temporal differences between frames (temporal information), which are a significant attribute of video sequences \cite{ghodrati2018video}.

\textbf{a) Block-based spatial feature extraction methods}: Frame subdivision divides the luminance component (Y) of video frames into a grid of smaller fixed-size blocks, often represented as a 4 $\times$ 4 grid. By subdividing the frames into smaller blocks, the generated fingerprints are robust to pixel value changes within the frames, and a compact and fixed-size frame fingerprint is produced for convenient subsequent processing and comparison. Block-based spatial signatures, such as ordinal signatures based on the ranking of block average brightness values \cite{hampapur2001comparison,hua2004robust,kim2005spatiotemporal} and differential signatures based on brightness differences between neighboring blocks \cite{oostveen2002feature,lee2006video,iwamoto2006image}, are vulnerable to geometric transformations like rotation, cropping, and scaling, which may alter the aspect ratio.

\textit{Ordinal signatures}: The steps of the ordinal signature include frame subdivision, calculation of average intensity, and block ranking \cite{bhat1998ordinal}. First, each frame image is divided into several small blocks (e.g., \( 8 \times 8 \)). These blocks are usually rectangular, each containing multiple pixels. For each block, the average intensity value of the pixels is calculated. The average intensity value reflects the brightness level within the block. The average intensity values of all blocks are then sorted in ascending order. Through this sorting process, each block is assigned a rank, indicating its position in the sorted intensity value sequence. The ranked sequence of the average intensity values is the ordinal signature of the frame. This rank sequence is used as the feature vector of the frame for subsequent video content recognition and matching \cite{mohan1998video}. The brightness sequence-based method is more robust than directly using pixel values to calculate fingerprints, but the downside is that the order between all blocks is interdependent, and when the brightness value of a block is changed, the entire order is disrupted.

\textit{Differential luminance signatures}: Differential luminance signatures are a method of generating signatures by calculating the brightness differences between video frames. This method has relatively strong robustness to local brightness changes and is applicable to video content recognition and matching. The steps of differential luminance signatures include frame subdivision, difference calculation, and difference quantization \cite{atrey2007scalable}. First, each frame image is divided into a grid of fixed-size blocks (e.g., \( 4 \times 4 \) or \( 8 \times 8 \)), ensuring that the pixel values within each block can be processed independently. For each block, the brightness difference between adjacent frames is calculated. Specifically, the average brightness value of each block in the adjacent frames is computed, and the difference between these values is obtained. The difference calculation can be performed in the horizontal, vertical, or diagonal direction to capture brightness changes in different directions. The calculated brightness difference values are then quantized into a few discrete levels. For example, the difference values can be divided into three cases: greater than, equal to, or less than a certain threshold. The purpose of quantization is to simplify the calculation and improve robustness to noise and small brightness changes. The quantized brightness difference values are used to generate the differential luminance signature for each block. These signatures can be combined to form the signature of the entire frame, used for subsequent matching and recognition.

\textbf{b) Points of interest-based spatial feature extraction methods}: Some feature points can remain stable under global geometric transformations (such as rotation and cropping) and can form effective video fingerprints to distinguish and recognize different video sequences. Representing the frame image as a set of feature points can result in a large number of feature points. Therefore, fingerprints can be generated by calculating spatial features around the points of interest in the video frames, including using Harris corner points \cite{law2006robust,joly2003robust}, scale-invariant feature points (SIFT) \cite{sarkar2008video}, and Difference-of-Gaussian scale-space feature points \cite{massoudi2006video}. Since the number and distribution of interest points in each frame are content-dependent, the point-of-interest-based spatial fingerprint methods can generate frame fingerprints of variable size, but their computational complexity is higher than the block-based methods.

\textbf{c) Color-based fingerprint construction methods}: Color is widely regarded as one of the most expressive visual features \cite{wang2013new}. Color-based features mostly come from color histograms \cite{datta2008image}, color correlograms \cite{rasheed2008image}, and dominant color descriptors (DCD) \cite{penatti2008color} of specific time and/or spatial regions in the video. RGB images are usually converted to YUV and LAB color spaces \cite{lu2009video}. Color histograms are highly sensitive to specific color spaces, which making them susceptible to variations in video formats. Additionally, color features are inapplicable to grayscale images. These histograms are also particularly sensitive to noise and saturation changes, potentially compromising the stability of the fingerprint. However, using local region color histogram features can enhance resistance to distortions and changes. In this approach, the image is divided into non-overlapping regions, with a color histogram calculated for each region. The frequencies of these histograms are then concatenated to create the feature vector \cite{lee2008robust1}. The color histogram-based method is computationally simple and suitable for fast processing of large-scale video databases. Besides, color histogram signatures can provide higher discrimination when processing video content with distinct color characteristics. The color correlogram captures the likelihood of encountering color pairs at a specific pixel distance, thereby incorporating spatial information into the analysis. As a result, it offers improved retrieval accuracy compared to traditional color histograms \cite{rasheed2008image}.

\begin{table*}[ht]
    \footnotesize
    \centering
    \caption{Video digital fingerprints}
    \label{table:video_fingerprint_domains}
    \begin{tabular}{|m{2.5cm}<{\centering}|m{5cm}<{\raggedright}|m{4cm}<{\raggedright}|m{4cm}<{\raggedright}|}
        \hline
        \multicolumn{1}{|c|}{\textbf{Type}} & 
        \multicolumn{1}{c|}{\textbf{Advantages}} & 
        \multicolumn{1}{c|}{\textbf{Disadvantages}} & 
        \multicolumn{1}{c|}{\textbf{Application scenarios}} \\
        \hline

        Based on spatial domain & Exhibits strong resilience to typical video processing operations, including compression and resizing & Difficult to select keyframes; requires large memory space & Video copy detection \\
        \hline
        Based on temporal domain & Suitable for long videos, capable of accurately locating actions & Not applicable to short video segments & Action localization, suitable for long video analysis \\
        \hline
        Based on spatio-temporal domain & More robust when handling frame rate changes and geometric transformations & Poor fault tolerance for operations like frame insertion & Advanced video analysis and copyright protection \\
        \hline
        Based on transform domain & Good adaptability to affine transformations such as translation and rotation & High computational complexity & Less suitable for large-scale video processing \\
        \hline
    \end{tabular}
\end{table*}

\subsubsection{Temporal Fingerprint Construction Methods}

Shivakumar at Stanford University pioneered the use of time as a video feature \cite{shivakumar1999detecting,indyk1999finding}. Temporal information typically includes the time difference between different shots, as well as the temporal relationships between adjacent video frames. Video hashing algorithms based on temporal features mainly utilize the temporal information of the video for hash construction and are therefore not suitable for short video clips. Temporal domain video fingerprint construction methods can be divided into key frame-based temporal feature extraction methods and adjacent frame-based temporal feature extraction methods. The earliest method for generating temporal feature-based video fingerprints involved dividing the video sequence into multiple shots and using the duration of each shot as a temporal signature. These shot durations were then concatenated in sequence to form the video fingerprint. Initially, several studies used the positions of video keyframes to propose a series of video perceptual hashing algorithms. Currently, more common temporal domain algorithms construct video perceptual hashes by the relationships between adjacent frames in the temporal domain. Chen and Stentiford proposed the use of temporal ordinal signatures \cite{chen2008video}, similar to the brightness order method in the spatial domain. The frames are divided into fixed-size blocks, and the average values of the same position blocks within a time window are ranked, as the temporal ordinal features. Another method is the temporal differential signature \cite{kim2005spatiotemporal,radhakrishnan2007content}, which calculates the brightness difference between adjacent frames or blocks, and quantizes it into a few discrete levels. Hampapur et al. \cite{hampapur2001comparison} used the quantization of block motion vectors as motion features, while Law-To et al. \cite{law2006robust} used the motion trajectories of points of interest as temporal features.

\subsubsection{Spatio-temporal Fingerprint Construction Methods}

In the field of video fingerprinting based on the spatio-temporal domain, video fingerprinting technology generates more complex hashes or feature vectors by combining the spatial and temporal features of the video, thereby improving the ability to identify video duplicates and handle video tampering. Spatio-temporal methods not only focus on the spatial information within a single frame, such as the pixel composition of the frame, but also analyze the changes between consecutive frames, i.e., the temporal information, which makes this method highly sensitive to changes in video content. Existing research has proposed some good methods, such as 3D-DCT \cite{coskun2006spatio}, temporal information representation of Images (TIRI) \cite{esmaeili2010robust1,esmaeili2010robust2,yuan2016shearlet}, video tomography and bag-of-visual-word \cite{min2012video}, Histogram of oriented gradients (HOG) and compression properties \cite{subramanyam2012video}, identifying shot-based semantic concepts along the temporal axis \cite{min2009near}, and self-similarity matrix (SSM) \cite{wu2009robust}, spatio-temporal triangle feature relationship (STTFR) \cite{sowmya2018video}. These methods utilize the spatial and temporal information in video clips or sequences to improve the performance of robust video copy or tampering detection.

\subsubsection{Transform Fingerprint Construction Methods}

Spatially partitioned features such as ordinal and differential features are easily affected by geometric transformations such as rotation, cropping, and aspect ratio changes. Moreover, in the actual application of video processing, when adjusting the frame size, it is often accompanied by translation, rotation, and scaling transformations. Some studies have been seeking features that are adaptive to geometric transformations. It is important to note that certain specialized image transformations, such as the Fourier transform, Radon transform, and SVD, exhibit strong adaptability to affine transformations like translation and rotation. Given that transform domain features offer distinct characteristics and representations of spatial and temporal features within the transform domain, some video fingerprints have leveraged these transform domain features. These features are typically derived from the coefficients of image and video wavelet transforms, discrete cosine transforms \cite{nie2010robust,sandeep2020detection}, and similar methods. Table \ref{table:video_fingerprint_domains} summarizes the characteristics of different video digital fingerprinting methods.

\section{Audio Digital Fingerprints} \label{sec:audiodf}

Audio is a very commonly used information transmission medium. However, certain challenges, such as audio copyright detection and music recognition, necessitate a robust, generic audio recognition system. The primary difficulty in developing such a system lies in the requirement to search through a vast database of audio tracks using only very short and noisy input audio. Therefore, the urgent task is to reduce this highly complex search problem, which has led to the concept of audio fingerprint recognition \cite{gupta2022audio}. Audio fingerprints are content-based and are compact digital signatures that represent the important acoustic features of a piece of music, obtained by applying a hash function to the digital audio signal.

\subsection{Concept and Application of Audio Fingerprint Technology}

In the audio fingerprint generation process, feature extraction and fingerprint modeling are key steps. First, the incoming digital audio signal is divided into overlapping frames, and a series of features are extracted from each frame using specific algorithms, such as fast Fourier coefficients (FFT), mel-frequency cepstral coefficients (MFCC) \cite{htun2019analytical}, etc. After extracting the audio fingerprint, fingerprint modeling usually uses classification techniques such as principal component analysis (PCA) \cite{martinez2001pca}, linear discriminant analysis (LDA), hidden Markov models (HMM), or quantization techniques to map the fingerprint features to a more concise representation for storage and retrieval. In the field of audio fingerprint feature extraction, improvements are often made to the traditional Philips method \cite{haitsma2002highly} and the Shazam method \cite{wang2003industrial}.

The most famous application of audio fingerprint technology is the music recognition system, which allows users to identify unknown songs through microphone input or real-time audio streams. In addition, its applications have also expanded to copyright detection, removal of duplicate audio files, broadcast content monitoring, and advertising placement tracking. The main advantage of audio fingerprint technology is its ability to efficiently and accurately identify audio content in large-scale audio libraries, and its strong robustness to handle a certain degree of audio deformation and noise interference. However, the recognition accuracy of audio fingerprint algorithms may decline under extreme conditions, and there are still issues of computational complexity in large-scale data processing. With the advancement of technology, machine learning, and deep learning are being integrated into audio fingerprint algorithms, using models such as CNNs to improve the accuracy of audio feature extraction, while also optimizing cross-platform compatibility and enhancing privacy protection to meet the diverse needs of the Internet of Things and mobile devices. Table \ref{table:audio_features} outlines various audio feature extraction techniques.

\subsection{Philips Audio Fingerprint Algorithm}

The Philips audio fingerprint algorithm proposed by Haitsma and Kalker \cite{haitsma2002highly} is one of the most classic audio fingerprint algorithms. The Philips algorithm is a frequency sub-band-based audio fingerprint extraction algorithm, which includes frame partitioning, Fourier transform, sub-band division, sub-band energy calculation and filtering, and finally hashing the sub-band energy differences of each signal frame into sub-fingerprints. The Philips audio fingerprint has good robustness to various signal distortions and can maintain retrieval performance even when the audio signal is interfered with. Moreover, its granularity is relatively short, about 3 seconds, which is the minimum length of the audio clip required for retrieval. Among the many audio fingerprint extraction algorithms, the Philips algorithm is widely regarded as one of the most reliable content-based fingerprint extraction algorithms after data analysis, but its performance is not ideal in high-noise environments and when the signal has relatively large linear speed changes.

Based on the Philips algorithm, existing research has made improvements to varying degrees. For example, Yang et al. \cite{yang2014efficient} performed secondary sampling on the original fingerprint, only retaining one-fourth of the original data, reducing memory requirements and improving search speed, while maintaining good robustness and reliability. Chu et al. \cite{chu2020peak} introduced a peak-based sub-band energy calculation method to improve the Philips fingerprint's resistance to pitch shifts, resulting in the Peak-based Philips Fingerprint (PPF). Chen et al. \cite{chen2017audio} proposed an enhanced version of the Philips fingerprint utilizing wavelet transform, which significantly reduces audio retrieval time to just 1 second while still maintaining a relatively high level of accuracy.

\subsection{Shazam Audio Fingerprint Algorithm}

Wang et al. \cite{wang2003industrial} proposed the Shazam fingerprint recognition, which is a spectral energy peak-based audio fingerprint extraction algorithm. The Shazam fingerprint extraction uses a sliding window to identify the spectral peaks in the audio clip's spectrogram and then combines the positions of the peaks in the frequency domain and time domain to form the fingerprint. Shazam fingerprints are robust to most attacks (including GSM distortion \cite{huerta2001distortion,}), but are sensitive to pitch shift attacks. However, the number and overall data volume of the fingerprints generated by this algorithm are relatively large. As the database size increases, its retrieval performance significantly decreases. Many scholars have made improvements based on the Shazam algorithm. Kishor et al. \cite{kishor2023audio} developed an efficient and robust audio fingerprint recognition method for song detection, drawing on the Shazam fingerprint algorithm. This approach combines spectral and temporal features extracted from the audio signal to create a compact and unique fingerprint for each song, demonstrating strong resistance to noise, foreground sounds, and audio compression. Sun et al. \cite{sun2018movie} proposed a feature point pair selection scheme based on dynamic regions, applying the Shazam algorithm to audio retrieval and improving it to better suit movie audio retrieval. Anguera et al. \cite{anguera2012mask} combined the Philips and Shazam methods and proposed an audio fingerprint called masked audio spectral key points (MASK). This method constructs the fingerprint calculation region centered on the spectral peaks and calculates the audio fingerprint based on the energy difference. This method has high robustness and is suitable for various audio mixtures such as music and speech. Gupta et al. \cite{gupta2022audio} introduced an audio fingerprint recognition method using two-stage feature extraction.  This method, derived from the Shazam algorithm, identifies the strongest peaks in the Mel-spectrogram, making it highly noise-resistant while reducing the audio representation from 3D to 2D by eliminating the amplitude factor.

\begin{table*}[ht]
    \footnotesize
    \centering
    \caption{Audio feature extraction techniques}
    \label{table:audio_features}
    \begin{tabular}{|m{2cm}<{\centering}|m{4cm}<{\raggedright}|m{10cm}<{\raggedright}|}   
        \hline
        \multicolumn{1}{|c|}{\textbf{Audio feature}} & 
        \multicolumn{1}{c|}{\textbf{Brief description}} & 
        \multicolumn{1}{c|}{\textbf{Methods}} \\ 
        \hline
       
        Time domain features & Time domain graph shows the signal variation w.r.t. time. & Zero crossing rate, modified zcr, amplitude-based features: amplitude descriptor, attach delay sustain release, log attack time, shimmer; energy-based features: short time energy, volume, temporal centroid; auto-correlation based features; rhythm-based features: speech duration, articulation rate, phoneme duration, pause ratio, beat histogram \\   \hline
        
        Frequency domain features  &  To analyze a signal's frequency, it can be transformed from the time domain to the frequency domain using Fourier transform or autoregression analysis. & Auto-regression based features(e.g., LPC, CELP, LSF); peak Frequency; STFT/time-frequency based: time-frequency matrix, sub-band energy ratio, spectrum envelope, SPSF, GDF; envelope modulation spectrum (EMS) based; long-term average spectrum (LTAS) based; chroma related features; tonality based features(e.g., FF, pitch histogram, pitch profile, harmonicity); spectrum shape based features(e.g., spectral centroid, spectral roll-off, spectral flatness) \\    \hline
        
        The cepstral domain & By applying the logarithm to the signal spectrum and then performing an inverse Fourier transform, a result known as the cepstrum is obtained. & MFCCs, linear prediction cepstral coefficients, perceptual linear prediction (PLP) cepstral coefficients, relative-spectral PLP (RASTA-PLP) feature, greenwood function cepstral coefficients (GFCCs), gammatone cepstral coefficients (GTCCs) \\    \hline

        Wavelet-based features & Wavelet transform converts time-domain audio signals to time-frequency representation by computing inner products with wavelet family members. & DWT-based features, CWT-based features, wavelet transform-based features, wavelet packet decomposition-based \\    \hline

        Image-Based Features& Convert the audio signal into a spectrogram or spectrogram and extract local features and texture information from image. & Local binary patterns, local ternary patterns, histogram of gradients (HOG) feature, SIFT \\    \hline
        
        Deep features & Deep learning is powerful for extracting high-level features from low-level information. & CNN, DNNs, RNNs, deep stacked auto-encoder (SAE), unidirectional LSTM, bi-directional long short-term memory (BLSTM) \\
        \hline
    \end{tabular}
\end{table*}

\section{Deep Hashing}  \label{sec:deephashing}

In recent years, deep neural networks have achieved widespread applications and breakthrough progress in fields such as natural language processing and computer vision, thanks to their powerful representational capacity. Compared to traditional feature extraction methods, deep learning models can automatically extract richer and more useful features from raw data, without the need for manual design of complex feature extraction pipelines. Therefore, more and more researchers have turned their attention to deep hashing, which uses deep learning to learn the hash feature representation of data. By extending traditional hash functions to DNNs, deep hashing implements the process of mapping data to binary codes (hash codes).

\subsection{Text}

The earliest research on deep hashing in text comes from the semantic hashing proposed by Hinton et al. \cite{salakhutdinov2009semantic} in 2009, the core idea of which is to map the vector based on word frequency statistics to hash codes through deep learning, using the obtained hash codes directly as the memory addresses of documents. Subsequently, Hinton et al. \cite{hinton2011discovering} further proposed deep generative models , building a four-layer neural network based on semantic hashing to learn the hash function through training. In 2013, Masci et al. \cite{masci2013multimodal} proposed a deep learning-based multimodal data hashing, where the model first uses the bag-of-words model to represent the text input, and then learns the text hash features through a deep learning model. In 2015, Xu et al. \cite{xu2015convolutional} proposed a new CNN-based text hashing framework to address the poor performance of existing methods in preserving semantic similarity, by combining keyword features and implicit features to generate compact binary codes, thereby better preserving semantic similarity. This method performs excellently in short text similarity search and is applicable to large-scale similarity search. In 2017, Zhang et al. \cite{zhang2015mixed} introduced semantic cross-modal hashing (SCMH) for approximate duplicate detection and cross-modal retrieval tasks. This approach utilizes continuous word representations to capture semantic text similarity and employs a deep belief network (DBN) to establish correlations across different modalities, effectively managing semantic-level text similarity. Chaidaroon et al. \cite{chaidaroon2017variational} then proposed three new deep document generation models, forming an encoder-decoder DNN for text hashing. Later, Chaidaroon et al. \cite{chaidaroon2018deep} proposed the neighborhood regression (NbrReg) model, which uses additional neighborhood information for generating hash codes. This method alleviates the problem of insufficient labeled data by utilizing unsupervised ranking to approximate the true text space. In 2020, Doan et al. \cite{doan2020efficient} introduced the denoising adversarial binary autoencoder (DABA) for efficient implicit unsupervised text hashing, thus directly learning the hash function from raw text data. This model effectively captures the semantic and syntactic dependencies of text documents, enabling it to encode a structured representation within the learned hash functions. In 2023, He et al. \cite{he2023efficient} proposed two robust text-based semantic hashing methods: MASH and SMASH, aimed at solving key problems in similar text search. These methods generate compact and balanced hash codes using an encoder-decoder framework and perform well in handling large-scale datasets and noisy conditions.

\subsection{Images}

Although learning-based methods can generate high-quality hash values, they are highly dependent on self-training or deep training, leading to high computational time costs, which limits their applicability in actual applications requiring rapid detection and classification \cite{roy2023various}. In addition, these methods typically require a large amount of raw images for training and still have shortcomings in resisting the diversity of transformations. The following introduces the use of several typical DNNs to realize deep hashing for image authentication.

\textbf{Convolutional neural networks (CNNs)-based deep image hashing}. CNNs \cite{li2021survey} reduce memory usage and the number of parameters through convolutional structures, alleviate overfitting problems, and improve feature extraction performance. CNNs are widely used in fields such as object detection and image classification. CNN-based image hashing schemes use the trained network model to extract feature matrices, and then generate quantized hash sequences for image authentication. In 2014, Xia et al. \cite{xia2014supervised} introduced convolutional neural network hashing (CNNH), the first method to use CNNs for simultaneous feature and hash function learning. It involves two stages: hash coding and hash function learning. While pioneering in using deep learning for perceptual hashing, this approach is not end-to-end, as the combination of multiple encoders and decoders is manually designed. In 2016, Liu et al. \cite{liu2016deep} developed a CNN architecture that learns compact binary codes by training on pairs of similar and dissimilar images, aiming to output discrete values (e.g., +1/-1). In 2018, Jiang et al. \cite{jiang2018perceptual} proposed a perceptual image hashing scheme using deep CNNs, training an AlexNet model to extract image feature matrices. By 2020, Qin et al. \cite{qin2020perceptual} had introduced a multi-constrained CNN-based perceptual image hashing scheme, where the network learns feature extraction automatically to generate the final hash sequence, focusing on image authentication rather than countering malicious editing. In 2022, Sun et al. \cite{sun2022deep} proposed a CNN-based perceptual image hashing method using hash centers and classification, introducing center quantization to reduce the required training images while balancing perceptual robustness and discriminability for image copyright protection. Despite advancements in enhancing perceptual robustness, new approaches are still needed to address increasingly complex operations. In 2023, Zhou et al. \cite{zhou2023perceptual} generated two new architectures called NASRes and NASCoNt, by designing a new dataset containing complex content-preserving operations and optimizing the ConvNeXt architecture using neural architecture search (NAS), significantly improving the robustness and discriminability of perceptual authentication hashing under complex operations. Meng et al. \cite{zhaoxiong2019perceptual} proposed an improved perceptual hashing method based on machine learning for digital rights management systems. By generating an image set through image pre-processing, extracting image features using a CNN, and computing perceptual hash values through machine learning, this method can handle various image transformations. Based on \cite{zhaoxiong2019perceptual}, a CNN-based image group perceptual hashing scheme was proposed \cite{yusei2023cnn}, which generates the same perceptual hash for all images in the group, making content management in actual applications easier. Compared to generating perceptual hashes separately for each image in the group, this method reduces the computational cost of fine-tuning the CNN.

\textbf{Convolutional autoencoder (CAE)-based deep image hashing}. The weight-sharing feature of CNNs makes them more robust to overfitting, while the CAE \cite{zhai2018autoencoder} achieves weight-sharing and two-dimensional spatial information preservation through local connections and convolutional operations. CAE integrates the strengths of both autoencoders and CNNs, enabling it to learn meaningful features from unlabeled data (or images) and produce a concise, effective representation (short-length array) of the input data. CAE-based image hashing schemes map the input image to low-dimensional hash codes through the trained encoder and then generate the image mapping through the decoder. By comparing the hash code mapping of the reference image and the received image, the tampered region in the image can be accurately located. In 2020, Li et al. \cite{li2019robust} first trained a hierarchical denoising autoencoder (DAE)-based image fingerprint computation network to gradually improve its robustness to content-preserving distortions. In 2021, Paul et al. \cite{paul2021robust} proposed an image hashing technique based on a convolutional stacked denoising autoencoder (CSDAE) to solve the problem of composite rotation-scale-translation (RST) distortion and tampering in color images.

\textbf{Generative adversarial networks (GANs)-based deep image hashing}. GANs \cite{borji2019pros} can generate clear and realistic images without complex computations. GAN-based image hashing schemes use real images and various images synthesized by the generative model to expand the training samples, further improving the model's generalization ability and learning compact binary hash values. GANs mainly consist of two parts: the generator and the discriminator. The generator, as a form of CAE, generates images by learning noisy inputs. The discriminator is a CNN-based discriminant network that is responsible for extracting data features to distinguish real images and GAN-synthesized images. Through the adversarial training between the two, the generator and discriminator compete and progress together, ultimately reaching a balanced state, thereby generating more realistic data for input. In 2018, Yarlagadda et al. \cite{yarlagadda2018satellite} introduced a GAN-based method for detecting and localizing satellite image forgeries, which also aids in feature representation learning of raw satellite images. In 2019, Wang et al. \cite{wang2019wegan} developed WeGAN, a deep image hashing method that uses a weighted GAN to generate hash codes, improving image retrieval performance by leveraging the uncertain relationship between images and labels. In 2019, Jin et al. \cite{jin2019deep} proposed a VAE-GAN-based hashing framework that combines variational autoencoder (VAE) and GANs to generate content-preserving images for pairwise hashing learning, for fast image retrieval. In 2022, Bin et al. \cite{bin2023image} extracted key information about the perceptual image content through a bidirectional GAN, thereby improving the robustness and fragility of the perceptual hash codes for the same image.

\subsection{Videos}

Neural network-based video fingerprinting methods \cite{kordopatis2017near,li2021video} initially focused mainly on single-faceted features, such as spatial features or temporal features, but have gradually shifted towards more holistic and comprehensive strategies in recent years. In addition to 3D CNN-based methods \cite{li2020two} that can simultaneously capture spatio-temporal integrated features, some transitional schemes have also emerged, which combine traditional image processing tools (such as SURF, TIRI, BoVW, etc.) with neural network methods. Neural network-based video fingerprint recognition methods rely on various types of neural networks, such as VGGNet \cite{muhammad2018pre}, AlexNet \cite{zhang2018video}, and GoogleNet \cite{zhong2015high} for extracting spatial features, and LSTM and RNN for extracting temporal features. More and more research is starting to focus on the combination of CNNs and other neural networks to extract integrated spatial and temporal features, thereby improving the accuracy and robustness of video fingerprint recognition.

\textbf{CNN-based deep video hashing}. Relying solely on CNNs to extract spatial features leverages the powerful visual capabilities of CNNs to capture the spatial characteristics of video frames. However, this approach primarily focuses on individual frames and tends to overlook the temporal features of the video. Lou et al. \cite{lou2017compact} introduced the nested invariant pooling (NIP) method to produce compact and robust CNN descriptors by using three different pooling operations to the output of the last convolutional layer of the input video frame. Kordopatis-Zilos et al. \cite{kordopatis2017near} proposed a layer-based CNN (CNN-L) feature aggregation scheme that uses the bag-of-words model to extract and aggregate histograms of intermediate CNN features. Zhang et al. \cite{zhang2018video} employed the output of the 6th fully connected (FC6) layer of the well-known AlexNet model as the key frame-level representation and then used an exhaustive search-based method to retrieve video copies from the database. Liu et al. \cite{mengyang2018content} used a CNN model to detect object regions within video frames and generated binary fingerprints from these regions for rapid copy detection. Li et al. \cite{li2020two} utilized a 3D-CNN model to directly extract features from the video stream, converting the multi-class classification problem into multiple binary classification tasks for copy detection.

\textbf{Combination of spatial and temporal features}  More and more research is focusing on the combination of CNNs and other neural networks to extract integrated spatial and temporal features. Li et al. \cite{xinwei2021compact} designed a capsule network with a 3D/2D hybrid convolutional module, which directly maps the raw data to a compact real-valued vector, with a 32-dimensional fingerprint as the output. The method proposed by Li et al. \cite{li2021video} is based on a quadruplet fully connected CNN, with four 3D ResNet-50 networks for extracting spatio-temporal features at its core, realizing an end-to-end mapping from the original video to binary code. Nie et al. \cite{nie2021classification} proposed a classification-enhancement deep hashing (CEDH) method. First, a CNN network is used to extract semantic features. It then uses an LSTM network to capture the temporal structure and relationships between different video frames. Finally, it adopts a classification module after the fully connected layer of the proposed LSTM to enhance the supervisory information. Anuranji et al. \cite{anuranji2020supervised} proposed a joint network model of a supervised stacked heterogeneous convolutional multi-kernel (Stacked HetConv-MK) - bidirectional long short-term memory (BiDLSTM) network model, which effectively encodes the rich structural and discriminative feature sequences in the video to estimate compact binary codes. Hu et al. \cite{hu2018learning} proposed a new video copy detection method based on spatio-temporal features, combining CNN and RNN. First, they use a residual CNN (ResNet) to extract frame-level content features, and then use a SiameseLSTM architecture for spatio-temporal fusion and sequence matching. Zhao et al. \cite{zhao2023tastnet} proposed a robust video content authentication fingerprinting scheme called TASTNet, which has a 2D attention mechanism and utilizes CNN and LSTM for spatio-temporal weighted fusion. This can automatically extract key spatio-temporal features from the input video and map them to the corresponding fingerprint. Zhang et al. \cite{zhang2023short} proposed to combine the R(2+1)D network and triplet network with shared weight parameters, fully exploring the spatio-temporal information of short videos and the correlation between videos, forming an effective visual fingerprint.

\subsection{Audio}

Deep features have been widely used in various audio signal processing fields such as acoustic scene classification, speaker recognition, and audio-visual analysis since 2010. DNNs can extract more complex and discriminative audio features, thereby providing more accurate recognition capabilities in complex audio environments (such as background noise, mixing, audio clipping, etc.).

Báez-Suárez et al. \cite{baez2020samaf} proposed a sequence-to-sequence autoencoder model (SAMAF) for audio fingerprint recognition. The deep learning framework SAMAF solves the audio recognition task through a sequence-to-sequence autoencoder model composed of two connected RNNs. Saravanos et al. \cite{saravanos2020audio} introduced an audio fingerprinting scheme that integrates sparse coding with deep learning techniques. This approach creates a unique and concise representation of the audio signal by applying a dictionary, which is learned using the K-SVD algorithm on a song database. Chang et al. \cite{chang2021neural} proposed a neural audio fingerprinting algorithm based on contrastive learning, for robust and high-accuracy audio retrieval. Zhang et al. \cite{zhang2023short} combined MASK and CRNN to extract audio features and generate audio fingerprints. The MASK method generates binary audio fingerprints by transforming the audio signal to the time-frequency domain, selecting the salient points of the spectrum, and encoding the regional energy. CRNN extracts high-level semantic features from audio using a combination of CNNs and RNNs. Kamuni et al. \cite{kamuni2024advancing} proposed an audio fingerprinting algorithm that integrates AI to improve accuracy. This study is based on the Dejavu project \footnote{https://github.com/worldveil/dejavu}, focusing on simulating real-world scenarios under various background noise and distortion conditions. Signal processing is the core of the Dejavu model, including fast Fourier transform (FFT), spectrogram, and peak extraction. The "constellation" concept and fingerprint hashing technology enable unique song recognition.

Deepsheka et al. \cite{deepsheka2020recurrent} proposed a method based on long short-term memory (LSTM) neural networks for cover song detection. This method is robust to time changes, key changes, and small local rhythmic deviations, and can effectively solve the problems faced by traditional music recognition systems due to different pitches and excessive noise. Wu et al. \cite{wu2022asymmetric} employed asymmetric contrastive learning to generate binary hash fingerprints, resulting in high recognition accuracy, rapid query speed, and minimal storage requirements. This scheme not only achieves a high top-1 hit rate on music and speech datasets but also offers faster query times and reduced storage needs. Arunakumari et al. \cite{arunakumari2023fingerprint} proposed using CNNs to process audio data to discover its unique features. This method converts the audio signal into a spectrogram and then extracts features robust to noise and compression through convolutional and pooling layers. Compared to other audio fingerprinting methods, this CNN-based approach has better scalability and accuracy when handling large-scale audio datasets and dealing with various audio distortions.

\section{Applications of Digital Fingerprints}
\label{section:applications}
\subsection{Content Recognition and Retrieval}

Perceptual hashing technology provides an effective means for the fast and reliable identification of multimedia content by generating short and robust digital fingerprints. Its core is to accurately retrieve relevant content by measuring the distance between the query fingerprint and the fingerprints in the database. Fingerprints, as a compressed representation of the original data, greatly reduce storage requirements and bandwidth consumption, while improving retrieval efficiency. Furthermore, the robustness of fingerprints ensures accurate and reliable retrieval even when the data is damaged or changed. This technology provides key support for solving the management and indexing problems of massive multimedia content.

(1) Image recognition and retrieval: The TinEye reverse image search engine based on image fingerprints provides image content identification services to help users find different versions of the same image or other visually similar images \cite{adrakatti2016search}. In comparison, Google's reverse image search \footnote{images.google.com} combines image content recognition and text-based search technology. It can not only find other images visually similar to the uploaded image, but also provide contextual information such as the web page content, image title, and metadata where the image is located.

\begin{figure}[ht]
    \centering
    \includegraphics[trim=135 0 135 0, clip, scale=0.35]{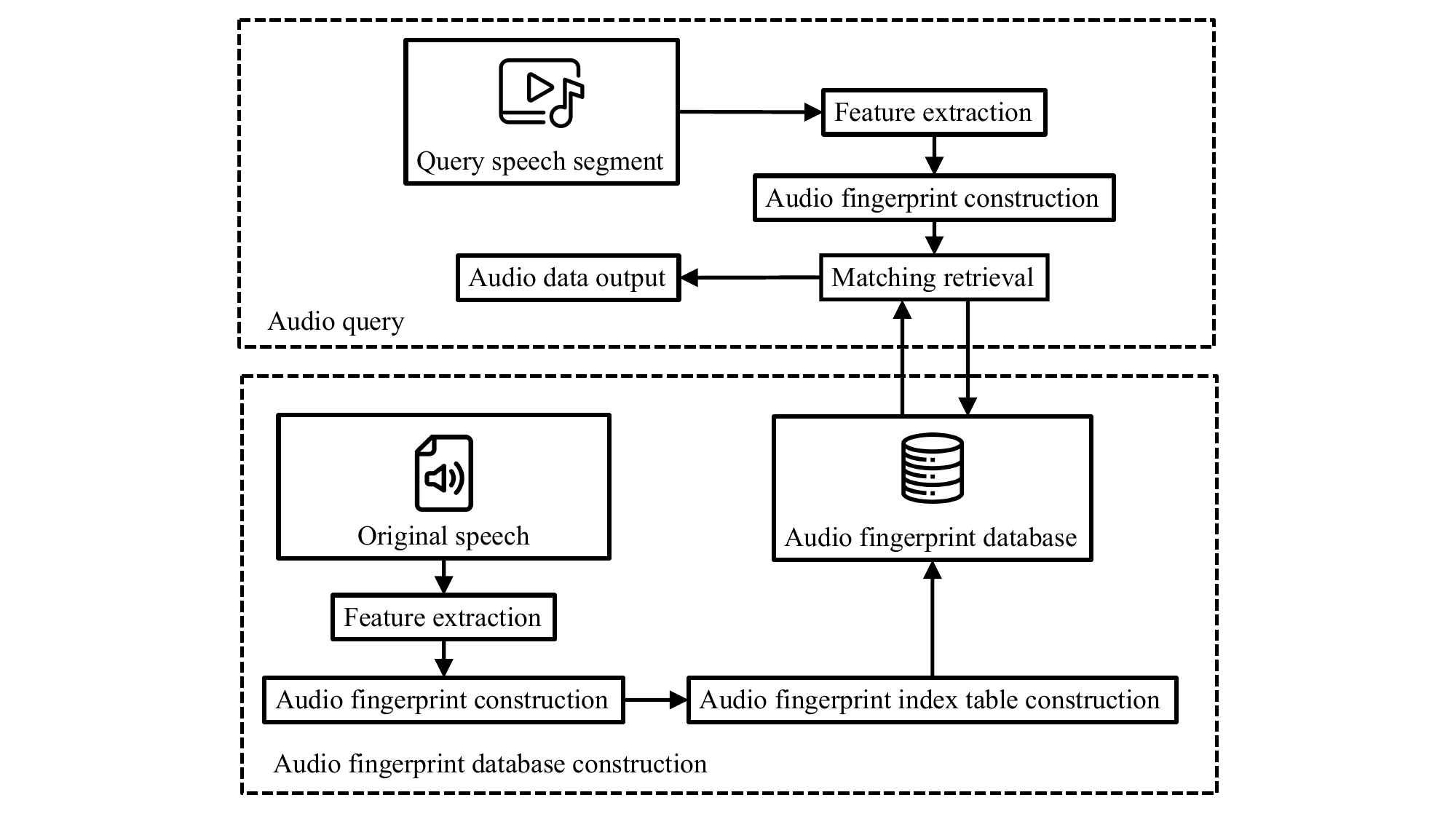}
    \caption{A fingerprint-based audio retrieval system.}
    \label{fig: AudioRetrievalSystemWorkflow}
\end{figure}

(2) Music recognition and retrieval: Industrial systems such as Music2Share \cite{kalker2004music2share}, Shazam \footnote{https://www.shazam.com/}, Midomi \footnote{https://www.midomi.com/}, and Snocap \cite{doets2008distortion} provide mature fingerprint technologies for music sharing, search, and automatic music recognition applications. In August 2002, Shazam Entertainment Ltd. in the UK launched an audio recognition service based on digital audio fingerprint technology \cite{wang2003industrial}. In 2004, Gracenote in the U.S. collaborated with Philips research to develop the "Gracenote Mobile" \footnote{https://gracenote.com/} that combines digital audio fingerprint matching and waveform fingerprint information database, establishing an audio retrieval service platform. Amena in Spain adopted Philips' digital audio fingerprint technology to launch the Musiwave audio recognition service \cite{haitsma2002highly}. Kuwo Technology Co., Ltd. in Beijing developed the "Kuwo MP3 Companion" \footnote{http://www.kuwo.cn} audio recognition software and established a digital audio fingerprint database, providing audio recognition services for users over the Internet. Figure \ref{fig: AudioRetrievalSystemWorkflow} illustrates the retrieval process of an audio system based on fingerprinting.

\subsection{Authentication}

Digital images and digital videos were once considered reliable, but with the widespread use of low-cost and user-friendly image and video editing software such as Adobe \footnote{https://www.adobe.com/} and Lightworks \footnote{https://lwks.com/}, even individuals without professional skills can easily modify the content. If malicious attackers tamper with users' sensitive or private images, it may have serious consequences. Therefore, in many practical application scenarios, verifying the authenticity of images and videos has become a critical issue \cite{qin2020perceptual}. To address this challenge, previous studies have proposed various digital watermarking methods for the authenticity verification of images and videos over the past two decades, but the distortion caused by watermark embedding has limited their application. In recent years, perceptual hashing technology \cite{farid2021overview}, as an emerging technology in the field of multimedia security, has gradually attracted widespread attention. Perceptual hashing algorithms generate fingerprints that can highly distinguish different content and are sensitive to minor modifications, providing an effective method for multimedia content authentication. In the content authentication process, the system extracts the fingerprint of the current multimedia content and compares it with the original fingerprint stored in the database. By calculating the similarity between the two, it can effectively detect whether the content has been tampered with or forged. Thus, perceptual hashing technology achieves efficient content authenticity verification without affecting the quality of multimedia, becoming an important technological means in multimedia content authentication.

\subsection{Broadcast Monitoring}

Broadcast monitoring is one of the earliest applications of fingerprint identification \cite{herre2003content}, a method of tracking and recording broadcast content, mainly for copyright fee collection, program verification, and audience statistics. This monitoring is passive, not modifying or interfering with the broadcast content itself, but rather collecting and analyzing the information of the broadcast content in real-time for subsequent analysis and reporting. In a broadcast monitoring system based on digital fingerprints, first, the system at the monitoring site extracts the digital fingerprints of the content from the locally received broadcast channels in real-time; then, the central site is responsible for collecting the extracted digital fingerprints from the various monitoring sites; next, the fingerprint server receives and analyzes the collected data, using the massive fingerprint information stored in its database to identify and classify the received fingerprints, and after comparison and analysis, the fingerprint server can generate playlists for each broadcast channel, providing accurate data support for copyright fee collection, program verification, and audience statistics \cite{oostveen2002feature}.

\subsection{Multimedia Data Filtering}

The internet is flooded with a large amount of content that is not appropriate to be publicly released and disseminated, with this problematic content mainly involving privacy, pornography, violence, and violations of other countries' customs, ethnic traditions, and ethical standards. To address this challenge, digital fingerprint technology has been widely applied as an important regulatory tool that can effectively distinguish and manage multimedia content.

In 2009, facing the detection of child sexual abuse material (CSAM), Microsoft developed the perceptual image hashing technology named PhotoDNA \cite{steinebach2023analysis} and tested and deployed it on Microsoft's Bing search engine and the then-called SkyDrive cloud service. The adoption of PhotoDNA significantly increased the number of reports of missing and exploited children. Subsequently, companies such as Meta, X, and Google also deployed this technology. In August 2019, Meta open-sourced the digital fingerprint-based photo or video matching algorithms (PDQ and TMK+PDQF)\footnote{https://github.com/facebook/ThreatExchange/tree/main/tmk} used to identify child sexual exploitation, terrorist propaganda, and image violence. Currently, PhotoDNA and PDQ are widely used, with reports indicating that over 70 companies are using PhotoDNA, and many other companies are using PDQ. In 2016, the Counter Extremism Project (CEP) developed eGlyph \cite{amit2021countering}, a perceptual hashing technology for audio, images, and videos, to analyze and identify relevant videos on YouTube for combating extremism. Through eGlyph, 1,348 ISIS videos uploaded by 278 accounts with a total of 163,391 views were discovered, revealing the deficiencies in YouTube's active review of terrorist content. In August 2021, Apple announced the development of a technology called NeuralHash \cite{struppek2022learning} to detect known CSAM in images stored in Apple's iCloud. Compared to PhotoDNA, which performs image hashing and comparison when an image is uploaded to the service, Apple's NeuralHash executes the image hashing and comparison directly on the device.

\subsection{Content-based Copy Detection}

Multimedia duplicate content detection \cite{wu2009real} based on digital fingerprints is mainly used to solve the problem of piracy and manage large multimedia databases \cite{wary2019review}. This type of technology generates compact signatures or hash codes of digital media content, allowing accurate identification of suspected content matching the multimedia files in the database without modifying the original content, thereby reliably detecting multimedia content duplication. Fingerprint-based anti-piracy search can prevent pirates from claiming ownership by modifying images \cite{monga2006perceptual}. Pirates may steal and publish images by compressing or geometrically distorting them. The true content owner can use web crawlers to calculate the hash values of webpage images and match them with their images to identify pirates. Compared to watermarking technology, digital fingerprint technology can be used for already published media content and is more effective in terms of distinctiveness and resistance to content transformations.

In addition, digital fingerprint technology also demonstrates strong capabilities in text plagiarism detection, helping to protect the rights of content creators and prevent academic plagiarism. This technology performs excellently in detecting local text reuse, reducing the search dimensions by using a selected set of fingerprints, and improving detection performance. This approach does not rely on the overall word frequency of the document but identifies text reuse based on the matching fingerprints. Some well-known text plagiarism detection systems that apply this technology include stiff \cite{manber1994finding}, COPS \cite{brin1995copy}, KOALA \cite{heintze1996scalable}, shingling \cite{broder1997syntactic}, MDR \cite{monostori2001parallel,monostori2001suffix}, I-Match \cite{chowdhury2002collection}, and Winnowing \cite{schleimer2003winnowing}. In 1996, Heintze developed the web-based plagiarism detection prototype KOALA using digital fingerprint technology and released it online. In 2003, Schleimer et al. \cite{schleimer2003winnowing} developed the online plagiarism detection service website MOSS using the digital fingerprint-based Winnowing algorithm.

\subsection{Data Deduplication}

With the development of the internet, the surge in user-generated content has led to the phenomenon of redundancy in network data, especially on video storage platforms, where videos in different encoding formats are seen as unique at the system level, but are redundant in terms of content \cite{katiyar2011videdup}. Traditional byte-level deduplication methods are ineffective in dealing with such situations, while content-aware deduplication technology, by understanding the characteristics and semantics of the data, not only detects byte-level similarities but also analyzes higher-level similarities, such as file format and content meaning. This technology is particularly suitable for multimedia file processing, effectively reducing storage space usage while maintaining data integrity and availability, by identifying similarities and retaining high-quality versions. The focus on identifying and reducing redundancy in highly similar videos or images based on visual content rather than binary data is increasingly attracting attention from academia and industry \cite{xia2016comprehensive}.

In cloud storage services, digital fingerprint technology allows clients to generate and upload fingerprints to the server to check for data duplication, ensuring that unique copies of the data are stored on the server. This method significantly reduces storage costs and the space and maintenance requirements associated with storing multiple copies of the same multimedia content. In a network environment, it is estimated that up to 40\% of pages are duplicate content of other pages. Digital fingerprint technologies such as Shingling \cite{broder1997syntactic,rajaraman2011mining}, Rabin \cite{rabin1981fingerprinting}, and SimHash \cite{charikar2002similarity} have been widely applied in search engines to efficiently perform web page deduplication. This not only improves search speed and saves storage space but also enhances search accuracy.

\subsection{Client-Side Scanning}

End-to-end encryption (E2EE) offers users robust technical protection, but it raises concerns among governments and law enforcement agencies about the potential for undetected sharing of illegal content. Client-side scanning (CSS) \cite{geierhaas2023attitudes} is proposed as a solution to this issue, using perceptual hashing to detect known illegal content before it is shared, thus preventing its spread while preserving encryption. CSS operates by scanning the user's device for illegal content, typically before the content is encrypted and transmitted. The goal of this technology is to identify copies of known illegal images, whether exact duplicates or altered versions, without exposing the database of illegal images \cite{jain2023deep}. In perceptual hashing-based CSS, the algorithm deploys a perceptual hashing function and a database of known illegal image hashes on the user's device. When a new image is checked, it is hashed using the perceptual hashing algorithm, and the resulting hash is compared against the database. If a match is found, the image is flagged as illegal \cite{jain2022adversarial}. Perceptual hashing is widely used in client-side scanning and has been adopted by several major tech companies.

\subsection{Large Model-based Application}

The recently emerged large language models (LLMs) \cite{gan2023model,zeng2023distributed} have demonstrated impressive language processing capabilities, especially in tasks such as generation, summarization, and question answering \cite{gan2023large,zeng2023large,lai2023large}. LLMs based on architectures like Transformer \cite{vaswani2017attention} and Mixture-of-Experts \cite{yuksel2012twenty}, leveraging large-scale pre-training and parameter scaling, can efficiently process natural language tasks and have been widely applied in various domains, including specialized fields like medicine. The advantages of LLMs lie not only in their powerful semantic understanding and generation abilities but also in their adaptability to diverse task requirements through fine-tuning and prompt engineering, enabling efficient and accurate performance across different application scenarios. The high development cost and vulnerability to theft of LLMs make companies highly protective of their intellectual property (IP). LLM-based fingerprinting has emerged as an important means to protect the IP of large-scale language models in recent years. LLMs possess unique writing styles, and large model fingerprinting techniques capture these distinctive lexical and syntactic features to identify and differentiate specific models or model families \cite{mcgovern2024your}. While large model watermarking and fingerprinting are often used interchangeably, the key difference lies in that model watermarking embeds identifiers in the generated content to trace the source, linking the output to the model, while model fingerprinting analyzes the model itself to determine if it is a variant of another model, establishing model-to-model relationships\cite{russinovich2024hey}.

\begin{figure}[ht]
    \centering
    \includegraphics[clip,scale=0.32]{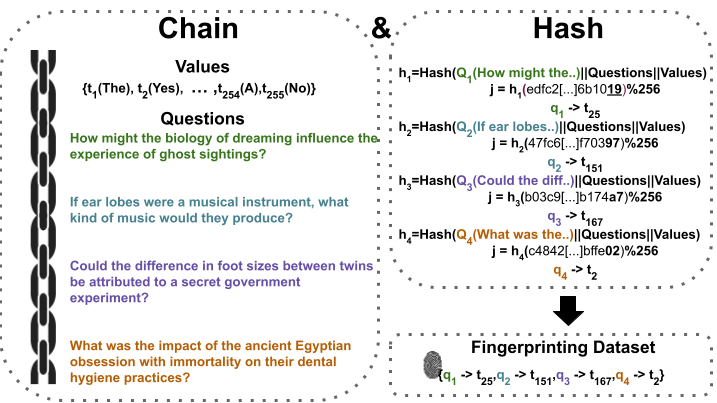}
    \caption{Generation of LLMs fingerprints.}
    \label{fig: chainAndHashOverview}
\end{figure}

Jin et al. \cite{jin2024proflingo} proposed ProFLingo, a black-box fingerprinting approach for LLMs. ProFLingo identifies models by generating queries and evaluating responses, particularly suited for IP protection without modifying the models. McGovern et al. \cite{mcgovern2024your} found that LLMs have unique fingerprints, manifested in subtle differences in the frequency of certain lexical and morphosyntactic features in the generated text. Experiments showed that n-gram and part-of-speech features can build effective classifiers to detect and identify model-generated text, and these fingerprints exhibit consistency within the same model family. Pasquini et al. \cite{pasquini2024llmmap} introduced LLMma, a fingerprinting method for LLMs that combines strategic questioning and machine learning analysis to identify the specific LLM used in an application, providing a lightweight and practical fingerprinting tool for AI red teaming. Russinovich et al. \cite{russinovich2024hey} proposed a novel fingerprinting technique called Chain \& Hash for LLMs. This method generates and chains multiple queries, and uses cryptographic hashing to select the corresponding responses, providing adversarially robust and forgery-resistant fingerprinting while maintaining model utility. An overview of the Chain \& Hash technique is shown in Figure \ref{fig: chainAndHashOverview}. In the era of large models, digital fingerprinting technology, with its non-invasive and efficient nature, has shown great application potential in model IP protection. As model scales continue to grow, effectively verifying and protecting model identity has become a critical issue. In the future, fingerprinting techniques will play an important role in ensuring the healthy development of AI technology and the security of the digital economy.

\section{Challenges and Opportunities} \label{sec:oppotrunities}
\subsection{Challenges}

With the continuous development of UGC websites like YouTube and AI-generated content, the volume of multimedia content is expected to continue growing. How to process, store, and retrieve such a large amount of content without affecting the user experience poses new challenges for digital fingerprint algorithms. Figure \ref{fig: ChallengesandOpportunities} highlights the challenges and opportunities in the field. The key challenges include, but are not limited to, the following:

\begin{figure}[ht]
    \centering
    \includegraphics[clip,scale=0.26]{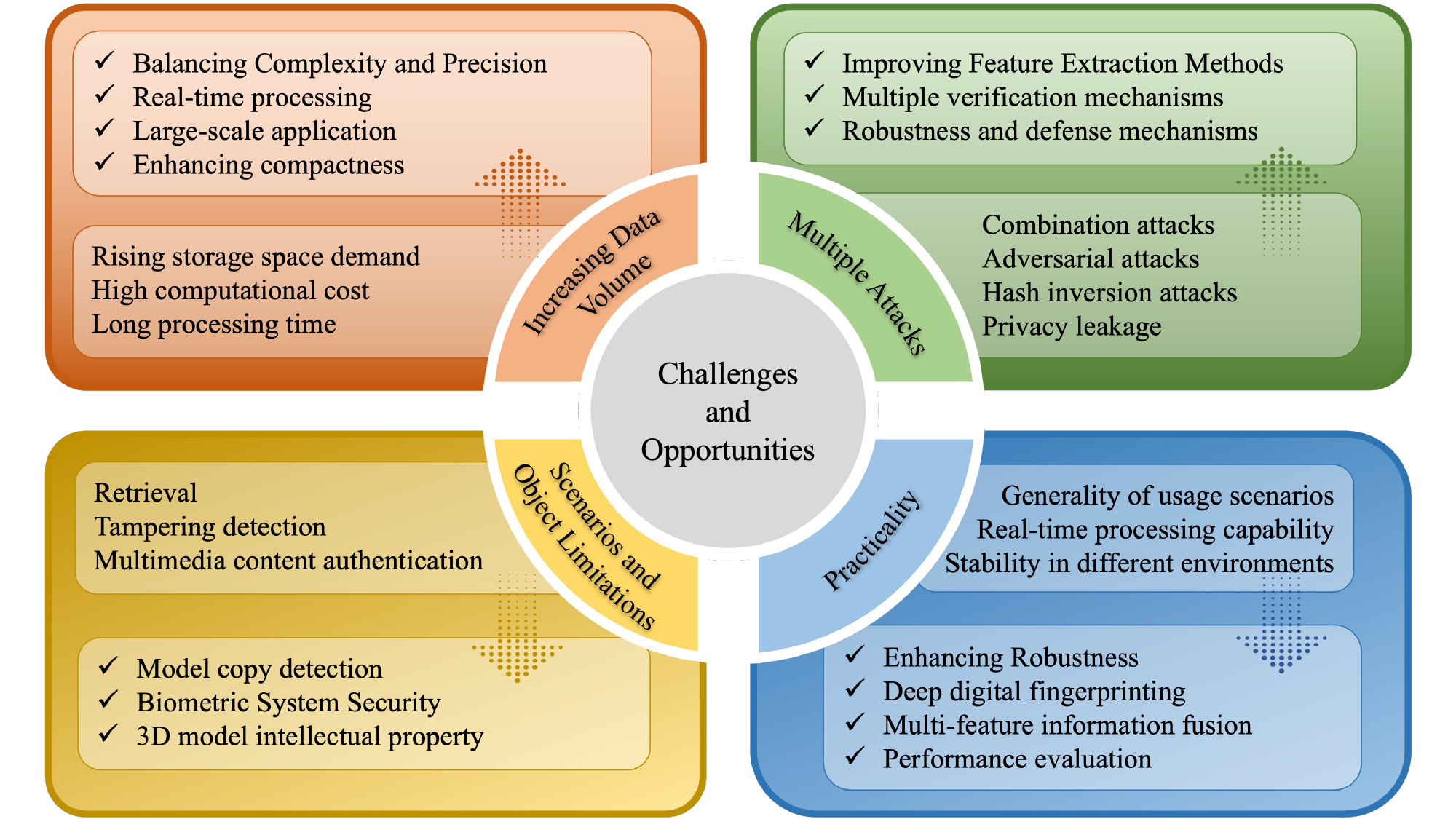}
    \caption{Challenges and opportunities.}
    \label{fig: ChallengesandOpportunities}
\end{figure}

\textbf{1. Reducing computational cost}: Existing fingerprint extraction and matching algorithms have high computational costs and long processing times when dealing with large-scale multimedia content. Although deep learning-based methods have achieved satisfactory results, they often face unacceptable memory consumption and computational requirements during the inference/training stage. To enable real-time detection on large-scale data, perceptual hashing algorithms need to strike a balance between computational complexity and detection accuracy. Moreover, most mainstream fingerprint algorithms are mainly designed for PC application scenarios, while mobile devices have limited computing and storage capabilities. How to efficiently run on resource-constrained devices is a problem that needs to be solved. To improve processing efficiency, the algorithms need to be optimized and the computational complexity reduced. For example, introducing more efficient feature extraction methods and matching algorithms can significantly reduce computational costs and achieve higher processing speeds. This is particularly important for real-time content processing, large-scale applications, and resource-constrained device scenarios.

\textbf{2. Reducing storage requirements}: With the increasing volume of multimedia content, the demand for storage space is increasing rapidly. The hash codes generated by current deep learning-based methods have not yet well achieved the goal of reducing storage consumption, and there is still an inevitable tradeoff between accuracy and storage \cite{singh2022learning}. To save resources, more compact and efficient fingerprint representation methods need to be developed to reduce the storage space of each fingerprint. For example, through compression techniques or more efficient encoding methods, the storage requirements can be significantly reduced without significantly compromising the fingerprint's discriminative capability. This not only helps reduce storage costs but also improves the overall system efficiency. In addition, data deduplication technology is also an effective method, as identifying and removing duplicate data can significantly reduce storage requirements and bandwidth consumption.

\textbf{3. Composite attacks} \cite{lin2020composite}: Multimedia content may face composite attacks or a combination of multiple attacks in real-world applications. For example, applying a combination of operations such as rotation, scaling, flipping, blurring, and noise to an image. Most current fingerprint technologies have good robustness against single-type attacks, but their performance is often poor when facing composite attacks. Composite attacks increase the difficulty of detection by combining multiple distortion operations. For example, the combination of rotation and scaling can change the geometric shape of the image, while the combination of blurring and noise can affect the details and texture of the image. The combined use of these attack methods greatly challenges the robustness of perceptual hashing algorithms. Therefore, improving the algorithm's robustness in complex attack environments has become an important research direction. Research has shown that by improving feature extraction methods and adopting multiple verification mechanisms, the algorithm's robustness to composite attacks can be enhanced.

\textbf{4. Adversarial attacks} \cite{goodfellow2015explaining,prokos2023squint}: Adversarial attacks can make small and imperceptible modifications to the content, making it visually similar to the original content but producing different hash values in the perceptual hashing system, thereby evading detection \cite{madden2024assessing} or interfering with image search and analysis applications based on perceptual hashing \cite{hao2021s}. Adversarial attacks use optimization techniques to generate adversarial samples that effectively bypass the detection of perceptual hashing systems, allowing illegal content to avoid the monitoring of the detection system and continue to spread, severely affecting the effectiveness and reliability of the perceptual hashing system. This highlights the vulnerability of these systems to adversarial samples. Adversarial methods may also be exploited by malicious actors to launch cyber attacks through perceptual hashing. One strategy of adversarial machine learning is to add carefully crafted noise to images, causing neural networks to fail to correctly identify image content. Malicious actors can start from harmless images, and add noise, so that the images are still harmless visually, but the machine's perceptual hashing will mark them as CSAM. If the malicious images are placed on the target device, when the target tries to share them, the device may be locked and the account may not be recoverable \cite{struppek2022learning}. Therefore, how to enhance the algorithm's robustness and develop effective defense mechanisms has become an important research direction.

\textbf{5. Privacy leakage} \cite{jain2023deep}: Digital fingerprints face multiple challenges in privacy protection. First, perceptual hash functions extract content features and convert them into hash values, which means that the hash values contain some content information, and hash inversion attacks allow attackers to extract partial information of the original content or even reconstruct partial content with the help of certain contextual information, posing a serious threat to user privacy \cite{twenning2023using}. Some studies have already found ways to perform reverse engineering. Second, many communication channels use end-to-end encryption technology to protect user privacy, but this encryption method hinders content scanning measures on the network and server-side, making it impossible to effectively detect illegal content. Therefore, some systems deploy perceptual hashes on client devices to scan before data encryption, which has raised widespread privacy concerns and criticism. It is believed that client-side scanning systems based on deep perceptual hashing algorithms may have hidden malicious functions that could lead to privacy leakage, large-scale surveillance, trust crises, legal and ethical issues, and technological abuse \cite{jain2023deep}. Although some research has proposed designing privacy-preserving perceptual hashing systems using cryptographic methods, these methods still face computational complexity and efficiency issues in practical applications. How to ensure the detection effect while guaranteeing privacy remains a challenge.

\textbf{6. Accuracy and robustness of algorithms}: Perceptual hashing algorithms need to maintain high accuracy and robustness in different environments and usage scenarios. For example, in image processing, common image editing operations such as brightness, contrast, and color adjustment can affect the consistency of hash values. How to improve the algorithm's robustness so that it can maintain consistency when faced with various image processing operations is an urgent problem to be solved. In addition, perceptual hashing algorithms also need to maintain stability under various environmental changes. Environmental factors such as weather changes, lighting conditions, and camera angle changes can all affect the hash values of images, so the algorithm must maintain stability and consistency under these environmental changes to ensure its reliability in practical applications. Perceptual hashing algorithms need to perform well in a single scenario and maintain consistency in various application scenarios. For example, in video surveillance, image search, and copyright protection, the algorithm needs to have sufficient adaptability to handle different types of image and video content. More and more images and videos are using HDR formats, so perceptual hashing algorithms need to adapt to HDR and SDR content and maintain consistency between HDR and standard dynamic range (SDR) content. Furthermore, in practical applications, images and videos are often edited or cropped, so perceptual hashing algorithms need to maintain high accuracy and robustness under these operations, ensuring correct identification and matching even when the content is partially modified, which poses higher requirements on the feature extraction and matching mechanism of the algorithm.

\subsection{Opportunities}

\textbf{1. Deep digital fingerprints} \cite{jiang2018perceptual}: In recent years, DNNs have shown excellent performance in multimedia content feature representation. Although deep learning-based methods take longer computation time, they can provide high-quality hash results through more complex and advanced feature extraction. Current methods do not fully utilize the prior information between the original multimedia content and the content after content-preserving operations during the training process, which is crucial for discriminating content. To cope with the dynamic growth of data, deep incremental hashing networks allow the hash function to be adjusted according to the updates of the dataset, but there is still a lot of work to be done in terms of accuracy and speed when processing large-scale databases \cite{singh2022learning}. Cross-modal hashing, by fusing data of different modalities (such as images and text), can improve the robustness and applicability of the hash representation, better addressing the complex multimodal data requirements. In terms of improving hash function design, the introduction of technologies such as GANs can further optimize the generation of hash functions, improving the quality of hash codes and retrieval accuracy.

\textbf{2. Fusion of multiple feature information}: A significant trend in current research is the development of more robust perceptual hashing methods by integrating multiple types of feature information. Traditional single-feature extraction techniques often lack the robustness needed to withstand complex composite attacks. By combining various feature sets (e.g., color, texture, shape features of images), along with structural information from videos (including shot number, shot switching frequency, switching methods, shot boundary frame positions, keyframe density, and keyframe positions) \cite{zhang2023short}, the resilience of fingerprinting technology against such attacks can be significantly improved.

\textbf{3. Robustness and attack-resistance}: In the future, enhancing the robustness and attack-resistance of perceptual hashing algorithms will be one of the research focus areas. More and more research is starting to introduce deep learning into perceptual hashing algorithms to improve the robustness of image transformations and attacks. For example, new deep perceptual hashing algorithms can more effectively handle complex changes and attacks in images. These algorithms can not only cope with conventional image transformations but also resist malicious attacks, ensuring the reliability of the hash values. In addition, designing new defense mechanisms to withstand black-box attacks targeting perceptual hashing is also an important direction for future research. By enhancing the defense capability of the algorithm, its security and reliability in practical applications can be ensured, better protecting user data.

\textbf{4. Practicality}: To enhance the practicality of perceptual hashing technology, improvements must be made in real-time processing, application performance, and generality. Particularly in the rapidly developing field of real-time visual tracking, these systems need to be able to process and generate hashes in real-time to support key applications such as video surveillance and real-time data analysis. Real-time processing capability is crucial for scenarios like video surveillance and data analysis. Furthermore, as data volumes continue to grow, perceptual hashing algorithms must run efficiently on large-scale datasets, requiring not only strong computational capabilities but also ensuring low resource consumption to meet the needs of the big data era. To further improve the practicality and wide application of perceptual hashing algorithms, future research also needs to focus on their generality, ensuring stable and efficient performance in different databases and application scenarios.

\textbf{5. Evaluation schemes of algorithm performance}: Currently, many perceptual hashing schemes have been designed, but the corresponding fair and unified hashing scheme evaluation methods still need further improvement. Quantitative evaluation through improved unified performance evaluation methods can not only verify the practicality and reliability of hashing schemes, but also provide a unified evaluation standard and tools to facilitate comparative analysis and optimization of algorithm performance \cite{li2022unified}. A comprehensive evaluation of hashing schemes from multiple perspectives can help optimize the parameter settings of perceptual hashing algorithms, further improving their performance and adaptability in different application scenarios to meet the needs of various complex applications.

\textbf{6. Emerging application scenarios}: The potential of perceptual hashing technology in emerging application fields should not be overlooked. The current applications of digital fingerprint technology are mainly limited to authentication, tampering detection, and retrieval. However, this technology has many unexplored application areas, such as shot boundary detection, model copy detection, visual tracking, and biometric system security issues \cite{hamadouche2023securing}. Biometric recognition systems are increasingly widely used in access control, identity authentication, and financial transactions. Due to the sensitive nature of biometric data and the emergence of sophisticated attack techniques, these systems face significant security challenges. Digital fingerprint technology, by generating compact and robust biometric image representations \cite{hamadouche2024replay}, can facilitate the secure storage, transmission, and comparison of biometric information, and can be applied in mobile devices, monitoring systems, automated fingerprint identification systems, and secure access control systems.

\textbf{7. Model copy detection} \cite{regazzoni2021protecting}: Pre-trained models have become core assets in research and applications, and protecting their intellectual property is crucial. Although traditional AI model watermarking techniques have made some progress in verifying ownership, efficiently detecting copied models in large-scale model libraries still faces huge challenges. There are several perceptual hashing-based DNN copy detection methods \cite{chen2023perceptual,zhao2020afa}.  \cite{chen2023perceptual} converts model weight features into fixed-length binary hash codes as the unique "fingerprints" of the model, and compares hash similarities to quickly retrieve potential copied models. This approach improves detection efficiency and effectively protects intellectual property while maintaining model performance. The future development directions of digital fingerprint technology for model copy detection include cross-task model copy detection and applications of large-scale model libraries. Existing algorithms mainly target classification tasks, and future work should expand to domains like object detection and semantic segmentation, and develop generic copy detection solutions. Meanwhile, optimizing the efficiency of hash algorithms to adapt to the needs of larger model libraries can provide a more comprehensive and efficient solution for protecting the intellectual property of deep learning models.

\textbf{8. Expansion of application objects}: In addition to traditional image and video authentication, digital fingerprint technology can also be applied to neural networks \cite{chen2023perceptual} and 3D models. For example, digital fingerprints can be used to protect the intellectual property of neural network models, which is particularly important in the context of the rapid development of deep learning technology. Furthermore, with the development of industrial metaverse and 3D printing technology, 3D models have been increasingly widely applied in various industries. However, the problem of 3D model intellectual property theft is also becoming increasingly serious. 3D perceptual hashing technology captures the geometric features of 3D models through multiple sphere slices, projecting these features onto a clustering distance to generate a compact and tamper-resistant fingerprint \cite{prummer2024onion}. These fingerprints can effectively detect tampering and forgery, thereby ensuring the intellectual property protection of 3D models in open ecosystems.

\section{Conclusion}  \label{sec:conclusion}

In this paper, we have systematically reviewed the development history, main algorithms, and applications of digital fingerprint technology in multimedia data management and copyright protection, exploring its key role in the modern information society. First, we introduce the basic concepts and development background of digital fingerprints, emphasizing their potential to address information floods and ensure content authenticity. With the advancement of technology, digital fingerprints, by forming a unique identifier for multimedia content, provide an effective means to authenticate and manage digital content, solving the problems of copyright protection and data management. We then analyze in-depth the digital fingerprint algorithms for several major multimedia types, including text, images, videos, and audio, and explore the application of neural networks in digital fingerprint extraction algorithms. Different feature extraction algorithms have their advantages and disadvantages in terms of accuracy, robustness, and computational efficiency. In practical applications, the choice of algorithms is determined by the scenario and requirements. Furthermore, we discuss the advanced practical applications of digital fingerprints in detail from different perspectives, especially large model-based applications. Finally, we highlight the main challenges and future research directions of digital fingerprints in the future. The main challenges currently include many issues such as attack resistance, robustness, and computational efficiency. In summary, digital fingerprints provide a viable path for data management and copyright protection in the modern information society. With the continuous progress of technology, we believe that digital fingerprints will play an increasingly important role in multimedia content management, contributing to the establishment of a more secure and trustworthy digital environment.

\bibliographystyle{elsarticle-num-names}

\bibliography{DigitalFingerprinting.bib}

\begin{thebibliography}{238}
\expandafter\ifx\csname natexlab\endcsname\relax\def\natexlab#1{#1}\fi
\providecommand{\url}[1]{\texttt{#1}}
\providecommand{\href}[2]{#2}
\providecommand{\path}[1]{#1}
\providecommand{\DOIprefix}{doi:}
\providecommand{\ArXivprefix}{arXiv:}
\providecommand{\URLprefix}{URL: }
\providecommand{\Pubmedprefix}{pmid:}
\providecommand{\doi}[1]{\href{http://dx.doi.org/#1}{\path{#1}}}
\providecommand{\Pubmed}[1]{\href{pmid:#1}{\path{#1}}}
\providecommand{\bibinfo}[2]{#2}
\ifx\xfnm\relax \def\xfnm[#1]{\unskip,\space#1}\fi
\bibitem[{Gan et~al.(2023)Gan, Ye, Wan, and Yu}]{gan2023web}
\bibinfo{author}{W.~Gan}, \bibinfo{author}{Z.~Ye}, \bibinfo{author}{S.~Wan}, \bibinfo{author}{P.~S. Yu},
\newblock \bibinfo{title}{Web 3.0: The future of {Internet}},
\newblock in: \bibinfo{booktitle}{Companion Proceedings of the ACM Web Conference}, \bibinfo{year}{2023}, pp. \bibinfo{pages}{1266--1275}.
\bibitem[{Wu et~al.(2023)Wu, Gan, Chen, Wan, and Yu}]{wu2023multimodal}
\bibinfo{author}{J.~Wu}, \bibinfo{author}{W.~Gan}, \bibinfo{author}{Z.~Chen}, \bibinfo{author}{S.~Wan}, \bibinfo{author}{P.~S. Yu},
\newblock \bibinfo{title}{Multimodal large language models: A survey},
\newblock in: \bibinfo{booktitle}{IEEE International Conference on Big Data}, \bibinfo{organization}{IEEE}, \bibinfo{year}{2023}, pp. \bibinfo{pages}{2247--2256}.
\bibitem[{Li et~al.(2007)Li, Chang, Lesk, Lienhart, Luo, and Smeulders}]{li2007new}
\bibinfo{author}{J.~Li}, \bibinfo{author}{S.-F. Chang}, \bibinfo{author}{M.~Lesk}, \bibinfo{author}{R.~Lienhart}, \bibinfo{author}{J.~Luo}, \bibinfo{author}{A.~W. Smeulders},
\newblock \bibinfo{title}{New challenges in multimedia research for the increasingly connected and fast-growing digital society},
\newblock in: \bibinfo{booktitle}{The International Workshop on Multimedia Information Retrieval}, \bibinfo{year}{2007}, pp. \bibinfo{pages}{3--10}.
\bibitem[{Sun et~al.(2023)Sun, Gan, Chao, Yu, and Ding}]{sun2023internet}
\bibinfo{author}{J.~Sun}, \bibinfo{author}{W.~Gan}, \bibinfo{author}{H.-C. Chao}, \bibinfo{author}{P.~S. Yu}, \bibinfo{author}{W.~Ding},
\newblock \bibinfo{title}{Internet of behaviors: A survey},
\newblock \bibinfo{journal}{IEEE Internet of Things Journal} \bibinfo{volume}{10} (\bibinfo{year}{2023}) \bibinfo{pages}{11117--11134}.
\bibitem[{Anantrasirichai and Bull(2022)}]{anantrasirichai2022artificial}
\bibinfo{author}{N.~Anantrasirichai}, \bibinfo{author}{D.~Bull},
\newblock \bibinfo{title}{Artificial intelligence in the creative industries: a review},
\newblock \bibinfo{journal}{Artificial Intelligence Review} \bibinfo{volume}{55} (\bibinfo{year}{2022}) \bibinfo{pages}{589--656}.
\bibitem[{Borkar et~al.(2021)Borkar, Patre, Khalsa, Kawale, and Chakurkar}]{borkar2021music}
\bibinfo{author}{N.~Borkar}, \bibinfo{author}{S.~Patre}, \bibinfo{author}{R.~S. Khalsa}, \bibinfo{author}{R.~Kawale}, \bibinfo{author}{P.~Chakurkar},
\newblock \bibinfo{title}{Music plagiarism detection using audio fingerprinting and segment matching},
\newblock in: \bibinfo{booktitle}{Smart Technologies, Communication and Robotics}, \bibinfo{organization}{IEEE}, \bibinfo{year}{2021}, pp. \bibinfo{pages}{1--4}.
\bibitem[{Chen et~al.(2022)Chen, Liu, Wang, Bakker, Georgiou, Fieguth, Liu, and Lew}]{chen2022deep}
\bibinfo{author}{W.~Chen}, \bibinfo{author}{Y.~Liu}, \bibinfo{author}{W.~Wang}, \bibinfo{author}{E.~M. Bakker}, \bibinfo{author}{T.~Georgiou}, \bibinfo{author}{P.~Fieguth}, \bibinfo{author}{L.~Liu}, \bibinfo{author}{M.~S. Lew},
\newblock \bibinfo{title}{Deep learning for instance retrieval: A survey},
\newblock \bibinfo{journal}{IEEE Transactions on Pattern Analysis and Machine Intelligence} \bibinfo{volume}{45} (\bibinfo{year}{2022}) \bibinfo{pages}{7270--7292}.
\bibitem[{Yildizer et~al.(2012)Yildizer, Balci, Hassan, and Alhajj}]{yildizer2012efficient}
\bibinfo{author}{E.~Yildizer}, \bibinfo{author}{A.~M. Balci}, \bibinfo{author}{M.~Hassan}, \bibinfo{author}{R.~Alhajj},
\newblock \bibinfo{title}{Efficient content-based image retrieval using multiple support vector machines ensemble},
\newblock \bibinfo{journal}{Expert Systems with Applications} \bibinfo{volume}{39} (\bibinfo{year}{2012}) \bibinfo{pages}{2385--2396}.
\bibitem[{Wang et~al.(2013)Wang, Yang, and Li}]{wang2013new}
\bibinfo{author}{X.-Y. Wang}, \bibinfo{author}{H.-Y. Yang}, \bibinfo{author}{D.-M. Li},
\newblock \bibinfo{title}{A new content-based image retrieval technique using color and texture information},
\newblock \bibinfo{journal}{Computers and Electrical Engineering} \bibinfo{volume}{39} (\bibinfo{year}{2013}) \bibinfo{pages}{746--761}.
\bibitem[{Zhang et~al.(2018)Zhang, Su, Yang, Zheng, Ren, and Zhao}]{zhang2018secure}
\bibinfo{author}{Y.~Zhang}, \bibinfo{author}{H.~Su}, \bibinfo{author}{M.~Yang}, \bibinfo{author}{D.~Zheng}, \bibinfo{author}{F.~Ren}, \bibinfo{author}{Q.~Zhao},
\newblock \bibinfo{title}{Secure deduplication based on rabin fingerprinting over wireless sensing data in cloud computing},
\newblock \bibinfo{journal}{Security and Communication Networks} \bibinfo{volume}{2018} (\bibinfo{year}{2018}) \bibinfo{pages}{9081814}.
\bibitem[{Xia et~al.(2016)Xia, Jiang, Feng, Douglis, Shilane, Hua, Fu, Zhang, and Zhou}]{xia2016comprehensive}
\bibinfo{author}{W.~Xia}, \bibinfo{author}{H.~Jiang}, \bibinfo{author}{D.~Feng}, \bibinfo{author}{F.~Douglis}, \bibinfo{author}{P.~Shilane}, \bibinfo{author}{Y.~Hua}, \bibinfo{author}{M.~Fu}, \bibinfo{author}{Y.~Zhang}, \bibinfo{author}{Y.~Zhou},
\newblock \bibinfo{title}{A comprehensive study of the past, present, and future of data deduplication},
\newblock \bibinfo{journal}{Proceedings of the IEEE} \bibinfo{volume}{104} (\bibinfo{year}{2016}) \bibinfo{pages}{1681--1710}.
\bibitem[{Bello-Orgaz et~al.(2016)Bello-Orgaz, Jung, and Camacho}]{bello2016social}
\bibinfo{author}{G.~Bello-Orgaz}, \bibinfo{author}{J.~J. Jung}, \bibinfo{author}{D.~Camacho},
\newblock \bibinfo{title}{Social big data: Recent achievements and new challenges},
\newblock \bibinfo{journal}{Information Fusion} \bibinfo{volume}{28} (\bibinfo{year}{2016}) \bibinfo{pages}{45--59}.
\bibitem[{Wu et~al.(2023)Wu, Gan, Chen, Wan, and Lin}]{wu2023ai}
\bibinfo{author}{J.~Wu}, \bibinfo{author}{W.~Gan}, \bibinfo{author}{Z.~Chen}, \bibinfo{author}{S.~Wan}, \bibinfo{author}{H.~Lin},
\newblock \bibinfo{title}{{AI}-generated content ({AIGC}): A survey},
\newblock \bibinfo{journal}{arXiv preprint arXiv:2304.06632}  (\bibinfo{year}{2023}).
\bibitem[{Subramanya and Yi(2006)}]{subramanya2006digital}
\bibinfo{author}{S.~Subramanya}, \bibinfo{author}{B.~K. Yi},
\newblock \bibinfo{title}{Digital rights management},
\newblock \bibinfo{journal}{IEEE Potentials} \bibinfo{volume}{25} (\bibinfo{year}{2006}) \bibinfo{pages}{31--34}.
\bibitem[{Lu(2009)}]{lu2009video}
\bibinfo{author}{J.~Lu},
\newblock \bibinfo{title}{Video fingerprinting for copy identification: from research to industry applications},
\newblock \bibinfo{journal}{Media Forensics and Security} \bibinfo{volume}{7254} (\bibinfo{year}{2009}) \bibinfo{pages}{725402}.
\bibitem[{Arunakumari et~al.(2023)Arunakumari, Shashidhar, Sahana, Jagadamba, Manjunath, and Roopa}]{arunakumari2023fingerprint}
\bibinfo{author}{B.~Arunakumari}, \bibinfo{author}{R.~Shashidhar}, \bibinfo{author}{B.~Sahana}, \bibinfo{author}{G.~Jagadamba}, \bibinfo{author}{A.~Manjunath}, \bibinfo{author}{M.~Roopa},
\newblock \bibinfo{title}{Fingerprint definition for song recognition using machine learning algorithm},
\newblock in: \bibinfo{booktitle}{International Conference on Smart Systems for Applications in Electrical Sciences}, \bibinfo{organization}{IEEE}, \bibinfo{year}{2023}, pp. \bibinfo{pages}{1--6}.
\bibitem[{Becker(2003)}]{becker2003digital}
\bibinfo{author}{E.~Becker}, \bibinfo{title}{Digital rights management: technological, economic, legal and political aspects}, volume \bibinfo{volume}{2770}, \bibinfo{publisher}{Springer Science \& Business Media}, \bibinfo{year}{2003}.
\bibitem[{Irtaza et~al.(2014)Irtaza, Jaffar, Aleisa, and Choi}]{irtaza2014embedding}
\bibinfo{author}{A.~Irtaza}, \bibinfo{author}{M.~A. Jaffar}, \bibinfo{author}{E.~Aleisa}, \bibinfo{author}{T.-S. Choi},
\newblock \bibinfo{title}{Embedding neural networks for semantic association in content based image retrieval},
\newblock \bibinfo{journal}{Multimedia Tools and Applications} \bibinfo{volume}{72} (\bibinfo{year}{2014}) \bibinfo{pages}{1911--1931}.
\bibitem[{Akg{\"u}l et~al.(2011)Akg{\"u}l, Rubin, Napel, Beaulieu, Greenspan, and Acar}]{akgul2011content}
\bibinfo{author}{C.~B. Akg{\"u}l}, \bibinfo{author}{D.~L. Rubin}, \bibinfo{author}{S.~Napel}, \bibinfo{author}{C.~F. Beaulieu}, \bibinfo{author}{H.~Greenspan}, \bibinfo{author}{B.~Acar},
\newblock \bibinfo{title}{Content-based image retrieval in radiology: current status and future directions},
\newblock \bibinfo{journal}{Journal of Digital Imaging} \bibinfo{volume}{24} (\bibinfo{year}{2011}) \bibinfo{pages}{208--222}.
\bibitem[{Chaudhari et~al.(2024)Chaudhari, Aparna, Anchalia, Somayaji, and Kumar}]{chaudhari2024hash}
\bibinfo{author}{S.~Chaudhari}, \bibinfo{author}{R.~Aparna}, \bibinfo{author}{A.~Anchalia}, \bibinfo{author}{A.~M. Somayaji}, \bibinfo{author}{A.~S. Kumar},
\newblock \bibinfo{title}{Hash overhead analysis for gop-level video deduplication in cloud storage environment},
\newblock in: \bibinfo{booktitle}{International Conference on Smart Systems for Applications in Electrical Sciences}, \bibinfo{organization}{IEEE}, \bibinfo{year}{2024}, pp. \bibinfo{pages}{1--6}.
\bibitem[{Schleimer et~al.(2003)Schleimer, Wilkerson, and Aiken}]{schleimer2003winnowing}
\bibinfo{author}{S.~Schleimer}, \bibinfo{author}{D.~S. Wilkerson}, \bibinfo{author}{A.~Aiken},
\newblock \bibinfo{title}{{Winnowing}: local algorithms for document fingerprinting},
\newblock in: \bibinfo{booktitle}{ACM SIGMOD International Conference on Management of Data}, \bibinfo{year}{2003}, pp. \bibinfo{pages}{76--85}.
\bibitem[{Rashid et~al.(2023)Rashid, Naseer, Khan, Khan, Ali, Ahmad, and Javed}]{rashid2023sampling}
\bibinfo{author}{U.~Rashid}, \bibinfo{author}{S.~Naseer}, \bibinfo{author}{A.~R. Khan}, \bibinfo{author}{M.~A. Khan}, \bibinfo{author}{G.~Ali}, \bibinfo{author}{N.~Ahmad}, \bibinfo{author}{Y.~Javed},
\newblock \bibinfo{title}{Sampling fingerprints from multimedia content resource clusters},
\newblock \bibinfo{journal}{IEEE Access}  (\bibinfo{year}{2023}).
\bibitem[{Ansori et~al.(2023)Ansori, Alief, Igboanusi, Lee, Kim et~al.}]{ansori2023hades}
\bibinfo{author}{M.~R.~R. Ansori}, \bibinfo{author}{R.~N. Alief}, \bibinfo{author}{I.~S. Igboanusi}, \bibinfo{author}{J.~M. Lee}, \bibinfo{author}{D.-S. Kim}, et~al.,
\newblock \bibinfo{title}{{HADES}: Hash-based audio copy detection system for copyright protection in decentralized music sharing},
\newblock \bibinfo{journal}{IEEE Transactions on Network and Service Management} \bibinfo{volume}{20} (\bibinfo{year}{2023}) \bibinfo{pages}{2845--2853}.
\bibitem[{Samanta and Jain(2021)}]{samanta2021analysis}
\bibinfo{author}{P.~Samanta}, \bibinfo{author}{S.~Jain},
\newblock \bibinfo{title}{Analysis of perceptual hashing algorithms in image manipulation detection},
\newblock \bibinfo{journal}{Procedia Computer Science} \bibinfo{volume}{185} (\bibinfo{year}{2021}) \bibinfo{pages}{203--212}.
\bibitem[{Winston(1984)}]{winston1984artificial}
\bibinfo{author}{P.~H. Winston}, \bibinfo{title}{Artificial intelligence}, \bibinfo{publisher}{Addison-Wesley Longman Publishing Co., Inc.}, \bibinfo{year}{1984}.
\bibitem[{Li et~al.(2019)Li, Wang, and Tang}]{li2019robust}
\bibinfo{author}{Y.~Li}, \bibinfo{author}{D.~Wang}, \bibinfo{author}{L.~Tang},
\newblock \bibinfo{title}{Robust and secure image fingerprinting learned by neural network},
\newblock \bibinfo{journal}{IEEE Transactions on Circuits and Systems for Video Technology} \bibinfo{volume}{30} (\bibinfo{year}{2019}) \bibinfo{pages}{362--375}.
\bibitem[{Esmaeili et~al.(2010)Esmaeili, Fatourechi, and Ward}]{esmaeili2010robust2}
\bibinfo{author}{M.~M. Esmaeili}, \bibinfo{author}{M.~Fatourechi}, \bibinfo{author}{R.~K. Ward},
\newblock \bibinfo{title}{A robust and fast video copy detection system using content-based fingerprinting},
\newblock \bibinfo{journal}{IEEE Transactions on Information Forensics and Security} \bibinfo{volume}{6} (\bibinfo{year}{2010}) \bibinfo{pages}{213--226}.
\bibitem[{Gan et~al.(2018)Gan, Lin, Chao, Wang, and Yu}]{gan2018privacy}
\bibinfo{author}{W.~Gan}, \bibinfo{author}{J.~C.-W. Lin}, \bibinfo{author}{H.~C. Chao}, \bibinfo{author}{S.~L. Wang}, \bibinfo{author}{P.~S. Yu},
\newblock \bibinfo{title}{Privacy preserving utility mining: a survey},
\newblock in: \bibinfo{booktitle}{IEEE International Conference on Big Data}, \bibinfo{organization}{IEEE}, \bibinfo{year}{2018}, pp. \bibinfo{pages}{2617--2626}.
\bibitem[{Li et~al.(2022)Li, Gan, Gui, Wu, and Yu}]{li2022frequent}
\bibinfo{author}{J.~Li}, \bibinfo{author}{W.~Gan}, \bibinfo{author}{Y.~Gui}, \bibinfo{author}{Y.~Wu}, \bibinfo{author}{P.~S. Yu},
\newblock \bibinfo{title}{Frequent itemset mining with local differential privacy},
\newblock in: \bibinfo{booktitle}{The 31st ACM International Conference on Information and Knowledge Management}, \bibinfo{year}{2022}, pp. \bibinfo{pages}{1146--1155}.
\bibitem[{Jain et~al.(2023)Jain, Cre{\c{t}}u, Cully, and de~Montjoye}]{jain2023deep}
\bibinfo{author}{S.~Jain}, \bibinfo{author}{A.-M. Cre{\c{t}}u}, \bibinfo{author}{A.~Cully}, \bibinfo{author}{Y.-A. de~Montjoye},
\newblock \bibinfo{title}{Deep perceptual hashing algorithms with hidden dual purpose: when client-side scanning does facial recognition},
\newblock in: \bibinfo{booktitle}{IEEE Symposium on Security and Privacy}, \bibinfo{organization}{IEEE}, \bibinfo{year}{2023}, pp. \bibinfo{pages}{234--252}.
\bibitem[{Varna et~al.(2009)Varna, He, Swaminathan, and Wu}]{varna2009fingerprinting}
\bibinfo{author}{A.~L. Varna}, \bibinfo{author}{S.~He}, \bibinfo{author}{A.~Swaminathan}, \bibinfo{author}{M.~Wu},
\newblock \bibinfo{title}{Fingerprinting compressed multimedia signals},
\newblock \bibinfo{journal}{IEEE Transactions on Information Forensics and Security} \bibinfo{volume}{4} (\bibinfo{year}{2009}) \bibinfo{pages}{330--345}.
\bibitem[{Herre(2003)}]{herre2003content}
\bibinfo{author}{J.~Herre},
\newblock \bibinfo{title}{Content based identification (fingerprinting)},
\newblock in: \bibinfo{booktitle}{Digital Rights Management: Technological, Economic, Legal and Political Aspects}, \bibinfo{publisher}{Springer}, \bibinfo{year}{2003}, pp. \bibinfo{pages}{93--100}.
\bibitem[{Farid(2021)}]{farid2021overview}
\bibinfo{author}{H.~Farid},
\newblock \bibinfo{title}{An overview of perceptual hashing},
\newblock \bibinfo{journal}{Journal of Online Trust and Safety} \bibinfo{volume}{1} (\bibinfo{year}{2021}).
\bibitem[{J{\=e}kabsons(2020)}]{jekabsons2020evaluation}
\bibinfo{author}{G.~J{\=e}kabsons},
\newblock \bibinfo{title}{Evaluation of fingerprint selection algorithms for local text reuse detection},
\newblock \bibinfo{journal}{Applied Computer Systems} \bibinfo{volume}{25} (\bibinfo{year}{2020}) \bibinfo{pages}{11--18}.
\bibitem[{Du et~al.(2020)Du, Ho, and Cong}]{du2020perceptual}
\bibinfo{author}{L.~Du}, \bibinfo{author}{A.~T. Ho}, \bibinfo{author}{R.~Cong},
\newblock \bibinfo{title}{Perceptual hashing for image authentication: A survey},
\newblock \bibinfo{journal}{Signal Processing: Image Communication} \bibinfo{volume}{81} (\bibinfo{year}{2020}) \bibinfo{pages}{115713}.
\bibitem[{Roy et~al.(2023)Roy, Thounaojam, and Pal}]{roy2023various}
\bibinfo{author}{M.~Roy}, \bibinfo{author}{D.~M. Thounaojam}, \bibinfo{author}{S.~Pal},
\newblock \bibinfo{title}{Various approaches to perceptual image hashing systems-a survey},
\newblock in: \bibinfo{booktitle}{International Conference on Intelligent Systems, Advanced Computing and Communication}, \bibinfo{organization}{IEEE}, \bibinfo{year}{2023}, pp. \bibinfo{pages}{1--9}.
\bibitem[{Allouche and Mitrea(2022)}]{allouche2022video}
\bibinfo{author}{M.~Allouche}, \bibinfo{author}{M.~Mitrea},
\newblock \bibinfo{title}{Video fingerprinting: Past, present, and future},
\newblock \bibinfo{journal}{Frontiers in Signal Processing} \bibinfo{volume}{2} (\bibinfo{year}{2022}) \bibinfo{pages}{984169}.
\bibitem[{Kekre et~al.(2013)Kekre, Bhandari, Nair, Padmanabhan, and Bhandari}]{kekre2013review}
\bibinfo{author}{H.~Kekre}, \bibinfo{author}{N.~Bhandari}, \bibinfo{author}{N.~Nair}, \bibinfo{author}{P.~Padmanabhan}, \bibinfo{author}{S.~Bhandari},
\newblock \bibinfo{title}{A review of audio fingerprinting and comparison of algorithms},
\newblock \bibinfo{journal}{International Journal of Computer Applications} \bibinfo{volume}{70} (\bibinfo{year}{2013}) \bibinfo{pages}{24--30}.
\bibitem[{Mohanarathinam et~al.(2020)Mohanarathinam, Kamalraj, Prasanna~Venkatesan, Ravi, and Manikandababu}]{mohanarathinam2020digital}
\bibinfo{author}{A.~Mohanarathinam}, \bibinfo{author}{S.~Kamalraj}, \bibinfo{author}{G.~Prasanna~Venkatesan}, \bibinfo{author}{R.~V. Ravi}, \bibinfo{author}{C.~Manikandababu},
\newblock \bibinfo{title}{Digital watermarking techniques for image security: a review},
\newblock \bibinfo{journal}{Journal of Ambient Intelligence and Humanized Computing} \bibinfo{volume}{11} (\bibinfo{year}{2020}) \bibinfo{pages}{3221--3229}.
\bibitem[{Wang et~al.(2015)Wang, Pang, Zhou, Zhou, Li, and Xue}]{wang2015visual}
\bibinfo{author}{X.~Wang}, \bibinfo{author}{K.~Pang}, \bibinfo{author}{X.~Zhou}, \bibinfo{author}{Y.~Zhou}, \bibinfo{author}{L.~Li}, \bibinfo{author}{J.~Xue},
\newblock \bibinfo{title}{A visual model-based perceptual image hash for content authentication},
\newblock \bibinfo{journal}{IEEE Transactions on Information Forensics and Security} \bibinfo{volume}{10} (\bibinfo{year}{2015}) \bibinfo{pages}{1336--1349}.
\bibitem[{Sun and Zhou(2022)}]{sun2022deep}
\bibinfo{author}{X.~Sun}, \bibinfo{author}{J.~Zhou},
\newblock \bibinfo{title}{Deep perceptual hash based on hash center for image copyright protection},
\newblock \bibinfo{journal}{IEEE Access} \bibinfo{volume}{10} (\bibinfo{year}{2022}) \bibinfo{pages}{120551--120562}.
\bibitem[{Steinebach(2023)}]{steinebach2023analysis}
\bibinfo{author}{M.~Steinebach},
\newblock \bibinfo{title}{An analysis of photodna},
\newblock in: \bibinfo{booktitle}{The 18th International Conference on Availability, Reliability and Security}, \bibinfo{year}{2023}, pp. \bibinfo{pages}{1--8}.
\bibitem[{Gabryel et~al.(2020)Gabryel, Grzanek, and Hayashi}]{gabryel2020browser}
\bibinfo{author}{M.~Gabryel}, \bibinfo{author}{K.~Grzanek}, \bibinfo{author}{Y.~Hayashi},
\newblock \bibinfo{title}{Browser fingerprint coding methods increasing the effectiveness of user identification in the web traffic},
\newblock \bibinfo{journal}{Journal of Artificial Intelligence and Soft Computing Research} \bibinfo{volume}{10} (\bibinfo{year}{2020}) \bibinfo{pages}{243--253}.
\bibitem[{Zhang et~al.(2015)Zhang, Chen, Ooi, Tan, and Zhang}]{zhang2015memory}
\bibinfo{author}{H.~Zhang}, \bibinfo{author}{G.~Chen}, \bibinfo{author}{B.~C. Ooi}, \bibinfo{author}{K.-L. Tan}, \bibinfo{author}{M.~Zhang},
\newblock \bibinfo{title}{In-memory big data management and processing: A survey},
\newblock \bibinfo{journal}{IEEE Transactions on Knowledge and Data Engineering} \bibinfo{volume}{27} (\bibinfo{year}{2015}) \bibinfo{pages}{1920--1948}.
\bibitem[{Monga et~al.(2006)Monga, Banerjee, and Evans}]{monga2006clustering}
\bibinfo{author}{V.~Monga}, \bibinfo{author}{A.~Banerjee}, \bibinfo{author}{B.~L. Evans},
\newblock \bibinfo{title}{A clustering based approach to perceptual image hashing},
\newblock \bibinfo{journal}{IEEE Transactions on Information Forensics and Security} \bibinfo{volume}{1} (\bibinfo{year}{2006}) \bibinfo{pages}{68--79}.
\bibitem[{Li et~al.(2022)Li, Qin, Wang, Qian, and Zhang}]{li2022unified}
\bibinfo{author}{X.~Li}, \bibinfo{author}{C.~Qin}, \bibinfo{author}{Z.~Wang}, \bibinfo{author}{Z.~Qian}, \bibinfo{author}{X.~Zhang},
\newblock \bibinfo{title}{Unified performance evaluation method for perceptual image hashing},
\newblock \bibinfo{journal}{IEEE Transactions on Information Forensics and Security} \bibinfo{volume}{17} (\bibinfo{year}{2022}) \bibinfo{pages}{1404--1419}.
\bibitem[{Monga and Evans(2006)}]{monga2006perceptual}
\bibinfo{author}{V.~Monga}, \bibinfo{author}{B.~L. Evans},
\newblock \bibinfo{title}{Perceptual image hashing via feature points: performance evaluation and tradeoffs},
\newblock \bibinfo{journal}{IEEE Transactions on Image Processing} \bibinfo{volume}{15} (\bibinfo{year}{2006}) \bibinfo{pages}{3452--3465}.
\bibitem[{Lulu et~al.(2016)Lulu, Belkhouche, and Harous}]{lulu2016overview}
\bibinfo{author}{L.~Lulu}, \bibinfo{author}{B.~Belkhouche}, \bibinfo{author}{S.~Harous},
\newblock \bibinfo{title}{Overview of fingerprinting methods for local text reuse detection},
\newblock in: \bibinfo{booktitle}{12th International Conference on Innovations in Information Technology}, \bibinfo{organization}{IEEE}, \bibinfo{year}{2016}, pp. \bibinfo{pages}{1--6}.
\bibitem[{Stein and zu~Eissen(2007)}]{stein2007fingerprint}
\bibinfo{author}{B.~Stein}, \bibinfo{author}{S.~M. zu~Eissen},
\newblock \bibinfo{title}{Fingerprint-based similarity search and its applications},
\newblock \bibinfo{journal}{Universit{\"a}t Weimar}  (\bibinfo{year}{2007}) \bibinfo{pages}{85--99}.
\bibitem[{Datta et~al.(2008)Datta, Joshi, Li, and Wang}]{datta2008image}
\bibinfo{author}{R.~Datta}, \bibinfo{author}{D.~Joshi}, \bibinfo{author}{J.~Li}, \bibinfo{author}{J.~Z. Wang},
\newblock \bibinfo{title}{Image retrieval: Ideas, influences, and trends of the new age},
\newblock \bibinfo{journal}{ACM Computing Surveys} \bibinfo{volume}{40} (\bibinfo{year}{2008}) \bibinfo{pages}{1--60}.
\bibitem[{Ong and Bober(2016)}]{ong2016improved}
\bibinfo{author}{E.-J. Ong}, \bibinfo{author}{M.~Bober},
\newblock \bibinfo{title}{Improved hamming distance search using variable length substrings},
\newblock in: \bibinfo{booktitle}{Proceedings of the IEEE Conference on Computer Vision and Pattern Recognition}, \bibinfo{year}{2016}, pp. \bibinfo{pages}{2000--2008}.
\bibitem[{Roche-Newton and Rudnev(2015)}]{roche2015minkowski}
\bibinfo{author}{O.~Roche-Newton}, \bibinfo{author}{M.~Rudnev},
\newblock \bibinfo{title}{On the minkowski distances and products of sum sets},
\newblock \bibinfo{journal}{Israel Journal of Mathematics} \bibinfo{volume}{209} (\bibinfo{year}{2015}) \bibinfo{pages}{507--526}.
\bibitem[{Rivest(1992)}]{rivest1992md5}
\bibinfo{author}{R.~Rivest}, \bibinfo{title}{The MD5 message-digest algorithm}, \bibinfo{type}{Technical Report}, \bibinfo{year}{1992}.
\bibitem[{Karp and Rabin(1987)}]{karp1987efficient}
\bibinfo{author}{R.~M. Karp}, \bibinfo{author}{M.~O. Rabin},
\newblock \bibinfo{title}{Efficient randomized pattern-matching algorithms},
\newblock \bibinfo{journal}{IBM Journal of Research and Development} \bibinfo{volume}{31} (\bibinfo{year}{1987}) \bibinfo{pages}{249--260}.
\bibitem[{Charikar(2002)}]{charikar2002similarity}
\bibinfo{author}{M.~S. Charikar},
\newblock \bibinfo{title}{Similarity estimation techniques from rounding algorithms},
\newblock in: \bibinfo{booktitle}{The thirty-fourth Annual ACM Symposium on Theory of Computing}, \bibinfo{year}{2002}, pp. \bibinfo{pages}{380--388}.
\bibitem[{Bondarenko and Janssen(2005)}]{bondarenko2005documents}
\bibinfo{author}{O.~Bondarenko}, \bibinfo{author}{R.~Janssen},
\newblock \bibinfo{title}{Documents at hand: Learning from paper to improve digital technologies},
\newblock in: \bibinfo{booktitle}{Proceedings of the SIGCHI conference on Human factors in computing systems}, \bibinfo{year}{2005}, pp. \bibinfo{pages}{121--130}.
\bibitem[{Manber et~al.(1994)}]{manber1994finding}
\bibinfo{author}{U.~Manber}, et~al.,
\newblock \bibinfo{title}{Finding similar files in a large file system.},
\newblock in: \bibinfo{booktitle}{Usenix Winter}, volume~\bibinfo{volume}{94}, \bibinfo{year}{1994}, pp. \bibinfo{pages}{1--10}.
\bibitem[{Sun et~al.(2013)Sun, Qin, and Wang}]{sun2013near}
\bibinfo{author}{Y.~Sun}, \bibinfo{author}{J.~Qin}, \bibinfo{author}{W.~Wang},
\newblock \bibinfo{title}{Near duplicate text detection using frequency-biased signatures},
\newblock in: \bibinfo{booktitle}{14th International Conference on Web Information Systems Engineering}, \bibinfo{organization}{Springer}, \bibinfo{year}{2013}, pp. \bibinfo{pages}{277--291}.
\bibitem[{Abdel~Hamid et~al.(2009)Abdel~Hamid, Behzadi, Christoph, and Henzinger}]{abdel2009detecting}
\bibinfo{author}{O.~Abdel~Hamid}, \bibinfo{author}{B.~Behzadi}, \bibinfo{author}{S.~Christoph}, \bibinfo{author}{M.~Henzinger},
\newblock \bibinfo{title}{Detecting the origin of text segments efficiently},
\newblock in: \bibinfo{booktitle}{The 18th International Conference on World Wide Web}, \bibinfo{year}{2009}, pp. \bibinfo{pages}{61--70}.
\bibitem[{Broder et~al.(1998)Broder, Charikar, Frieze, and Mitzenmacher}]{broder1998min}
\bibinfo{author}{A.~Z. Broder}, \bibinfo{author}{M.~Charikar}, \bibinfo{author}{A.~M. Frieze}, \bibinfo{author}{M.~Mitzenmacher},
\newblock \bibinfo{title}{Min-wise independent permutations},
\newblock in: \bibinfo{booktitle}{The thirtieth Annual ACM Symposium on Theory of Computing}, \bibinfo{year}{1998}, pp. \bibinfo{pages}{327--336}.
\bibitem[{Hassanian-esfahani and Kargar(2019)}]{hassanian2019pruning}
\bibinfo{author}{R.~Hassanian-esfahani}, \bibinfo{author}{M.-j. Kargar},
\newblock \bibinfo{title}{A pruning strategy to improve pairwise comparison-based near-duplicate detection},
\newblock \bibinfo{journal}{Knowledge and Information Systems} \bibinfo{volume}{61} (\bibinfo{year}{2019}) \bibinfo{pages}{931--963}.
\bibitem[{Rabin(1981)}]{rabin1981fingerprinting}
\bibinfo{author}{M.~O. Rabin},
\newblock \bibinfo{title}{Fingerprinting by random polynomials},
\newblock \bibinfo{journal}{Technical report}  (\bibinfo{year}{1981}).
\bibitem[{Brin et~al.(1995)Brin, Davis, and Garcia-Molina}]{brin1995copy}
\bibinfo{author}{S.~Brin}, \bibinfo{author}{J.~Davis}, \bibinfo{author}{H.~Garcia-Molina},
\newblock \bibinfo{title}{Copy detection mechanisms for digital documents},
\newblock in: \bibinfo{booktitle}{ACM SIGMOD International Conference on Management of Data}, \bibinfo{year}{1995}, pp. \bibinfo{pages}{398--409}.
\bibitem[{Seo and Croft(2008)}]{seo2008local}
\bibinfo{author}{J.~Seo}, \bibinfo{author}{W.~B. Croft},
\newblock \bibinfo{title}{Local text reuse detection},
\newblock in: \bibinfo{booktitle}{The 31st Annual International ACM SIGIR Conference on Research and Development in Information Retrieval}, \bibinfo{year}{2008}, pp. \bibinfo{pages}{571--578}.
\bibitem[{Kulis and Grauman(2011)}]{kulis2011kernelized}
\bibinfo{author}{B.~Kulis}, \bibinfo{author}{K.~Grauman},
\newblock \bibinfo{title}{Kernelized locality-sensitive hashing},
\newblock \bibinfo{journal}{IEEE Transactions on Pattern Analysis and Machine Intelligence} \bibinfo{volume}{34} (\bibinfo{year}{2011}) \bibinfo{pages}{1092--1104}.
\bibitem[{Wu et~al.(2020)Wu, Li, Chen, Gao, and Zhang}]{wu2020review}
\bibinfo{author}{W.~Wu}, \bibinfo{author}{B.~Li}, \bibinfo{author}{L.~Chen}, \bibinfo{author}{J.~Gao}, \bibinfo{author}{C.~Zhang},
\newblock \bibinfo{title}{A review for weighted minhash algorithms},
\newblock \bibinfo{journal}{IEEE Transactions on Knowledge and Data Engineering} \bibinfo{volume}{34} (\bibinfo{year}{2020}) \bibinfo{pages}{2553--2573}.
\bibitem[{Gao et~al.(2010)Gao, Lu, Tao, and Li}]{gao2010image}
\bibinfo{author}{X.~Gao}, \bibinfo{author}{W.~Lu}, \bibinfo{author}{D.~Tao}, \bibinfo{author}{X.~Li},
\newblock \bibinfo{title}{Image quality assessment and human visual system},
\newblock in: \bibinfo{booktitle}{Visual Communications and Image Processing}, volume \bibinfo{volume}{7744}, \bibinfo{organization}{SPIE}, \bibinfo{year}{2010}, pp. \bibinfo{pages}{316--325}.
\bibitem[{Zhao et~al.(2012)Zhao, Wang, Zhang, and Yao}]{zhao2012robust}
\bibinfo{author}{Y.~Zhao}, \bibinfo{author}{S.~Wang}, \bibinfo{author}{X.~Zhang}, \bibinfo{author}{H.~Yao},
\newblock \bibinfo{title}{Robust hashing for image authentication using zernike moments and local features},
\newblock \bibinfo{journal}{IEEE Transactions on Information Forensics and Security} \bibinfo{volume}{8} (\bibinfo{year}{2012}) \bibinfo{pages}{55--63}.
\bibitem[{Xia et~al.(2023)Xia, Li, Chen, and Yang}]{xia2023perceptual}
\bibinfo{author}{M.~Xia}, \bibinfo{author}{S.~Li}, \bibinfo{author}{W.~Chen}, \bibinfo{author}{G.~Yang},
\newblock \bibinfo{title}{Perceptual image hashing using rotation invariant uniform local binary patterns and color feature},
\newblock in: \bibinfo{booktitle}{Advances in Computers}, volume \bibinfo{volume}{130}, \bibinfo{publisher}{Elsevier}, \bibinfo{year}{2023}, pp. \bibinfo{pages}{163--205}.
\bibitem[{Tang et~al.(2015)Tang, Zhang, Li, and Zhang}]{tang2015robust2}
\bibinfo{author}{Z.~Tang}, \bibinfo{author}{X.~Zhang}, \bibinfo{author}{X.~Li}, \bibinfo{author}{S.~Zhang},
\newblock \bibinfo{title}{Robust image hashing with ring partition and invariant vector distance},
\newblock \bibinfo{journal}{IEEE Transactions on Information Forensics and Security} \bibinfo{volume}{11} (\bibinfo{year}{2015}) \bibinfo{pages}{200--214}.
\bibitem[{Huang and Liu(2018)}]{huang2018robustness}
\bibinfo{author}{Z.~Huang}, \bibinfo{author}{S.~Liu},
\newblock \bibinfo{title}{Robustness and discrimination oriented hashing combining texture and invariant vector distance},
\newblock in: \bibinfo{booktitle}{The 26th ACM International Conference on Multimedia}, \bibinfo{year}{2018}, pp. \bibinfo{pages}{1389--1397}.
\bibitem[{Hosny et~al.(2018)Hosny, Khedr, Khedr, and Mohamed}]{hosny2018robust}
\bibinfo{author}{K.~M. Hosny}, \bibinfo{author}{Y.~M. Khedr}, \bibinfo{author}{W.~I. Khedr}, \bibinfo{author}{E.~R. Mohamed},
\newblock \bibinfo{title}{Robust image hashing using exact gaussian--hermite moments},
\newblock \bibinfo{journal}{IET Image Processing} \bibinfo{volume}{12} (\bibinfo{year}{2018}) \bibinfo{pages}{2178--2185}.
\bibitem[{Tang et~al.(2018)Tang, Li, Zhang, Zhang, and Dai}]{tang2018image}
\bibinfo{author}{Z.~Tang}, \bibinfo{author}{X.~Li}, \bibinfo{author}{X.~Zhang}, \bibinfo{author}{S.~Zhang}, \bibinfo{author}{Y.~Dai},
\newblock \bibinfo{title}{Image hashing with color vector angle},
\newblock \bibinfo{journal}{Neurocomputing} \bibinfo{volume}{308} (\bibinfo{year}{2018}) \bibinfo{pages}{147--158}.
\bibitem[{Tang et~al.(2019)Tang, Yu, Zhang, Yu, Yu, and Zhang}]{tang2019robust}
\bibinfo{author}{Z.~Tang}, \bibinfo{author}{Y.~Yu}, \bibinfo{author}{H.~Zhang}, \bibinfo{author}{M.~Yu}, \bibinfo{author}{C.~Yu}, \bibinfo{author}{X.~Zhang},
\newblock \bibinfo{title}{Robust image hashing via visual attention model and ring partition},
\newblock \bibinfo{journal}{Mathematical Biosciences and Engineering} \bibinfo{volume}{16} (\bibinfo{year}{2019}) \bibinfo{pages}{6103--6120}.
\bibitem[{Zhao and Yuan(2020)}]{zhao2020perceptual}
\bibinfo{author}{Y.~Zhao}, \bibinfo{author}{X.~Yuan},
\newblock \bibinfo{title}{Perceptual image hashing based on color structure and intensity gradient},
\newblock \bibinfo{journal}{IEEE Access} \bibinfo{volume}{8} (\bibinfo{year}{2020}) \bibinfo{pages}{26041--26053}.
\bibitem[{Roy and Shukla(2013)}]{roy2013spatial}
\bibinfo{author}{V.~Roy}, \bibinfo{author}{S.~Shukla},
\newblock \bibinfo{title}{Spatial and transform domain filtering method for image de-noising: a review},
\newblock \bibinfo{journal}{International Journal of Modern Education and Computer Science} \bibinfo{volume}{5} (\bibinfo{year}{2013}) \bibinfo{pages}{41}.
\bibitem[{Tang et~al.(2014)Tang, Yang, Huang, and Zhang}]{tang2014robust2}
\bibinfo{author}{Z.~Tang}, \bibinfo{author}{F.~Yang}, \bibinfo{author}{L.~Huang}, \bibinfo{author}{X.~Zhang},
\newblock \bibinfo{title}{Robust image hashing with dominant {DCT} coefficients},
\newblock \bibinfo{journal}{International Journal for Light and Electron Optics} \bibinfo{volume}{125} (\bibinfo{year}{2014}) \bibinfo{pages}{5102--5107}.
\bibitem[{Liu and Huang(2019)}]{liu2019efficient}
\bibinfo{author}{S.~Liu}, \bibinfo{author}{Z.~Huang},
\newblock \bibinfo{title}{Efficient image hashing with geometric invariant vector distance for copy detection},
\newblock \bibinfo{journal}{ACM Transactions on Multimedia Computing, Communications, and Applications} \bibinfo{volume}{15} (\bibinfo{year}{2019}) \bibinfo{pages}{1--22}.
\bibitem[{Tang et~al.(2014)Tang, Dai, Zhang, Huang, and Yang}]{tang2014robust1}
\bibinfo{author}{Z.~Tang}, \bibinfo{author}{Y.~Dai}, \bibinfo{author}{X.~Zhang}, \bibinfo{author}{L.~Huang}, \bibinfo{author}{F.~Yang},
\newblock \bibinfo{title}{Robust image hashing via colour vector angles and discrete wavelet transform},
\newblock \bibinfo{journal}{IET Image Processing} \bibinfo{volume}{8} (\bibinfo{year}{2014}) \bibinfo{pages}{142--149}.
\bibitem[{Tang et~al.(2018)Tang, Huang, Yao, Zhang, Chen, and Yu}]{tang2018perceptual}
\bibinfo{author}{Z.~Tang}, \bibinfo{author}{Z.~Huang}, \bibinfo{author}{H.~Yao}, \bibinfo{author}{X.~Zhang}, \bibinfo{author}{L.~Chen}, \bibinfo{author}{C.~Yu},
\newblock \bibinfo{title}{Perceptual image hashing with weighted {DWT} features for reduced-reference image quality assessment},
\newblock \bibinfo{journal}{The Computer Journal} \bibinfo{volume}{61} (\bibinfo{year}{2018}) \bibinfo{pages}{1695--1709}.
\bibitem[{Ahmed et~al.(2010)Ahmed, Siyal, and Abbas}]{ahmed2010secure}
\bibinfo{author}{F.~Ahmed}, \bibinfo{author}{M.~Y. Siyal}, \bibinfo{author}{V.~U. Abbas},
\newblock \bibinfo{title}{A secure and robust hash-based scheme for image authentication},
\newblock \bibinfo{journal}{Signal Processing} \bibinfo{volume}{90} (\bibinfo{year}{2010}) \bibinfo{pages}{1456--1470}.
\bibitem[{Lu and Hsu(2005)}]{lu2005geometric}
\bibinfo{author}{C.-S. Lu}, \bibinfo{author}{C.-Y. Hsu},
\newblock \bibinfo{title}{Geometric distortion-resilient image hashing scheme and its applications on copy detection and authentication},
\newblock \bibinfo{journal}{Multimedia Systems} \bibinfo{volume}{11} (\bibinfo{year}{2005}) \bibinfo{pages}{159--173}.
\bibitem[{Lu et~al.(2004)Lu, Hsu, Sun, and Chang}]{lu2004robust}
\bibinfo{author}{C.-S. Lu}, \bibinfo{author}{C.-Y. Hsu}, \bibinfo{author}{S.-W. Sun}, \bibinfo{author}{P.-C. Chang},
\newblock \bibinfo{title}{Robust mesh-based hashing for copy detection and tracing of images},
\newblock in: \bibinfo{booktitle}{IEEE International Conference on Multimedia and Expo}, volume~\bibinfo{volume}{1}, \bibinfo{organization}{IEEE}, \bibinfo{year}{2004}, pp. \bibinfo{pages}{731--734}.
\bibitem[{Karsh et~al.(2017)Karsh, Laskar, and Aditi}]{karsh2017robust}
\bibinfo{author}{R.~K. Karsh}, \bibinfo{author}{R.~H. Laskar}, \bibinfo{author}{Aditi},
\newblock \bibinfo{title}{Robust image hashing through dwt-svd and spectral residual method},
\newblock \bibinfo{journal}{EURASIP Journal on Image and Video Processing} \bibinfo{volume}{2017} (\bibinfo{year}{2017}) \bibinfo{pages}{1--17}.
\bibitem[{Tang et~al.(2016)Tang, Lao, Zhang, and Liu}]{tang2016robust}
\bibinfo{author}{Z.~Tang}, \bibinfo{author}{H.~Lao}, \bibinfo{author}{X.~Zhang}, \bibinfo{author}{K.~Liu},
\newblock \bibinfo{title}{Robust image hashing via {DCT} and {LLE}},
\newblock \bibinfo{journal}{Computers and Security} \bibinfo{volume}{62} (\bibinfo{year}{2016}) \bibinfo{pages}{133--148}.
\bibitem[{De~Roover et~al.(2005)De~Roover, De~Vleeschouwer, Lefebvre, and Macq}]{de2005robust}
\bibinfo{author}{C.~De~Roover}, \bibinfo{author}{C.~De~Vleeschouwer}, \bibinfo{author}{F.~Lefebvre}, \bibinfo{author}{B.~Macq},
\newblock \bibinfo{title}{Robust video hashing based on radial projections of key frames},
\newblock \bibinfo{journal}{IEEE Transactions on Signal Processing} \bibinfo{volume}{53} (\bibinfo{year}{2005}) \bibinfo{pages}{4020--4037}.
\bibitem[{Tang et~al.(2011)Tang, Wang, Zhang, Wei, and Zhao}]{tang2011lexicographical}
\bibinfo{author}{Z.~Tang}, \bibinfo{author}{S.~Wang}, \bibinfo{author}{X.~Zhang}, \bibinfo{author}{W.~Wei}, \bibinfo{author}{Y.~Zhao},
\newblock \bibinfo{title}{Lexicographical framework for image hashing with implementation based on {DCT} and {NMF}},
\newblock \bibinfo{journal}{Multimedia Tools and Applications} \bibinfo{volume}{52} (\bibinfo{year}{2011}) \bibinfo{pages}{325--345}.
\bibitem[{Paul et~al.(2020)Paul, Tulshan, Karsh, and Talukdar}]{paul2020image}
\bibinfo{author}{M.~Paul}, \bibinfo{author}{S.~Tulshan}, \bibinfo{author}{R.~K. Karsh}, \bibinfo{author}{F.~Talukdar},
\newblock \bibinfo{title}{Image authentication using radon transform and local features},
\newblock in: \bibinfo{booktitle}{Advanced Communication Technologies and Signal Processing}, \bibinfo{organization}{IEEE}, \bibinfo{year}{2020}, pp. \bibinfo{pages}{1--6}.
\bibitem[{Lei et~al.(2011)Lei, Wang, and Huang}]{lei2011robust}
\bibinfo{author}{Y.~Lei}, \bibinfo{author}{Y.~Wang}, \bibinfo{author}{J.~Huang},
\newblock \bibinfo{title}{Robust image hash in radon transform domain for authentication},
\newblock \bibinfo{journal}{Signal Processing: Image Communication} \bibinfo{volume}{26} (\bibinfo{year}{2011}) \bibinfo{pages}{280--288}.
\bibitem[{Seo et~al.(2004)Seo, Haitsma, Kalker, and Yoo}]{seo2004robust}
\bibinfo{author}{J.~S. Seo}, \bibinfo{author}{J.~Haitsma}, \bibinfo{author}{T.~Kalker}, \bibinfo{author}{C.~D. Yoo},
\newblock \bibinfo{title}{A robust image fingerprinting system using the radon transform},
\newblock \bibinfo{journal}{Signal Processing: Image Communication} \bibinfo{volume}{19} (\bibinfo{year}{2004}) \bibinfo{pages}{325--339}.
\bibitem[{Wu et~al.(2009)Wu, Zhou, and Niu}]{wu2009novel}
\bibinfo{author}{D.~Wu}, \bibinfo{author}{X.~Zhou}, \bibinfo{author}{X.~Niu},
\newblock \bibinfo{title}{A novel image hash algorithm resistant to print--scan},
\newblock \bibinfo{journal}{Signal Processing} \bibinfo{volume}{89} (\bibinfo{year}{2009}) \bibinfo{pages}{2415--2424}.
\bibitem[{Qin et~al.(2013)Qin, Chang, and Tsou}]{qin2013robust}
\bibinfo{author}{C.~Qin}, \bibinfo{author}{C.-C. Chang}, \bibinfo{author}{P.-L. Tsou},
\newblock \bibinfo{title}{Robust image hashing using non-uniform sampling in discrete fourier domain},
\newblock \bibinfo{journal}{Digital Signal Processing} \bibinfo{volume}{23} (\bibinfo{year}{2013}) \bibinfo{pages}{578--585}.
\bibitem[{Swaminathan et~al.(2004)Swaminathan, Mao, and Wu}]{swaminathan2004image}
\bibinfo{author}{A.~Swaminathan}, \bibinfo{author}{Y.~Mao}, \bibinfo{author}{M.~Wu},
\newblock \bibinfo{title}{Image hashing resilient to geometric and filtering operations},
\newblock in: \bibinfo{booktitle}{IEEE 6th Workshop on Multimedia Signal Processing}, \bibinfo{organization}{IEEE}, \bibinfo{year}{2004}, pp. \bibinfo{pages}{355--358}.
\bibitem[{Swaminathan et~al.(2006)Swaminathan, Mao, and Wu}]{swaminathan2006robust}
\bibinfo{author}{A.~Swaminathan}, \bibinfo{author}{Y.~Mao}, \bibinfo{author}{M.~Wu},
\newblock \bibinfo{title}{Robust and secure image hashing},
\newblock \bibinfo{journal}{IEEE Transactions on Information Forensics and Security} \bibinfo{volume}{1} (\bibinfo{year}{2006}) \bibinfo{pages}{215--230}.
\bibitem[{Wang et~al.(2012)Wang, Xue, Zheng, Liu, and Li}]{wang2012image}
\bibinfo{author}{X.~Wang}, \bibinfo{author}{J.~Xue}, \bibinfo{author}{Z.~Zheng}, \bibinfo{author}{Z.~Liu}, \bibinfo{author}{N.~Li},
\newblock \bibinfo{title}{Image forensic signature for content authenticity analysis},
\newblock \bibinfo{journal}{Journal of Visual Communication and Image Representation} \bibinfo{volume}{23} (\bibinfo{year}{2012}) \bibinfo{pages}{782--797}.
\bibitem[{Lv and Wang(2012)}]{lv2012perceptual}
\bibinfo{author}{X.~Lv}, \bibinfo{author}{Z.~J. Wang},
\newblock \bibinfo{title}{Perceptual image hashing based on shape contexts and local feature points},
\newblock \bibinfo{journal}{IEEE Transactions on Information Forensics and Security} \bibinfo{volume}{7} (\bibinfo{year}{2012}) \bibinfo{pages}{1081--1093}.
\bibitem[{Yan et~al.(2016)Yan, Pun, and Yuan}]{yan2016multi}
\bibinfo{author}{C.-P. Yan}, \bibinfo{author}{C.-M. Pun}, \bibinfo{author}{X.-C. Yuan},
\newblock \bibinfo{title}{Multi-scale image hashing using adaptive local feature extraction for robust tampering detection},
\newblock \bibinfo{journal}{Signal Processing} \bibinfo{volume}{121} (\bibinfo{year}{2016}) \bibinfo{pages}{1--16}.
\bibitem[{Pun et~al.(2018)Pun, Yan, and Yuan}]{pun2018robust}
\bibinfo{author}{C.-M. Pun}, \bibinfo{author}{C.-P. Yan}, \bibinfo{author}{X.-C. Yuan},
\newblock \bibinfo{title}{Robust image hashing using progressive feature selection for tampering detection},
\newblock \bibinfo{journal}{Multimedia Tools and Applications} \bibinfo{volume}{77} (\bibinfo{year}{2018}) \bibinfo{pages}{11609--11633}.
\bibitem[{Paul et~al.(2019)Paul, Karsh, and Talukdar}]{paul2019image}
\bibinfo{author}{M.~Paul}, \bibinfo{author}{R.~K. Karsh}, \bibinfo{author}{F.~A. Talukdar},
\newblock \bibinfo{title}{Image hashing based on shape context and speeded up robust features ({SURF})},
\newblock in: \bibinfo{booktitle}{International Conference on Automation, Computational and Technology Management}, \bibinfo{organization}{IEEE}, \bibinfo{year}{2019}, pp. \bibinfo{pages}{464--468}.
\bibitem[{Singh et~al.(2021)Singh, Bhatnagar, and Singh}]{singh2021new}
\bibinfo{author}{S.~P. Singh}, \bibinfo{author}{G.~Bhatnagar}, \bibinfo{author}{A.~K. Singh},
\newblock \bibinfo{title}{A new robust reference image hashing system},
\newblock \bibinfo{journal}{IEEE Transactions on Dependable and Secure Computing} \bibinfo{volume}{19} (\bibinfo{year}{2021}) \bibinfo{pages}{2211--2225}.
\bibitem[{M{\i}h{\c{c}}ak and Venkatesan(2001)}]{mihccak2001new}
\bibinfo{author}{M.~K. M{\i}h{\c{c}}ak}, \bibinfo{author}{R.~Venkatesan},
\newblock \bibinfo{title}{New iterative geometric methods for robust perceptual image hashing},
\newblock in: \bibinfo{booktitle}{ACM Workshop on Digital Rights Management}, \bibinfo{organization}{Springer}, \bibinfo{year}{2001}, pp. \bibinfo{pages}{13--21}.
\bibitem[{Monga and Mih{\c{c}}ak(2007)}]{monga2007robust}
\bibinfo{author}{V.~Monga}, \bibinfo{author}{M.~K. Mih{\c{c}}ak},
\newblock \bibinfo{title}{Robust and secure image hashing via non-negative matrix factorizations},
\newblock \bibinfo{journal}{IEEE Transactions on Information Forensics and Security} \bibinfo{volume}{2} (\bibinfo{year}{2007}) \bibinfo{pages}{376--390}.
\bibitem[{Lv and Wang(2009)}]{lv2009extended}
\bibinfo{author}{X.~Lv}, \bibinfo{author}{Z.~Wang},
\newblock \bibinfo{title}{An extended image hashing concept: content-based fingerprinting using fjlt},
\newblock \bibinfo{journal}{EURASIP Journal on Information Security} \bibinfo{volume}{2009} (\bibinfo{year}{2009}) \bibinfo{pages}{1--16}.
\bibitem[{Tang et~al.(2015)Tang, Ruan, Qin, Zhang, and Yu}]{tang2015robust1}
\bibinfo{author}{Z.~Tang}, \bibinfo{author}{L.~Ruan}, \bibinfo{author}{C.~Qin}, \bibinfo{author}{X.~Zhang}, \bibinfo{author}{C.~Yu},
\newblock \bibinfo{title}{Robust image hashing with embedding vector variance of {LLE}},
\newblock \bibinfo{journal}{Digital Signal Processing} \bibinfo{volume}{43} (\bibinfo{year}{2015}) \bibinfo{pages}{17--27}.
\bibitem[{Sun and Zeng(2014)}]{sun2014secure}
\bibinfo{author}{R.~Sun}, \bibinfo{author}{W.~Zeng},
\newblock \bibinfo{title}{Secure and robust image hashing via compressive sensing},
\newblock \bibinfo{journal}{Multimedia Tools and Applications} \bibinfo{volume}{70} (\bibinfo{year}{2014}) \bibinfo{pages}{1651--1665}.
\bibitem[{Ghouti(2014)}]{ghouti2014robust}
\bibinfo{author}{L.~Ghouti},
\newblock \bibinfo{title}{Robust perceptual color image hashing using quaternion singular value decomposition},
\newblock in: \bibinfo{booktitle}{IEEE International Conference on Acoustics, Speech and Signal Processing}, \bibinfo{organization}{IEEE}, \bibinfo{year}{2014}, pp. \bibinfo{pages}{3794--3798}.
\bibitem[{Tagliasacchi et~al.(2009)Tagliasacchi, Valenzise, and Tubaro}]{tagliasacchi2009hash}
\bibinfo{author}{M.~Tagliasacchi}, \bibinfo{author}{G.~Valenzise}, \bibinfo{author}{S.~Tubaro},
\newblock \bibinfo{title}{Hash-based identification of sparse image tampering},
\newblock \bibinfo{journal}{IEEE Transactions on Image Processing} \bibinfo{volume}{18} (\bibinfo{year}{2009}) \bibinfo{pages}{2491--2504}.
\bibitem[{Tang et~al.(2017)Tang, Huang, Zhang, and Lao}]{tang2017robust}
\bibinfo{author}{Z.~Tang}, \bibinfo{author}{Z.~Huang}, \bibinfo{author}{X.~Zhang}, \bibinfo{author}{H.~Lao},
\newblock \bibinfo{title}{Robust image hashing with multidimensional scaling},
\newblock \bibinfo{journal}{Signal Processing} \bibinfo{volume}{137} (\bibinfo{year}{2017}) \bibinfo{pages}{240--250}.
\bibitem[{Qin et~al.(2018)Qin, Sun, and Chang}]{qin2018perceptual}
\bibinfo{author}{C.~Qin}, \bibinfo{author}{M.~Sun}, \bibinfo{author}{C.-C. Chang},
\newblock \bibinfo{title}{Perceptual hashing for color images based on hybrid extraction of structural features},
\newblock \bibinfo{journal}{Signal Processing} \bibinfo{volume}{142} (\bibinfo{year}{2018}) \bibinfo{pages}{194--205}.
\bibitem[{Indyk et~al.(1999)Indyk, Iyengar, and Shivakumar}]{indyk1999finding}
\bibinfo{author}{P.~Indyk}, \bibinfo{author}{G.~Iyengar}, \bibinfo{author}{N.~Shivakumar}, \bibinfo{title}{Finding pirated video sequences on the internet}, \bibinfo{type}{Technical Report}, Technical report, Stanford University, \bibinfo{year}{1999}.
\bibitem[{Lee and Yoo(2008)}]{lee2008robust}
\bibinfo{author}{S.~Lee}, \bibinfo{author}{C.~D. Yoo},
\newblock \bibinfo{title}{Robust video fingerprinting for content-based video identification},
\newblock \bibinfo{journal}{IEEE Transactions on Circuits and Systems for Video Technology} \bibinfo{volume}{18} (\bibinfo{year}{2008}) \bibinfo{pages}{983--988}.
\bibitem[{Hampapur et~al.(2001)Hampapur, Hyun, and Bolle}]{hampapur2001comparison}
\bibinfo{author}{A.~Hampapur}, \bibinfo{author}{K.~Hyun}, \bibinfo{author}{R.~M. Bolle},
\newblock \bibinfo{title}{Comparison of sequence matching techniques for video copy detection},
\newblock in: \bibinfo{booktitle}{Storage and Retrieval for Media Databases}, volume \bibinfo{volume}{4676}, \bibinfo{organization}{SPIE}, \bibinfo{year}{2001}, pp. \bibinfo{pages}{194--201}.
\bibitem[{Neelima and Singh(2017)}]{neelima2017collusion}
\bibinfo{author}{A.~Neelima}, \bibinfo{author}{K.~Singh},
\newblock \bibinfo{title}{Collusion and rotation resilient video hashing based on scale invariant feature transform},
\newblock \bibinfo{journal}{The Imaging Science Journal} \bibinfo{volume}{65} (\bibinfo{year}{2017}) \bibinfo{pages}{62--74}.
\bibitem[{Wary and Neelima(2019)}]{wary2019review}
\bibinfo{author}{A.~Wary}, \bibinfo{author}{A.~Neelima},
\newblock \bibinfo{title}{A review on robust video copy detection},
\newblock \bibinfo{journal}{International Journal of Multimedia Information Retrieval} \bibinfo{volume}{8} (\bibinfo{year}{2019}) \bibinfo{pages}{61--78}.
\bibitem[{Ghodrati et~al.(2018)Ghodrati, Gavves, Snoek et~al.}]{ghodrati2018video}
\bibinfo{author}{A.~Ghodrati}, \bibinfo{author}{E.~Gavves}, \bibinfo{author}{C.~Snoek}, et~al.,
\newblock \bibinfo{title}{Video time: Properties, encoders and evaluation}  (\bibinfo{year}{2018}).
\bibitem[{Hua et~al.(2004)Hua, Chen, and Zhang}]{hua2004robust}
\bibinfo{author}{X.-S. Hua}, \bibinfo{author}{X.~Chen}, \bibinfo{author}{H.-J. Zhang},
\newblock \bibinfo{title}{Robust video signature based on ordinal measure},
\newblock in: \bibinfo{booktitle}{International Conference on Image Processing}, volume~\bibinfo{volume}{1}, \bibinfo{organization}{IEEE}, \bibinfo{year}{2004}, pp. \bibinfo{pages}{685--688}.
\bibitem[{Kim and Vasudev(2005)}]{kim2005spatiotemporal}
\bibinfo{author}{C.~Kim}, \bibinfo{author}{B.~Vasudev},
\newblock \bibinfo{title}{Spatiotemporal sequence matching for efficient video copy detection},
\newblock \bibinfo{journal}{IEEE Transactions on Circuits and Systems for Video Technology} \bibinfo{volume}{15} (\bibinfo{year}{2005}) \bibinfo{pages}{127--132}.
\bibitem[{Oostveen et~al.(2002)Oostveen, Kalker, and Haitsma}]{oostveen2002feature}
\bibinfo{author}{J.~Oostveen}, \bibinfo{author}{T.~Kalker}, \bibinfo{author}{J.~Haitsma},
\newblock \bibinfo{title}{Feature extraction and a database strategy for video fingerprinting},
\newblock in: \bibinfo{booktitle}{5th International Conference on Recent Advances in Visual Information Systems}, \bibinfo{organization}{Springer}, \bibinfo{year}{2002}, pp. \bibinfo{pages}{117--128}.
\bibitem[{Lee and Yoo(2006)}]{lee2006video}
\bibinfo{author}{S.~Lee}, \bibinfo{author}{C.~D. Yoo},
\newblock \bibinfo{title}{Video fingerprinting based on centroids of gradient orientations},
\newblock in: \bibinfo{booktitle}{International Conference on Acoustics Speech and Signal Processing Proceedings}, volume~\bibinfo{volume}{2}, \bibinfo{organization}{IEEE}, \bibinfo{year}{2006}, pp. \bibinfo{pages}{II--II}.
\bibitem[{Iwamoto et~al.(2006)Iwamoto, Kasutani, and Yamada}]{iwamoto2006image}
\bibinfo{author}{K.~Iwamoto}, \bibinfo{author}{E.~Kasutani}, \bibinfo{author}{A.~Yamada},
\newblock \bibinfo{title}{Image signature robust to caption superimposition for video sequence identification},
\newblock in: \bibinfo{booktitle}{International Conference on Image Processing}, \bibinfo{organization}{IEEE}, \bibinfo{year}{2006}, pp. \bibinfo{pages}{3185--3188}.
\bibitem[{Bhat and Nayar(1998)}]{bhat1998ordinal}
\bibinfo{author}{D.~N. Bhat}, \bibinfo{author}{S.~K. Nayar},
\newblock \bibinfo{title}{Ordinal measures for image correspondence},
\newblock \bibinfo{journal}{IEEE Transactions on Pattern Analysis and Machine Intelligence} \bibinfo{volume}{20} (\bibinfo{year}{1998}) \bibinfo{pages}{415--423}.
\bibitem[{Mohan(1998)}]{mohan1998video}
\bibinfo{author}{R.~Mohan},
\newblock \bibinfo{title}{Video sequence matching},
\newblock in: \bibinfo{booktitle}{IEEE International Conference on Acoustics, Speech and Signal Processing}, volume~\bibinfo{volume}{6}, \bibinfo{organization}{IEEE}, \bibinfo{year}{1998}, pp. \bibinfo{pages}{3697--3700}.
\bibitem[{Atrey et~al.(2007)Atrey, Yan, and Kankanhalli}]{atrey2007scalable}
\bibinfo{author}{P.~K. Atrey}, \bibinfo{author}{W.-Q. Yan}, \bibinfo{author}{M.~S. Kankanhalli},
\newblock \bibinfo{title}{A scalable signature scheme for video authentication},
\newblock \bibinfo{journal}{Multimedia Tools and Applications} \bibinfo{volume}{34} (\bibinfo{year}{2007}) \bibinfo{pages}{107--135}.
\bibitem[{Law-To et~al.(2006)Law-To, Buisson, Gouet-Brunet, and Boujemaa}]{law2006robust}
\bibinfo{author}{J.~Law-To}, \bibinfo{author}{O.~Buisson}, \bibinfo{author}{V.~Gouet-Brunet}, \bibinfo{author}{N.~Boujemaa},
\newblock \bibinfo{title}{Robust voting algorithm based on labels of behavior for video copy detection},
\newblock in: \bibinfo{booktitle}{The 14th ACM international conference on Multimedia}, \bibinfo{year}{2006}, pp. \bibinfo{pages}{835--844}.
\bibitem[{Joly et~al.(2003)Joly, Fr{\'e}licot, and Buisson}]{joly2003robust}
\bibinfo{author}{A.~Joly}, \bibinfo{author}{C.~Fr{\'e}licot}, \bibinfo{author}{O.~Buisson},
\newblock \bibinfo{title}{Robust content-based video copy identification in a large reference database},
\newblock in: \bibinfo{booktitle}{Second International Conference on Image and Video Retrieval}, \bibinfo{organization}{Springer}, \bibinfo{year}{2003}, pp. \bibinfo{pages}{414--424}.
\bibitem[{Sarkar et~al.(2008)Sarkar, Ghosh, Moxley, and Manjunath}]{sarkar2008video}
\bibinfo{author}{A.~Sarkar}, \bibinfo{author}{P.~Ghosh}, \bibinfo{author}{E.~Moxley}, \bibinfo{author}{B.~Manjunath},
\newblock \bibinfo{title}{Video fingerprinting: features for duplicate and similar video detection and query-based video retrieval},
\newblock in: \bibinfo{booktitle}{Multimedia Content Access: Algorithms and Systems II}, volume \bibinfo{volume}{6820}, \bibinfo{organization}{SPIE}, \bibinfo{year}{2008}, pp. \bibinfo{pages}{153--164}.
\bibitem[{Massoudi et~al.(2006)Massoudi, Lefebvre, Demarty, Oisel, and Chupeau}]{massoudi2006video}
\bibinfo{author}{A.~Massoudi}, \bibinfo{author}{F.~Lefebvre}, \bibinfo{author}{C.-H. Demarty}, \bibinfo{author}{L.~Oisel}, \bibinfo{author}{B.~Chupeau},
\newblock \bibinfo{title}{A video fingerprint based on visual digest and local fingerprints},
\newblock in: \bibinfo{booktitle}{International Conference on Image Processing}, \bibinfo{organization}{IEEE}, \bibinfo{year}{2006}, pp. \bibinfo{pages}{2297--2300}.
\bibitem[{Rasheed et~al.(2008)Rasheed, An, Pan, Jeong, Park, and Kang}]{rasheed2008image}
\bibinfo{author}{W.~Rasheed}, \bibinfo{author}{Y.~An}, \bibinfo{author}{S.~Pan}, \bibinfo{author}{I.~Jeong}, \bibinfo{author}{J.~Park}, \bibinfo{author}{J.~Kang},
\newblock \bibinfo{title}{Image retrieval using maximum frequency of local histogram based color correlogram},
\newblock in: \bibinfo{booktitle}{International Conference on Multimedia and Ubiquitous Engineering}, \bibinfo{organization}{IEEE}, \bibinfo{year}{2008}, pp. \bibinfo{pages}{62--66}.
\bibitem[{Penatti and da~Silva~Torres(2008)}]{penatti2008color}
\bibinfo{author}{O.~A.~B. Penatti}, \bibinfo{author}{R.~da~Silva~Torres},
\newblock \bibinfo{title}{Color descriptors for web image retrieval: a comparative study},
\newblock in: \bibinfo{booktitle}{XXI Brazilian Symposium on Computer Graphics and Image Processing}, \bibinfo{organization}{IEEE}, \bibinfo{year}{2008}, pp. \bibinfo{pages}{163--170}.
\bibitem[{Lee and Yoo(2008)}]{lee2008robust1}
\bibinfo{author}{S.~Lee}, \bibinfo{author}{C.~D. Yoo},
\newblock \bibinfo{title}{Robust video fingerprinting based on affine covariant regions},
\newblock in: \bibinfo{booktitle}{IEEE International Conference on Acoustics, Speech and Signal Processing}, \bibinfo{organization}{IEEE}, \bibinfo{year}{2008}, pp. \bibinfo{pages}{1237--1240}.
\bibitem[{Shivakumar(1999)}]{shivakumar1999detecting}
\bibinfo{author}{N.~Shivakumar}, \bibinfo{title}{Detecting digital copyright violations on the Internet}, \bibinfo{publisher}{Stanford University}, \bibinfo{year}{1999}.
\bibitem[{Chen and Stentiford(2008)}]{chen2008video}
\bibinfo{author}{L.~Chen}, \bibinfo{author}{F.~Stentiford},
\newblock \bibinfo{title}{Video sequence matching based on temporal ordinal measurement},
\newblock \bibinfo{journal}{Pattern Recognition Letters} \bibinfo{volume}{29} (\bibinfo{year}{2008}) \bibinfo{pages}{1824--1831}.
\bibitem[{Radhakrishnan and Bauer(2007)}]{radhakrishnan2007content}
\bibinfo{author}{R.~Radhakrishnan}, \bibinfo{author}{C.~Bauer},
\newblock \bibinfo{title}{Content-based video signatures based on projections of difference images},
\newblock in: \bibinfo{booktitle}{IEEE 9th Workshop on Multimedia Signal Processing}, \bibinfo{organization}{IEEE}, \bibinfo{year}{2007}, pp. \bibinfo{pages}{341--344}.
\bibitem[{Coskun et~al.(2006)Coskun, Sankur, and Memon}]{coskun2006spatio}
\bibinfo{author}{B.~Coskun}, \bibinfo{author}{B.~Sankur}, \bibinfo{author}{N.~Memon},
\newblock \bibinfo{title}{Spatio--temporal transform based video hashing},
\newblock \bibinfo{journal}{IEEE Transactions on Multimedia} \bibinfo{volume}{8} (\bibinfo{year}{2006}) \bibinfo{pages}{1190--1208}.
\bibitem[{Esmaeili and Ward(2010)}]{esmaeili2010robust1}
\bibinfo{author}{M.~M. Esmaeili}, \bibinfo{author}{R.~K. Ward},
\newblock \bibinfo{title}{Robust video hashing based on temporally informative representative images},
\newblock in: \bibinfo{booktitle}{Digest of Technical Papers International Conference on Consumer Electronics}, \bibinfo{organization}{IEEE}, \bibinfo{year}{2010}, pp. \bibinfo{pages}{179--180}.
\bibitem[{Yuan et~al.(2016)Yuan, Po, Liu, Xu, Jian, Wong, and Cheung}]{yuan2016shearlet}
\bibinfo{author}{F.~Yuan}, \bibinfo{author}{L.-M. Po}, \bibinfo{author}{M.~Liu}, \bibinfo{author}{X.~Xu}, \bibinfo{author}{W.~Jian}, \bibinfo{author}{K.~Wong}, \bibinfo{author}{K.~W. Cheung},
\newblock \bibinfo{title}{Shearlet based video fingerprint for content-based copy detection},
\newblock \bibinfo{journal}{Journal of Signal and Information Processing} \bibinfo{volume}{7} (\bibinfo{year}{2016}) \bibinfo{pages}{84--97}.
\bibitem[{Min et~al.(2012)Min, Kim, De~Neve, and Ro}]{min2012video}
\bibinfo{author}{H.-s. Min}, \bibinfo{author}{S.~M. Kim}, \bibinfo{author}{W.~De~Neve}, \bibinfo{author}{Y.~M. Ro},
\newblock \bibinfo{title}{Video copy detection using inclined video tomography and bag-of-visual-words},
\newblock in: \bibinfo{booktitle}{IEEE International Conference on Multimedia and Expo}, \bibinfo{organization}{IEEE}, \bibinfo{year}{2012}, pp. \bibinfo{pages}{562--567}.
\bibitem[{Subramanyam and Emmanuel(2012)}]{subramanyam2012video}
\bibinfo{author}{A.~V. Subramanyam}, \bibinfo{author}{S.~Emmanuel},
\newblock \bibinfo{title}{Video forgery detection using hog features and compression properties},
\newblock in: \bibinfo{booktitle}{IEEE 14th International Workshop on Multimedia Signal Processing}, \bibinfo{organization}{IEEE}, \bibinfo{year}{2012}, pp. \bibinfo{pages}{89--94}.
\bibitem[{Min et~al.(2009)Min, Choi, De~Neve, and Ro}]{min2009near}
\bibinfo{author}{H.-s. Min}, \bibinfo{author}{J.~Choi}, \bibinfo{author}{W.~De~Neve}, \bibinfo{author}{Y.~M. Ro},
\newblock \bibinfo{title}{Near-duplicate video detection using temporal patterns of semantic concepts},
\newblock in: \bibinfo{booktitle}{11th IEEE International Symposium on Multimedia}, \bibinfo{organization}{IEEE}, \bibinfo{year}{2009}, pp. \bibinfo{pages}{65--71}.
\bibitem[{Wu et~al.(2009)Wu, Huang, and Jiang}]{wu2009robust}
\bibinfo{author}{Z.~Wu}, \bibinfo{author}{Q.~Huang}, \bibinfo{author}{S.~Jiang},
\newblock \bibinfo{title}{Robust copy detection by mining temporal self-similarities},
\newblock in: \bibinfo{booktitle}{IEEE International Conference on Multimedia and Expo}, \bibinfo{organization}{IEEE}, \bibinfo{year}{2009}, pp. \bibinfo{pages}{554--557}.
\bibitem[{Sowmya et~al.(2018)Sowmya, Chennamma, and Rangarajan}]{sowmya2018video}
\bibinfo{author}{K.~Sowmya}, \bibinfo{author}{H.~Chennamma}, \bibinfo{author}{L.~Rangarajan},
\newblock \bibinfo{title}{Video authentication using spatiotemporal relationship for tampering detection},
\newblock \bibinfo{journal}{Journal of Information Security and Applications} \bibinfo{volume}{41} (\bibinfo{year}{2018}) \bibinfo{pages}{159--169}.
\bibitem[{Nie et~al.(2010)Nie, Liu, and Sun}]{nie2010robust}
\bibinfo{author}{X.~Nie}, \bibinfo{author}{J.~Liu}, \bibinfo{author}{J.~Sun},
\newblock \bibinfo{title}{Robust video hashing for identification based on mds},
\newblock in: \bibinfo{booktitle}{IEEE International Conference on Acoustics, Speech and Signal Processing}, \bibinfo{organization}{IEEE}, \bibinfo{year}{2010}, pp. \bibinfo{pages}{1834--1837}.
\bibitem[{Sandeep and Bora(2020)}]{sandeep2020detection}
\bibinfo{author}{R.~Sandeep}, \bibinfo{author}{P.~K. Bora},
\newblock \bibinfo{title}{Detection of malicious video modifications using perceptual video hashing},
\newblock in: \bibinfo{booktitle}{5th International Conference on Computing, Communication and Security}, \bibinfo{organization}{IEEE}, \bibinfo{year}{2020}, pp. \bibinfo{pages}{1--5}.
\bibitem[{Gupta et~al.(2022)Gupta, Rahman, and Yasmin}]{gupta2022audio}
\bibinfo{author}{A.~Gupta}, \bibinfo{author}{A.~Rahman}, \bibinfo{author}{G.~Yasmin},
\newblock \bibinfo{title}{Audio fingerprinting using high-level feature extraction},
\newblock in: \bibinfo{booktitle}{Computational Intelligence in Pattern Recognition}, \bibinfo{organization}{Springer}, \bibinfo{year}{2022}, pp. \bibinfo{pages}{281--291}.
\bibitem[{Htun(2019)}]{htun2019analytical}
\bibinfo{author}{M.~T. Htun}, \bibinfo{title}{Analytical approach to MFCC based space-saving audio fingerprinting system}, Ph.D. thesis, MERAL Portal, \bibinfo{year}{2019}.
\bibitem[{Martinez and Kak(2001)}]{martinez2001pca}
\bibinfo{author}{A.~M. Martinez}, \bibinfo{author}{A.~C. Kak},
\newblock \bibinfo{title}{Pca versus lda},
\newblock \bibinfo{journal}{IEEE Transactions on Pattern Analysis and Machine Intelligence} \bibinfo{volume}{23} (\bibinfo{year}{2001}) \bibinfo{pages}{228--233}.
\bibitem[{Haitsma and Kalker(2002)}]{haitsma2002highly}
\bibinfo{author}{J.~Haitsma}, \bibinfo{author}{T.~Kalker},
\newblock \bibinfo{title}{A highly robust audio fingerprinting system},
\newblock in: \bibinfo{booktitle}{The International Society for Music Information Retrieval}, volume \bibinfo{volume}{2002}, \bibinfo{year}{2002}, pp. \bibinfo{pages}{107--115}.
\bibitem[{Wang et~al.(2003)}]{wang2003industrial}
\bibinfo{author}{A.~Wang}, et~al.,
\newblock \bibinfo{title}{An industrial strength audio search algorithm},
\newblock in: \bibinfo{booktitle}{The International Society for Music Information Retrieval}, volume \bibinfo{volume}{2003}, \bibinfo{organization}{Washington, DC}, \bibinfo{year}{2003}, pp. \bibinfo{pages}{7--13}.
\bibitem[{Yang et~al.(2014)Yang, Chen, and Yang}]{yang2014efficient}
\bibinfo{author}{G.~Yang}, \bibinfo{author}{X.~Chen}, \bibinfo{author}{D.~Yang},
\newblock \bibinfo{title}{Efficient music identification by utilizing space-saving audio fingerprinting system},
\newblock in: \bibinfo{booktitle}{IEEE International Conference on Multimedia and Expo}, \bibinfo{organization}{IEEE}, \bibinfo{year}{2014}, pp. \bibinfo{pages}{1--6}.
\bibitem[{Chu et~al.(2020)Chu, Niu, Yao, and Liu}]{chu2020peak}
\bibinfo{author}{R.~Chu}, \bibinfo{author}{B.~Niu}, \bibinfo{author}{S.~Yao}, \bibinfo{author}{J.~Liu},
\newblock \bibinfo{title}{Peak-based philips fingerprint robust to pitch-shift for audio identification},
\newblock \bibinfo{journal}{IEEE MultiMedia} \bibinfo{volume}{28} (\bibinfo{year}{2020}) \bibinfo{pages}{74--82}.
\bibitem[{Chen et~al.(2017)Chen, Zhang, Zhang, Huang, and Ao}]{chen2017audio}
\bibinfo{author}{D.~Chen}, \bibinfo{author}{W.~Zhang}, \bibinfo{author}{Z.~Zhang}, \bibinfo{author}{W.~Huang}, \bibinfo{author}{J.~Ao},
\newblock \bibinfo{title}{Audio retrieval based on wavelet transform},
\newblock in: \bibinfo{booktitle}{IEEE/ACIS 16th International Conference on Computer and Information Science}, \bibinfo{organization}{IEEE}, \bibinfo{year}{2017}, pp. \bibinfo{pages}{531--534}.
\bibitem[{Huerta and Stern(2001)}]{huerta2001distortion}
\bibinfo{author}{J.~M. Huerta}, \bibinfo{author}{R.~M. Stern},
\newblock \bibinfo{title}{Distortion-class modeling for robust speech recognition under gsm rpe-ltp coding},
\newblock \bibinfo{journal}{Speech Communication} \bibinfo{volume}{34} (\bibinfo{year}{2001}) \bibinfo{pages}{213--225}.
\bibitem[{Kishor et~al.(2023)Kishor, Venkatesh, and Koolagudi}]{kishor2023audio}
\bibinfo{author}{K.~Kishor}, \bibinfo{author}{S.~Venkatesh}, \bibinfo{author}{S.~G. Koolagudi},
\newblock \bibinfo{title}{Audio fingerprinting system to detect and match audio recordings},
\newblock in: \bibinfo{booktitle}{International Conference on Pattern Recognition and Machine Intelligence}, \bibinfo{organization}{Springer}, \bibinfo{year}{2023}, pp. \bibinfo{pages}{683--690}.
\bibitem[{Sun et~al.(2018)Sun, Zhang, and Chen}]{sun2018movie}
\bibinfo{author}{X.~Sun}, \bibinfo{author}{W.~Zhang}, \bibinfo{author}{D.~Chen},
\newblock \bibinfo{title}{Movie retrieval based on shazam algorithm},
\newblock in: \bibinfo{booktitle}{IEEE 4th Information Technology and Mechatronics Engineering Conference}, \bibinfo{organization}{IEEE}, \bibinfo{year}{2018}, pp. \bibinfo{pages}{1129--1133}.
\bibitem[{Anguera et~al.(2012)Anguera, Garzon, and Adamek}]{anguera2012mask}
\bibinfo{author}{X.~Anguera}, \bibinfo{author}{A.~Garzon}, \bibinfo{author}{T.~Adamek},
\newblock \bibinfo{title}{{MASK}: Robust local features for audio fingerprinting},
\newblock in: \bibinfo{booktitle}{IEEE International Conference on Multimedia and Expo}, \bibinfo{organization}{IEEE}, \bibinfo{year}{2012}, pp. \bibinfo{pages}{455--460}.
\bibitem[{Salakhutdinov and Hinton(2009)}]{salakhutdinov2009semantic}
\bibinfo{author}{R.~Salakhutdinov}, \bibinfo{author}{G.~Hinton},
\newblock \bibinfo{title}{Semantic hashing},
\newblock \bibinfo{journal}{International Journal of Approximate Reasoning} \bibinfo{volume}{50} (\bibinfo{year}{2009}) \bibinfo{pages}{969--978}.
\bibitem[{Hinton and Salakhutdinov(2011)}]{hinton2011discovering}
\bibinfo{author}{G.~Hinton}, \bibinfo{author}{R.~Salakhutdinov},
\newblock \bibinfo{title}{Discovering binary codes for documents by learning deep generative models},
\newblock \bibinfo{journal}{Topics in Cognitive Science} \bibinfo{volume}{3} (\bibinfo{year}{2011}) \bibinfo{pages}{74--91}.
\bibitem[{Masci et~al.(2013)Masci, Bronstein, Bronstein, and Schmidhuber}]{masci2013multimodal}
\bibinfo{author}{J.~Masci}, \bibinfo{author}{M.~M. Bronstein}, \bibinfo{author}{A.~M. Bronstein}, \bibinfo{author}{J.~Schmidhuber},
\newblock \bibinfo{title}{Multimodal similarity-preserving hashing},
\newblock \bibinfo{journal}{IEEE Transactions on Pattern Analysis and Machine Intelligence} \bibinfo{volume}{36} (\bibinfo{year}{2013}) \bibinfo{pages}{824--830}.
\bibitem[{Xu et~al.(2015)Xu, Wang, Tian, Xu, Zhao, Wang, and Hao}]{xu2015convolutional}
\bibinfo{author}{J.~Xu}, \bibinfo{author}{P.~Wang}, \bibinfo{author}{G.~Tian}, \bibinfo{author}{B.~Xu}, \bibinfo{author}{J.~Zhao}, \bibinfo{author}{F.~Wang}, \bibinfo{author}{H.~Hao},
\newblock \bibinfo{title}{Convolutional neural networks for text hashing},
\newblock in: \bibinfo{booktitle}{Twenty-Fourth International Joint Conference on Artificial Intelligence}, \bibinfo{year}{2015}.
\bibitem[{Zhang et~al.(2015)Zhang, Wang, Qian, and Huang}]{zhang2015mixed}
\bibinfo{author}{Q.~Zhang}, \bibinfo{author}{Y.~Wang}, \bibinfo{author}{J.~Qian}, \bibinfo{author}{X.~Huang},
\newblock \bibinfo{title}{A mixed generative-discriminative based hashing method},
\newblock \bibinfo{journal}{IEEE Transactions on Knowledge and Data Engineering} \bibinfo{volume}{28} (\bibinfo{year}{2015}) \bibinfo{pages}{845--857}.
\bibitem[{Chaidaroon and Fang(2017)}]{chaidaroon2017variational}
\bibinfo{author}{S.~Chaidaroon}, \bibinfo{author}{Y.~Fang},
\newblock \bibinfo{title}{Variational deep semantic hashing for text documents},
\newblock in: \bibinfo{booktitle}{The 40th International ACM SIGIR Conference on Research and Development in Information Retrieval}, \bibinfo{year}{2017}, pp. \bibinfo{pages}{75--84}.
\bibitem[{Chaidaroon et~al.(2018)Chaidaroon, Ebesu, and Fang}]{chaidaroon2018deep}
\bibinfo{author}{S.~Chaidaroon}, \bibinfo{author}{T.~Ebesu}, \bibinfo{author}{Y.~Fang},
\newblock \bibinfo{title}{Deep semantic text hashing with weak supervision},
\newblock in: \bibinfo{booktitle}{The 41st international acm sigir Conference on Research and Development in Information Retrieval}, \bibinfo{year}{2018}, pp. \bibinfo{pages}{1109--1112}.
\bibitem[{Doan and Reddy(2020)}]{doan2020efficient}
\bibinfo{author}{K.~D. Doan}, \bibinfo{author}{C.~K. Reddy},
\newblock \bibinfo{title}{Efficient implicit unsupervised text hashing using adversarial autoencoder},
\newblock in: \bibinfo{booktitle}{The Web Conference}, \bibinfo{year}{2020}, pp. \bibinfo{pages}{684--694}.
\bibitem[{He et~al.(2023)He, Huang, Chen, Liu, Tong, Wang, Lian, and Wang}]{he2023efficient}
\bibinfo{author}{L.~He}, \bibinfo{author}{Z.~Huang}, \bibinfo{author}{E.~Chen}, \bibinfo{author}{Q.~Liu}, \bibinfo{author}{S.~Tong}, \bibinfo{author}{H.~Wang}, \bibinfo{author}{D.~Lian}, \bibinfo{author}{S.~Wang},
\newblock \bibinfo{title}{An efficient and robust semantic hashing framework for similar text search},
\newblock \bibinfo{journal}{ACM Transactions on Information Systems} \bibinfo{volume}{41} (\bibinfo{year}{2023}) \bibinfo{pages}{1--31}.
\bibitem[{Li et~al.(2021)Li, Liu, Yang, Peng, and Zhou}]{li2021survey}
\bibinfo{author}{Z.~Li}, \bibinfo{author}{F.~Liu}, \bibinfo{author}{W.~Yang}, \bibinfo{author}{S.~Peng}, \bibinfo{author}{J.~Zhou},
\newblock \bibinfo{title}{A survey of convolutional neural networks: analysis, applications, and prospects},
\newblock \bibinfo{journal}{IEEE Transactions on Neural Networks and Learning Systems} \bibinfo{volume}{33} (\bibinfo{year}{2021}) \bibinfo{pages}{6999--7019}.
\bibitem[{Xia et~al.(2014)Xia, Pan, Lai, Liu, and Yan}]{xia2014supervised}
\bibinfo{author}{R.~Xia}, \bibinfo{author}{Y.~Pan}, \bibinfo{author}{H.~Lai}, \bibinfo{author}{C.~Liu}, \bibinfo{author}{S.~Yan},
\newblock \bibinfo{title}{Supervised hashing for image retrieval via image representation learning},
\newblock in: \bibinfo{booktitle}{The AAAI Conference on Artificial Intelligence}, volume~\bibinfo{volume}{28}, \bibinfo{year}{2014}.
\bibitem[{Liu et~al.(2016)Liu, Wang, Shan, and Chen}]{liu2016deep}
\bibinfo{author}{H.~Liu}, \bibinfo{author}{R.~Wang}, \bibinfo{author}{S.~Shan}, \bibinfo{author}{X.~Chen},
\newblock \bibinfo{title}{Deep supervised hashing for fast image retrieval},
\newblock in: \bibinfo{booktitle}{The IEEE Conference on Computer Vision and Pattern Recognition}, \bibinfo{year}{2016}, pp. \bibinfo{pages}{2064--2072}.
\bibitem[{Jiang and Pang(2018)}]{jiang2018perceptual}
\bibinfo{author}{C.~Jiang}, \bibinfo{author}{Y.~Pang},
\newblock \bibinfo{title}{Perceptual image hashing based on a deep convolution neural network for content authentication},
\newblock \bibinfo{journal}{Journal of Electronic Imaging} \bibinfo{volume}{27} (\bibinfo{year}{2018}) \bibinfo{pages}{043055--043055}.
\bibitem[{Qin et~al.(2020)Qin, Liu, Feng, and Zhang}]{qin2020perceptual}
\bibinfo{author}{C.~Qin}, \bibinfo{author}{E.~Liu}, \bibinfo{author}{G.~Feng}, \bibinfo{author}{X.~Zhang},
\newblock \bibinfo{title}{Perceptual image hashing for content authentication based on convolutional neural network with multiple constraints},
\newblock \bibinfo{journal}{IEEE Transactions on Circuits and Systems for Video Technology} \bibinfo{volume}{31} (\bibinfo{year}{2020}) \bibinfo{pages}{4523--4537}.
\bibitem[{Zhou et~al.(2023)Zhou, Li, Fang, and Qin}]{zhou2023perceptual}
\bibinfo{author}{Y.~Zhou}, \bibinfo{author}{X.~Li}, \bibinfo{author}{Y.~Fang}, \bibinfo{author}{C.~Qin},
\newblock \bibinfo{title}{When perceptual authentication hashing meets neural architecture search},
\newblock in: \bibinfo{booktitle}{The 31st ACM International Conference on Multimedia}, \bibinfo{year}{2023}, pp. \bibinfo{pages}{8975--8983}.
\bibitem[{Zhaoxiong et~al.(2019)Zhaoxiong, Tetsuya, Sumiko, and Hirotsugu}]{zhaoxiong2019perceptual}
\bibinfo{author}{M.~Zhaoxiong}, \bibinfo{author}{M.~Tetsuya}, \bibinfo{author}{M.~Sumiko}, \bibinfo{author}{K.~Hirotsugu},
\newblock \bibinfo{title}{Perceptual hashing based on machine learning for blockchain and digital watermarking},
\newblock in: \bibinfo{booktitle}{Third World Conference on Smart Trends in Systems Security and Sustainablity}, \bibinfo{organization}{IEEE}, \bibinfo{year}{2019}, pp. \bibinfo{pages}{193--198}.
\bibitem[{Yusei et~al.(2023)Yusei, Zhaoxiong, Tetsuya, Sumiko, Kaito, and Hirotsugu}]{yusei2023cnn}
\bibinfo{author}{S.~Yusei}, \bibinfo{author}{M.~Zhaoxiong}, \bibinfo{author}{M.~Tetsuya}, \bibinfo{author}{M.~Sumiko}, \bibinfo{author}{H.~Kaito}, \bibinfo{author}{K.~Hirotsugu},
\newblock \bibinfo{title}{{CNN}-based perceptual hashing scheme for image groups suitable for security systems},
\newblock in: \bibinfo{booktitle}{IEEE 47th Annual Computers, Software, and Applications Conference}, \bibinfo{organization}{IEEE}, \bibinfo{year}{2023}, pp. \bibinfo{pages}{1231--1236}.
\bibitem[{Zhai et~al.(2018)Zhai, Zhang, Chen, and He}]{zhai2018autoencoder}
\bibinfo{author}{J.~Zhai}, \bibinfo{author}{S.~Zhang}, \bibinfo{author}{J.~Chen}, \bibinfo{author}{Q.~He},
\newblock \bibinfo{title}{Autoencoder and its various variants},
\newblock in: \bibinfo{booktitle}{IEEE International Conference on Systems, Man, and Cybernetics}, \bibinfo{organization}{IEEE}, \bibinfo{year}{2018}, pp. \bibinfo{pages}{415--419}.
\bibitem[{Paul et~al.(2021)Paul, Thakuria, Karsh, and Talukdar}]{paul2021robust}
\bibinfo{author}{M.~Paul}, \bibinfo{author}{A.~J. Thakuria}, \bibinfo{author}{R.~K. Karsh}, \bibinfo{author}{F.~A. Talukdar},
\newblock \bibinfo{title}{Robust color image hashing using convolutional stacked denoising auto-encoders for image authentication},
\newblock \bibinfo{journal}{Neural Computing and Applications} \bibinfo{volume}{33} (\bibinfo{year}{2021}) \bibinfo{pages}{13317--13331}.
\bibitem[{Borji(2019)}]{borji2019pros}
\bibinfo{author}{A.~Borji},
\newblock \bibinfo{title}{Pros and cons of gan evaluation measures},
\newblock \bibinfo{journal}{Computer Vision and Image Understanding} \bibinfo{volume}{179} (\bibinfo{year}{2019}) \bibinfo{pages}{41--65}.
\bibitem[{Yarlagadda et~al.(2018)Yarlagadda, G{\"u}era, Bestagini, Zhu, Tubaro, and Delp}]{yarlagadda2018satellite}
\bibinfo{author}{S.~K. Yarlagadda}, \bibinfo{author}{D.~G{\"u}era}, \bibinfo{author}{P.~Bestagini}, \bibinfo{author}{F.~M. Zhu}, \bibinfo{author}{S.~Tubaro}, \bibinfo{author}{E.~J. Delp},
\newblock \bibinfo{title}{Satellite image forgery detection and localization using {GAN} and one-class classifier},
\newblock \bibinfo{journal}{Electronic Imaging} \bibinfo{volume}{30} (\bibinfo{year}{2018}) \bibinfo{pages}{1--9}.
\bibitem[{Wang et~al.(2019)Wang, Zhang, Nie, Li, Chen, and Wang}]{wang2019wegan}
\bibinfo{author}{Y.~Wang}, \bibinfo{author}{L.~Zhang}, \bibinfo{author}{F.~Nie}, \bibinfo{author}{X.~Li}, \bibinfo{author}{Z.~Chen}, \bibinfo{author}{F.~Wang},
\newblock \bibinfo{title}{{WeGAN}: Deep image hashing with weighted generative adversarial networks},
\newblock \bibinfo{journal}{IEEE Transactions on Multimedia} \bibinfo{volume}{22} (\bibinfo{year}{2019}) \bibinfo{pages}{1458--1469}.
\bibitem[{Jin et~al.(2019)Jin, Zhang, and Lu}]{jin2019deep}
\bibinfo{author}{G.~Jin}, \bibinfo{author}{Y.~Zhang}, \bibinfo{author}{K.~Lu},
\newblock \bibinfo{title}{Deep hashing based on {VAE-GAN} for efficient similarity retrieval},
\newblock \bibinfo{journal}{Chinese Journal of Electronics} \bibinfo{volume}{28} (\bibinfo{year}{2019}) \bibinfo{pages}{1191--1197}.
\bibitem[{Bin et~al.(2023)Bin, Yi-Li, Jian, Chun-Peng, Jian, and Lin-Na}]{bin2023image}
\bibinfo{author}{M.~Bin}, \bibinfo{author}{W.~Yi-Li}, \bibinfo{author}{X.~Jian}, \bibinfo{author}{W.~Chun-Peng}, \bibinfo{author}{L.~Jian}, \bibinfo{author}{Z.~Lin-Na},
\newblock \bibinfo{title}{An image perceptual hash algorithm based on bidirectional generative adversarial network},
\newblock \bibinfo{journal}{Acta Electronica Sinica} \bibinfo{volume}{51} (\bibinfo{year}{2023}) \bibinfo{pages}{1405--1412}.
\bibitem[{Kordopatis-Zilos et~al.(2017)Kordopatis-Zilos, Papadopoulos, Patras, and Kompatsiaris}]{kordopatis2017near}
\bibinfo{author}{G.~Kordopatis-Zilos}, \bibinfo{author}{S.~Papadopoulos}, \bibinfo{author}{I.~Patras}, \bibinfo{author}{Y.~Kompatsiaris},
\newblock \bibinfo{title}{Near-duplicate video retrieval by aggregating intermediate {CNN} layers},
\newblock in: \bibinfo{booktitle}{23rd International Conference on MultiMedia Modeling}, \bibinfo{organization}{Springer}, \bibinfo{year}{2017}, pp. \bibinfo{pages}{251--263}.
\bibitem[{Li et~al.(2021)Li, Guo, Yang, and Xu}]{li2021video}
\bibinfo{author}{X.~Li}, \bibinfo{author}{C.~Guo}, \bibinfo{author}{Y.~Yang}, \bibinfo{author}{L.~Xu},
\newblock \bibinfo{title}{Video fingerprinting based on quadruplet convolutional neural network},
\newblock \bibinfo{journal}{Systems Science \& Control Engineering} \bibinfo{volume}{9} (\bibinfo{year}{2021}) \bibinfo{pages}{131--141}.
\bibitem[{Li et~al.(2020)Li, Zhang, Wan, and Sun}]{li2020two}
\bibinfo{author}{J.~Li}, \bibinfo{author}{H.~Zhang}, \bibinfo{author}{W.~Wan}, \bibinfo{author}{J.~Sun},
\newblock \bibinfo{title}{Two-class {3D-CNN} classifiers combination for video copy detection},
\newblock \bibinfo{journal}{Multimedia Tools and Applications} \bibinfo{volume}{79} (\bibinfo{year}{2020}) \bibinfo{pages}{4749--4761}.
\bibitem[{Muhammad et~al.(2018)Muhammad, Wang, Chattha, and Ali}]{muhammad2018pre}
\bibinfo{author}{U.~Muhammad}, \bibinfo{author}{W.~Wang}, \bibinfo{author}{S.~P. Chattha}, \bibinfo{author}{S.~Ali},
\newblock \bibinfo{title}{Pre-trained vggnet architecture for remote-sensing image scene classification},
\newblock in: \bibinfo{booktitle}{24th International Conference on Pattern Recognition}, \bibinfo{organization}{IEEE}, \bibinfo{year}{2018}, pp. \bibinfo{pages}{1622--1627}.
\bibitem[{Zhang et~al.(2018)Zhang, Xie, Luan, He, Zhang, and Wu}]{zhang2018video}
\bibinfo{author}{X.~Zhang}, \bibinfo{author}{Y.~Xie}, \bibinfo{author}{X.~Luan}, \bibinfo{author}{J.~He}, \bibinfo{author}{L.~Zhang}, \bibinfo{author}{L.~Wu},
\newblock \bibinfo{title}{Video copy detection based on deep {CNN} features and graph-based sequence matching},
\newblock \bibinfo{journal}{Wireless Personal Communications} \bibinfo{volume}{103} (\bibinfo{year}{2018}) \bibinfo{pages}{401--416}.
\bibitem[{Zhong et~al.(2015)Zhong, Jin, and Xie}]{zhong2015high}
\bibinfo{author}{Z.~Zhong}, \bibinfo{author}{L.~Jin}, \bibinfo{author}{Z.~Xie},
\newblock \bibinfo{title}{High performance offline handwritten chinese character recognition using googlenet and directional feature maps},
\newblock in: \bibinfo{booktitle}{13th International Conference on Document Analysis and Recognition}, \bibinfo{organization}{IEEE}, \bibinfo{year}{2015}, pp. \bibinfo{pages}{846--850}.
\bibitem[{Lou et~al.(2017)Lou, Bai, Lin, Wang, Chen, Chandrasekhar, Duan, Huang, Kot, and Gao}]{lou2017compact}
\bibinfo{author}{Y.~Lou}, \bibinfo{author}{Y.~Bai}, \bibinfo{author}{J.~Lin}, \bibinfo{author}{S.~Wang}, \bibinfo{author}{J.~Chen}, \bibinfo{author}{V.~Chandrasekhar}, \bibinfo{author}{L.-Y. Duan}, \bibinfo{author}{T.~Huang}, \bibinfo{author}{A.~C. Kot}, \bibinfo{author}{W.~Gao},
\newblock \bibinfo{title}{Compact deep invariant descriptors for video retrieval},
\newblock in: \bibinfo{booktitle}{Data Compression Conference}, \bibinfo{organization}{IEEE}, \bibinfo{year}{2017}, pp. \bibinfo{pages}{420--429}.
\bibitem[{Mengyang et~al.(2018)Mengyang, Po, Chang, Yuen, Cheung, Peter, Luk, and Lau}]{mengyang2018content}
\bibinfo{author}{L.~Mengyang}, \bibinfo{author}{L.-M. Po}, \bibinfo{author}{Z.~Chang}, \bibinfo{author}{W.~Y. Yuen}, \bibinfo{author}{H.-K. Cheung}, \bibinfo{author}{H.~Peter}, \bibinfo{author}{H.-T. Luk}, \bibinfo{author}{K.-W. Lau},
\newblock \bibinfo{title}{Content-based video copy detection using binary object fingerprints},
\newblock in: \bibinfo{booktitle}{IEEE International Conference on Signal Processing, Communications and Computing}, \bibinfo{organization}{IEEE}, \bibinfo{year}{2018}, pp. \bibinfo{pages}{1--6}.
\bibitem[{Xinwei et~al.(2021)Xinwei, Lianghao, and Yi}]{xinwei2021compact}
\bibinfo{author}{L.~Xinwei}, \bibinfo{author}{X.~Lianghao}, \bibinfo{author}{Y.~Yi},
\newblock \bibinfo{title}{Compact video fingerprinting via an improved capsule net},
\newblock \bibinfo{journal}{Systems Science \& Control Engineering} \bibinfo{volume}{9} (\bibinfo{year}{2021}) \bibinfo{pages}{122--130}.
\bibitem[{Nie et~al.(2021)Nie, Zhou, Shi, Sun, and Yin}]{nie2021classification}
\bibinfo{author}{X.~Nie}, \bibinfo{author}{X.~Zhou}, \bibinfo{author}{Y.~Shi}, \bibinfo{author}{J.~Sun}, \bibinfo{author}{Y.~Yin},
\newblock \bibinfo{title}{Classification-enhancement deep hashing for large-scale video retrieval},
\newblock \bibinfo{journal}{Applied Soft Computing} \bibinfo{volume}{109} (\bibinfo{year}{2021}) \bibinfo{pages}{107467}.
\bibitem[{Anuranji and Srimathi(2020)}]{anuranji2020supervised}
\bibinfo{author}{R.~Anuranji}, \bibinfo{author}{H.~Srimathi},
\newblock \bibinfo{title}{A supervised deep convolutional based bidirectional long short term memory video hashing for large scale video retrieval applications},
\newblock \bibinfo{journal}{Digital Signal Processing} \bibinfo{volume}{102} (\bibinfo{year}{2020}) \bibinfo{pages}{102729}.
\bibitem[{Hu and Lu(2018)}]{hu2018learning}
\bibinfo{author}{Y.~Hu}, \bibinfo{author}{X.~Lu},
\newblock \bibinfo{title}{Learning spatial-temporal features for video copy detection by the combination of {CNN} and {RNN}},
\newblock \bibinfo{journal}{Journal of Visual Communication and Image Representation} \bibinfo{volume}{55} (\bibinfo{year}{2018}) \bibinfo{pages}{21--29}.
\bibitem[{Zhao et~al.(2023)Zhao, Li, Yao, and Qin}]{zhao2023tastnet}
\bibinfo{author}{G.~Zhao}, \bibinfo{author}{F.~Li}, \bibinfo{author}{H.~Yao}, \bibinfo{author}{C.~Qin},
\newblock \bibinfo{title}{{TASTNet}: An end-to-end deep fingerprinting net with two-dimensional attention mechanism and spatio-temporal weighted fusion for video content authentication},
\newblock \bibinfo{journal}{Journal of Visual Communication and Image Representation} \bibinfo{volume}{96} (\bibinfo{year}{2023}) \bibinfo{pages}{103913}.
\bibitem[{Zhang et~al.(2023)Zhang, Zhang, Wang, and Zhuo}]{zhang2023short}
\bibinfo{author}{S.~Zhang}, \bibinfo{author}{J.~Zhang}, \bibinfo{author}{Y.~Wang}, \bibinfo{author}{L.~Zhuo},
\newblock \bibinfo{title}{Short video fingerprint extraction: from audio--visual fingerprint fusion to multi-index hashing},
\newblock \bibinfo{journal}{Multimedia Systems} \bibinfo{volume}{29} (\bibinfo{year}{2023}) \bibinfo{pages}{981--1000}.
\bibitem[{B{\'a}ez-Su{\'a}rez et~al.(2020)B{\'a}ez-Su{\'a}rez, Shah, Nolazco-Flores, Huang, Gnawali, and Shi}]{baez2020samaf}
\bibinfo{author}{A.~B{\'a}ez-Su{\'a}rez}, \bibinfo{author}{N.~Shah}, \bibinfo{author}{J.~A. Nolazco-Flores}, \bibinfo{author}{S.-H.~S. Huang}, \bibinfo{author}{O.~Gnawali}, \bibinfo{author}{W.~Shi},
\newblock \bibinfo{title}{{SAMAF}: Sequence-to-sequence autoencoder model for audio fingerprinting},
\newblock \bibinfo{journal}{ACM Transactions on Multimedia Computing, Communications, and Applications} \bibinfo{volume}{16} (\bibinfo{year}{2020}) \bibinfo{pages}{1--23}.
\bibitem[{Saravanos et~al.(2020)Saravanos, Ampeliotis, and Berberidis}]{saravanos2020audio}
\bibinfo{author}{C.~Saravanos}, \bibinfo{author}{D.~Ampeliotis}, \bibinfo{author}{K.~Berberidis},
\newblock \bibinfo{title}{Audio-fingerprinting via dictionary learning},
\newblock in: \bibinfo{booktitle}{IEEE 22nd International Workshop on Multimedia Signal Processing}, \bibinfo{organization}{IEEE}, \bibinfo{year}{2020}, pp. \bibinfo{pages}{1--7}.
\bibitem[{Chang et~al.(2021)Chang, Lee, Park, Lim, Lee, Ko, and Han}]{chang2021neural}
\bibinfo{author}{S.~Chang}, \bibinfo{author}{D.~Lee}, \bibinfo{author}{J.~Park}, \bibinfo{author}{H.~Lim}, \bibinfo{author}{K.~Lee}, \bibinfo{author}{K.~Ko}, \bibinfo{author}{Y.~Han},
\newblock \bibinfo{title}{Neural audio fingerprint for high-specific audio retrieval based on contrastive learning},
\newblock in: \bibinfo{booktitle}{IEEE International Conference on Acoustics, Speech and Signal Processing}, \bibinfo{organization}{IEEE}, \bibinfo{year}{2021}, pp. \bibinfo{pages}{3025--3029}.
\bibitem[{Kamuni et~al.(2024)Kamuni, Chintala, Kunchakuri, Narasimharaju, and Kumar}]{kamuni2024advancing}
\bibinfo{author}{N.~Kamuni}, \bibinfo{author}{S.~Chintala}, \bibinfo{author}{N.~Kunchakuri}, \bibinfo{author}{J.~S.~A. Narasimharaju}, \bibinfo{author}{V.~Kumar},
\newblock \bibinfo{title}{Advancing audio fingerprinting accuracy with {AI} and {ML}: Addressing background noise and distortion challenges},
\newblock in: \bibinfo{booktitle}{IEEE 18th International Conference on Semantic Computing}, \bibinfo{organization}{IEEE}, \bibinfo{year}{2024}, pp. \bibinfo{pages}{341--345}.
\bibitem[{Deepsheka et~al.(2020)Deepsheka, Kheerthana, Mourina, and Bharathi}]{deepsheka2020recurrent}
\bibinfo{author}{G.~Deepsheka}, \bibinfo{author}{R.~Kheerthana}, \bibinfo{author}{M.~Mourina}, \bibinfo{author}{B.~Bharathi},
\newblock \bibinfo{title}{Recurrent neural network based music recognition using audio fingerprinting},
\newblock in: \bibinfo{booktitle}{Third International Conference on Smart Systems and Inventive Technology}, \bibinfo{organization}{IEEE}, \bibinfo{year}{2020}, pp. \bibinfo{pages}{1--6}.
\bibitem[{Wu and Wang(2022)}]{wu2022asymmetric}
\bibinfo{author}{X.~Wu}, \bibinfo{author}{H.~Wang},
\newblock \bibinfo{title}{Asymmetric contrastive learning for audio fingerprinting},
\newblock \bibinfo{journal}{IEEE Signal Processing Letters} \bibinfo{volume}{29} (\bibinfo{year}{2022}) \bibinfo{pages}{1873--1877}.
\bibitem[{Adrakatti et~al.(2016)Adrakatti, Wodeyar, and Mulla}]{adrakatti2016search}
\bibinfo{author}{A.~Adrakatti}, \bibinfo{author}{R.~Wodeyar}, \bibinfo{author}{K.~Mulla},
\newblock \bibinfo{title}{Search by image: a novel approach to content based image retrieval system},
\newblock \bibinfo{journal}{International Journal of Library Science} \bibinfo{volume}{14} (\bibinfo{year}{2016}) \bibinfo{pages}{41--47}.
\bibitem[{Kalker et~al.(2004)Kalker, Epema, Hartel, Lagendijk, and Van~Steen}]{kalker2004music2share}
\bibinfo{author}{T.~Kalker}, \bibinfo{author}{D.~H. Epema}, \bibinfo{author}{P.~H. Hartel}, \bibinfo{author}{R.~L. Lagendijk}, \bibinfo{author}{M.~Van~Steen},
\newblock \bibinfo{title}{Music2share-copyright-compliant music sharing in p2p systems},
\newblock \bibinfo{journal}{Proceedings of the IEEE} \bibinfo{volume}{92} (\bibinfo{year}{2004}) \bibinfo{pages}{961--970}.
\bibitem[{Doets and Lagendijk(2008)}]{doets2008distortion}
\bibinfo{author}{P.~J.~O. Doets}, \bibinfo{author}{R.~L. Lagendijk},
\newblock \bibinfo{title}{Distortion estimation in compressed music using only audio fingerprints},
\newblock \bibinfo{journal}{IEEE Transactions on Audio, Speech, and Language Processing} \bibinfo{volume}{16} (\bibinfo{year}{2008}) \bibinfo{pages}{302--317}.
\bibitem[{Amit et~al.(2021)Amit, Barua, and Kafy}]{amit2021countering}
\bibinfo{author}{S.~Amit}, \bibinfo{author}{L.~Barua}, \bibinfo{author}{A.-A. Kafy},
\newblock \bibinfo{title}{Countering violent extremism using social media and preventing implementable strategies for bangladesh},
\newblock \bibinfo{journal}{Heliyon} \bibinfo{volume}{7} (\bibinfo{year}{2021}).
\bibitem[{Struppek et~al.(2022)Struppek, Hintersdorf, Neider, and Kersting}]{struppek2022learning}
\bibinfo{author}{L.~Struppek}, \bibinfo{author}{D.~Hintersdorf}, \bibinfo{author}{D.~Neider}, \bibinfo{author}{K.~Kersting},
\newblock \bibinfo{title}{Learning to break deep perceptual hashing: The use case neuralhash},
\newblock in: \bibinfo{booktitle}{ACM Conference on Fairness, Accountability, and Transparency}, \bibinfo{year}{2022}, pp. \bibinfo{pages}{58--69}.
\bibitem[{Wu et~al.(2009)Wu, Ngo, Hauptmann, and Tan}]{wu2009real}
\bibinfo{author}{X.~Wu}, \bibinfo{author}{C.-W. Ngo}, \bibinfo{author}{A.~G. Hauptmann}, \bibinfo{author}{H.-K. Tan},
\newblock \bibinfo{title}{Real-time near-duplicate elimination for web video search with content and context},
\newblock \bibinfo{journal}{IEEE Transactions on Multimedia} \bibinfo{volume}{11} (\bibinfo{year}{2009}) \bibinfo{pages}{196--207}.
\bibitem[{Heintze et~al.(1996)}]{heintze1996scalable}
\bibinfo{author}{N.~Heintze}, et~al.,
\newblock \bibinfo{title}{Scalable document fingerprinting},
\newblock in: \bibinfo{booktitle}{USENIX Workshop on Electronic Commerce}, volume~\bibinfo{volume}{3}, \bibinfo{organization}{Citeseer}, \bibinfo{year}{1996}.
\bibitem[{Broder et~al.(1997)Broder, Glassman, Manasse, and Zweig}]{broder1997syntactic}
\bibinfo{author}{A.~Z. Broder}, \bibinfo{author}{S.~C. Glassman}, \bibinfo{author}{M.~S. Manasse}, \bibinfo{author}{G.~Zweig},
\newblock \bibinfo{title}{Syntactic clustering of the web},
\newblock \bibinfo{journal}{Computer Networks and ISDN Systems} \bibinfo{volume}{29} (\bibinfo{year}{1997}) \bibinfo{pages}{1157--1166}.
\bibitem[{Monostori et~al.(2001{\natexlab{a}})Monostori, Zaslavsky, and Schmidt}]{monostori2001parallel}
\bibinfo{author}{K.~Monostori}, \bibinfo{author}{A.~Zaslavsky}, \bibinfo{author}{H.~Schmidt},
\newblock \bibinfo{title}{Parallel and distributed document overlap detection on the web},
\newblock in: \bibinfo{booktitle}{5th International Workshop on Applied Parallel Computing}, \bibinfo{organization}{Springer}, \bibinfo{year}{2001}{\natexlab{a}}, pp. \bibinfo{pages}{206--214}.
\bibitem[{Monostori et~al.(2001{\natexlab{b}})Monostori, Zaslavsky, and Vajk}]{monostori2001suffix}
\bibinfo{author}{K.~Monostori}, \bibinfo{author}{A.~Zaslavsky}, \bibinfo{author}{I.~Vajk},
\newblock \bibinfo{title}{Suffix vector: A space-efficient suffix tree representation},
\newblock in: \bibinfo{booktitle}{12th International Symposium on Algorithms and Computation}, \bibinfo{organization}{Springer}, \bibinfo{year}{2001}{\natexlab{b}}, pp. \bibinfo{pages}{707--718}.
\bibitem[{Chowdhury et~al.(2002)Chowdhury, Frieder, Grossman, and McCabe}]{chowdhury2002collection}
\bibinfo{author}{A.~Chowdhury}, \bibinfo{author}{O.~Frieder}, \bibinfo{author}{D.~Grossman}, \bibinfo{author}{M.~C. McCabe},
\newblock \bibinfo{title}{Collection statistics for fast duplicate document detection},
\newblock \bibinfo{journal}{ACM Transactions on Information Systems} \bibinfo{volume}{20} (\bibinfo{year}{2002}) \bibinfo{pages}{171--191}.
\bibitem[{Katiyar and Weissman(2011)}]{katiyar2011videdup}
\bibinfo{author}{A.~Katiyar}, \bibinfo{author}{J.~Weissman},
\newblock \bibinfo{title}{{ViDeDup}: An application-aware framework for video de-duplication},
\newblock in: \bibinfo{booktitle}{3rd Workshop on Hot Topics in Storage and File Systems}, \bibinfo{year}{2011}.
\bibitem[{Rajaraman and Ullman(2011)}]{rajaraman2011mining}
\bibinfo{author}{A.~Rajaraman}, \bibinfo{author}{J.~D. Ullman}, \bibinfo{title}{Mining of massive datasets}, \bibinfo{publisher}{Autoedicion}, \bibinfo{year}{2011}.
\bibitem[{Geierhaas et~al.(2023)Geierhaas, Otto, H{\"a}ring, and Smith}]{geierhaas2023attitudes}
\bibinfo{author}{L.~Geierhaas}, \bibinfo{author}{F.~Otto}, \bibinfo{author}{M.~H{\"a}ring}, \bibinfo{author}{M.~Smith},
\newblock \bibinfo{title}{Attitudes towards client-side scanning for csam, terrorism, drug trafficking, drug use and tax evasion in germany},
\newblock in: \bibinfo{booktitle}{IEEE Symposium on Security and Privacy}, \bibinfo{organization}{IEEE}, \bibinfo{year}{2023}, pp. \bibinfo{pages}{217--233}.
\bibitem[{Jain et~al.(2022)Jain, Crețu, and de~Montjoye}]{jain2022adversarial}
\bibinfo{author}{S.~Jain}, \bibinfo{author}{A.-M. Crețu}, \bibinfo{author}{Y.-A. de~Montjoye},
\newblock \bibinfo{title}{Adversarial detection avoidance attacks: Evaluating the robustness of perceptual hashing-based client-side scanning},
\newblock in: \bibinfo{booktitle}{31st USENIX Security Symposium}, \bibinfo{year}{2022}, pp. \bibinfo{pages}{2317--2334}.
\bibitem[{Gan et~al.(2023)Gan, Wan, and Yu}]{gan2023model}
\bibinfo{author}{W.~Gan}, \bibinfo{author}{S.~Wan}, \bibinfo{author}{P.~Yu},
\newblock \bibinfo{title}{Model-as-a-service ({MaaS}): A survey},
\newblock in: \bibinfo{booktitle}{IEEE International Conference on Big Data}, \bibinfo{organization}{IEEE}, \bibinfo{year}{2023}, pp. \bibinfo{pages}{4636--4645}.
\bibitem[{Zeng et~al.(2023)Zeng, Gan, Wang, and Yu}]{zeng2023distributed}
\bibinfo{author}{F.~Zeng}, \bibinfo{author}{W.~Gan}, \bibinfo{author}{Y.~Wang}, \bibinfo{author}{P.~S. Yu},
\newblock \bibinfo{title}{Distributed training of large language models},
\newblock in: \bibinfo{booktitle}{IEEE 29th International Conference on Parallel and Distributed Systems}, \bibinfo{organization}{IEEE}, \bibinfo{year}{2023}, pp. \bibinfo{pages}{840--847}.
\bibitem[{Gan et~al.(2023)Gan, Qi, Wu, and Lin}]{gan2023large}
\bibinfo{author}{W.~Gan}, \bibinfo{author}{Z.~Qi}, \bibinfo{author}{J.~Wu}, \bibinfo{author}{J.~C.-W. Lin},
\newblock \bibinfo{title}{Large language models in education: Vision and opportunities},
\newblock in: \bibinfo{booktitle}{IEEE international conference on big data}, \bibinfo{organization}{IEEE}, \bibinfo{year}{2023}, pp. \bibinfo{pages}{4776--4785}.
\bibitem[{Zeng et~al.(2023)Zeng, Gan, Wang, Liu, and Yu}]{zeng2023large}
\bibinfo{author}{F.~Zeng}, \bibinfo{author}{W.~Gan}, \bibinfo{author}{Y.~Wang}, \bibinfo{author}{N.~Liu}, \bibinfo{author}{P.~S. Yu},
\newblock \bibinfo{title}{Large language models for robotics: A survey},
\newblock \bibinfo{journal}{arXiv preprint arXiv:2311.07226}  (\bibinfo{year}{2023}).
\bibitem[{Lai et~al.(2023)Lai, Gan, Wu, Qi, and Yu}]{lai2023large}
\bibinfo{author}{J.~Lai}, \bibinfo{author}{W.~Gan}, \bibinfo{author}{J.~Wu}, \bibinfo{author}{Z.~Qi}, \bibinfo{author}{P.~S. Yu},
\newblock \bibinfo{title}{Large language models in law: A survey},
\newblock \bibinfo{journal}{arXiv preprint arXiv:2312.03718}  (\bibinfo{year}{2023}).
\bibitem[{Vaswani(2017)}]{vaswani2017attention}
\bibinfo{author}{A.~Vaswani},
\newblock \bibinfo{title}{Attention is all you need},
\newblock \bibinfo{journal}{arXiv preprint arXiv:1706.03762}  (\bibinfo{year}{2017}).
\bibitem[{Yuksel et~al.(2012)Yuksel, Wilson, and Gader}]{yuksel2012twenty}
\bibinfo{author}{S.~E. Yuksel}, \bibinfo{author}{J.~N. Wilson}, \bibinfo{author}{P.~D. Gader},
\newblock \bibinfo{title}{Twenty years of mixture of experts},
\newblock \bibinfo{journal}{IEEE transactions on neural networks and learning systems} \bibinfo{volume}{23} (\bibinfo{year}{2012}) \bibinfo{pages}{1177--1193}.
\bibitem[{McGovern et~al.(2024)McGovern, Stureborg, Suhara, and Alikaniotis}]{mcgovern2024your}
\bibinfo{author}{H.~McGovern}, \bibinfo{author}{R.~Stureborg}, \bibinfo{author}{Y.~Suhara}, \bibinfo{author}{D.~Alikaniotis},
\newblock \bibinfo{title}{Your large language models are leaving fingerprints},
\newblock \bibinfo{journal}{arXiv preprint arXiv:2405.14057}  (\bibinfo{year}{2024}).
\bibitem[{Russinovich and Salem(2024)}]{russinovich2024hey}
\bibinfo{author}{M.~Russinovich}, \bibinfo{author}{A.~Salem},
\newblock \bibinfo{title}{Hey, that's my model! introducing chain \& hash, an llm fingerprinting technique},
\newblock \bibinfo{journal}{arXiv preprint arXiv:2407.10887}  (\bibinfo{year}{2024}).
\bibitem[{Jin et~al.(2024)Jin, Zhang, Shi, Lou, and Hou}]{jin2024proflingo}
\bibinfo{author}{H.~Jin}, \bibinfo{author}{C.~Zhang}, \bibinfo{author}{S.~Shi}, \bibinfo{author}{W.~Lou}, \bibinfo{author}{Y.~T. Hou},
\newblock \bibinfo{title}{{ProFLingo}: A fingerprinting-based copyright protection scheme for large language models},
\newblock \bibinfo{journal}{arXiv preprint arXiv:2405.02466}  (\bibinfo{year}{2024}).
\bibitem[{Pasquini et~al.(2024)Pasquini, Kornaropoulos, and Ateniese}]{pasquini2024llmmap}
\bibinfo{author}{D.~Pasquini}, \bibinfo{author}{E.~M. Kornaropoulos}, \bibinfo{author}{G.~Ateniese},
\newblock \bibinfo{title}{{LLMmap}: Fingerprinting for large language models},
\newblock \bibinfo{journal}{arXiv preprint arXiv:2407.15847}  (\bibinfo{year}{2024}).
\bibitem[{Singh and Gupta(2022)}]{singh2022learning}
\bibinfo{author}{A.~Singh}, \bibinfo{author}{S.~Gupta},
\newblock \bibinfo{title}{Learning to hash: a comprehensive survey of deep learning-based hashing methods},
\newblock \bibinfo{journal}{Knowledge and Information Systems} \bibinfo{volume}{64} (\bibinfo{year}{2022}) \bibinfo{pages}{2565--2597}.
\bibitem[{Lin et~al.(2020)Lin, Xu, Liu, and Zhang}]{lin2020composite}
\bibinfo{author}{J.~Lin}, \bibinfo{author}{L.~Xu}, \bibinfo{author}{Y.~Liu}, \bibinfo{author}{X.~Zhang},
\newblock \bibinfo{title}{Composite backdoor attack for deep neural network by mixing existing benign features},
\newblock in: \bibinfo{booktitle}{Proceedings of the ACM SIGSAC Conference on Computer and Communications Security}, \bibinfo{year}{2020}, pp. \bibinfo{pages}{113--131}.
\bibitem[{Goodfellow et~al.(2015)Goodfellow, Shlens, and Szegedy}]{goodfellow2015explaining}
\bibinfo{author}{I.~J. Goodfellow}, \bibinfo{author}{J.~Shlens}, \bibinfo{author}{C.~Szegedy},
\newblock \bibinfo{title}{Explaining and harnessing adversarial examples},
\newblock \bibinfo{journal}{stat} \bibinfo{volume}{1050} (\bibinfo{year}{2015}) \bibinfo{pages}{20}.
\bibitem[{Prokos et~al.(2023)Prokos, Fendley, Green, Schuster, Tromer, Jois, and Cao}]{prokos2023squint}
\bibinfo{author}{J.~Prokos}, \bibinfo{author}{N.~Fendley}, \bibinfo{author}{M.~Green}, \bibinfo{author}{R.~Schuster}, \bibinfo{author}{E.~Tromer}, \bibinfo{author}{T.~Jois}, \bibinfo{author}{Y.~Cao},
\newblock \bibinfo{title}{Squint hard enough: Attacking perceptual hashing with adversarial machine learning},
\newblock in: \bibinfo{booktitle}{32nd USENIX Security Symposium}, \bibinfo{year}{2023}, pp. \bibinfo{pages}{211--228}.
\bibitem[{Madden et~al.(2024)Madden, Bhavsar, Dorje, and Li}]{madden2024assessing}
\bibinfo{author}{J.~Madden}, \bibinfo{author}{M.~Bhavsar}, \bibinfo{author}{L.~Dorje}, \bibinfo{author}{X.~Li},
\newblock \bibinfo{title}{Assessing the adversarial security of perceptual hashing algorithms},
\newblock \bibinfo{journal}{arXiv preprint arXiv:2406.00918}  (\bibinfo{year}{2024}).
\bibitem[{Hao et~al.(2021)Hao, Luo, Jan, and Wang}]{hao2021s}
\bibinfo{author}{Q.~Hao}, \bibinfo{author}{L.~Luo}, \bibinfo{author}{S.~T. Jan}, \bibinfo{author}{G.~Wang},
\newblock \bibinfo{title}{It's not what it looks like: Manipulating perceptual hashing based applications},
\newblock in: \bibinfo{booktitle}{ACM SIGSAC Conference on Computer and Communications Security}, \bibinfo{year}{2021}, pp. \bibinfo{pages}{69--85}.
\bibitem[{Twenning et~al.(2023)Twenning, Baier, and G{\"o}bel}]{twenning2023using}
\bibinfo{author}{L.~Twenning}, \bibinfo{author}{H.~Baier}, \bibinfo{author}{T.~G{\"o}bel},
\newblock \bibinfo{title}{Using perceptual hashing for targeted content scanning},
\newblock in: \bibinfo{booktitle}{IFIP International Conference on Digital Forensics}, \bibinfo{organization}{Springer}, \bibinfo{year}{2023}, pp. \bibinfo{pages}{125--142}.
\bibitem[{Hamadouche et~al.(2023)Hamadouche, Zebbiche, and Zafoune}]{hamadouche2023securing}
\bibinfo{author}{M.~Hamadouche}, \bibinfo{author}{K.~Zebbiche}, \bibinfo{author}{Y.~Zafoune},
\newblock \bibinfo{title}{Securing biometric systems by using perceptual hashing techniques},
\newblock in: \bibinfo{booktitle}{20th ACS/IEEE International Conference on Computer Systems and Applications}, \bibinfo{organization}{IEEE}, \bibinfo{year}{2023}, pp. \bibinfo{pages}{1--4}.
\bibitem[{Hamadouche et~al.(2024)Hamadouche, Khalil, Tebbi, Guerroumi, and Zafoune}]{hamadouche2024replay}
\bibinfo{author}{M.~Hamadouche}, \bibinfo{author}{Z.~Khalil}, \bibinfo{author}{H.~Tebbi}, \bibinfo{author}{M.~Guerroumi}, \bibinfo{author}{Y.~Zafoune},
\newblock \bibinfo{title}{A replay attack detection scheme based on perceptual image hashing},
\newblock \bibinfo{journal}{Multimedia Tools and Applications} \bibinfo{volume}{83} (\bibinfo{year}{2024}) \bibinfo{pages}{8999--9031}.
\bibitem[{Regazzoni et~al.(2021)Regazzoni, Palmieri, Smailbegovic, Cammarota, and Polian}]{regazzoni2021protecting}
\bibinfo{author}{F.~Regazzoni}, \bibinfo{author}{P.~Palmieri}, \bibinfo{author}{F.~Smailbegovic}, \bibinfo{author}{R.~Cammarota}, \bibinfo{author}{I.~Polian},
\newblock \bibinfo{title}{Protecting artificial intelligence ips: a survey of watermarking and fingerprinting for machine learning},
\newblock \bibinfo{journal}{CAAI Transactions on Intelligence Technology} \bibinfo{volume}{6} (\bibinfo{year}{2021}) \bibinfo{pages}{180--191}.
\bibitem[{Chen et~al.(2023)Chen, Zhou, Zhang, Chen, Zhang, Chen, Hua, and Yu}]{chen2023perceptual}
\bibinfo{author}{H.~Chen}, \bibinfo{author}{H.~Zhou}, \bibinfo{author}{J.~Zhang}, \bibinfo{author}{D.~Chen}, \bibinfo{author}{W.~Zhang}, \bibinfo{author}{K.~Chen}, \bibinfo{author}{G.~Hua}, \bibinfo{author}{N.~Yu},
\newblock \bibinfo{title}{Perceptual hashing of deep convolutional neural networks for model copy detection},
\newblock \bibinfo{journal}{ACM Transactions on Multimedia Computing, Communications and Applications} \bibinfo{volume}{19} (\bibinfo{year}{2023}) \bibinfo{pages}{1--20}.
\bibitem[{Zhao et~al.(2020)Zhao, Hu, Liu, Ma, Chen, and Hassan}]{zhao2020afa}
\bibinfo{author}{J.~Zhao}, \bibinfo{author}{Q.~Hu}, \bibinfo{author}{G.~Liu}, \bibinfo{author}{X.~Ma}, \bibinfo{author}{F.~Chen}, \bibinfo{author}{M.~M. Hassan},
\newblock \bibinfo{title}{{AFA}: Adversarial fingerprinting authentication for deep neural networks},
\newblock \bibinfo{journal}{Computer Communications} \bibinfo{volume}{150} (\bibinfo{year}{2020}) \bibinfo{pages}{488--497}.
\bibitem[{Prummer et~al.(2024)Prummer, Regnath, and Kosch}]{prummer2024onion}
\bibinfo{author}{M.~Prummer}, \bibinfo{author}{E.~Regnath}, \bibinfo{author}{H.~Kosch},
\newblock \bibinfo{title}{{Onion-Hash}: A compact and robust {3D} perceptual hash for asset authentication},
\newblock \bibinfo{journal}{Computer-Aided Design}  (\bibinfo{year}{2024}) \bibinfo{pages}{103752}.

\end{thebibliography}


\end{document}